\documentclass[prb,
twocolumn,
nopacs,
nofootinbib,
superscriptaddress]{revtex4-2}
\usepackage{comment}
\usepackage{pifont}
\usepackage{amsmath}
\usepackage{amssymb}
\usepackage{graphicx}
\usepackage{braket}
\usepackage[colorlinks,
    linkcolor=blue,
    anchorcolor=red,
    citecolor=gray
]{hyperref}
\usepackage{xcolor}

\renewcommand{\v}[1]{{\mathbf{#1}}}
\newcommand{\tr}[0]{{\text{tr}}}

\newcommand\ptwiddle[1]{\mathord{\mathop{#1}\limits^{\scriptscriptstyle(\sim)}}}

\begin{document}

\title{Topological Floquet Green's function zeros}
\author{Elio J. K\"onig}
\affiliation{
Department of Physics, University of Wisconsin-Madison, Madison, Wisconsin 53706, USA
}
\author{Aditi Mitra}
\affiliation{Center for Quantum Phenomena, Department of Physics,
New York University, 726 Broadway, New York, New York, 10003, USA}

\begin{abstract}
Motivated by recent advances in digital quantum emulation using noisy intermediate-scale quantum (NISQ) devices and an increased interest in topological Green's function zeros in condensed matter systems, we here study Green's function zeros in topological Floquet systems. We concentrate on interacting Kitaev-like Floquet chains (or equivalently transverse field Ising circuits) and introduce Floquet Green's-function-based topological invariants for the corresponding symmetry class BDI. In the vicinity of special points in the free fermion phase diagram and using tailor-made interactions which lead to the Floquet version of symmetric mass generation, we analytically calculate both edge and bulk Green's functions. Just as in the case of continuum time evolution, topological bands of Green's function zeros may also contribute to the topological invariant. However, contrary to the case of continuum time evolution, Floquet Green's functions can have zeros even in the absence of interactions. 
Finally, we also discuss an implementation of this Floquet system in a digital quantum emulator: We present a circuit which encodes the interaction under consideration and pinpoint the observables carrying information about the topological Green's function boundary zeros. 
\end{abstract}

\maketitle

\section{Introduction}

Noisy intermediate-scale quantum (NISQ)~\cite{Preskill2018} devices open up an exciting route for highly designable quantum systems. Besides analogue emulators, gate-based quantum simulations with up to $\sim$ 100 qubits are nowadays realized and offer a promising route to a better understanding of the quantum-many-body problem and its dynamics~\cite{MorvanRoushan2022,RosenbergRoushan2024,EcksteinHolmes2024}. Thereby, quantum information science is on the verge of producing first results of major impact in materials science, solid state physics and quantum chemistry. 

The latter areas of research represent instances of strongly correlated electronic systems, where the energetically low-lying excitations behave in a fundamentally different way than the underlying electrons and, at the same time, the quantum many-body ground state is fundamentally more complex than simple Slater-determinant free fermion states~\cite{LiuVerstraete2007,GiorgadzeVayrynen2025}.  
This should be contrasted to weakly correlated electronic systems, e.g.~Landau Fermi liquids~\cite{Landau1957}, in which energetically low-lying degrees of freedom carry the same quantum numbers (spin, charge, statistical angle) as the microscopic electronic constituents, and the ground state is adiabatically connected to the Slater state.

As practical means to computationally account for strong electron-electron correlations, field-theoretical methods~\cite{AGD} are a prime tool, both in analytical calculations and in numerics (e.g.~in dynamical mean-field theory and diagrammatic Monte Carlo). A central role is played by the two-point correlator of fermions, e.g. the retarded and advanced Green's functions (denoted by roman type R/A)
\begin{equation} \label{eq:GreenDef}
    G_{\rm R/A}^{AB} (t) = \mp i \theta(\pm t) \langle \lbrace \gamma_A(t), \gamma_B (0) \rbrace \rangle,
\end{equation}
where  -- to be concrete -- we here consider Majorana fermions $\gamma_A, \gamma_B$ which anticommute and square to unity (italic $A,B$ are multiindices encompassing spatial position and flavor of fermions). The average $\langle \dots \rangle$ denotes a quantum expectation with respect to an appropriate (e.g.~equilibrium) density matrix, see details below. 

Poles in the Green's function may exist even in interacting systems and signal well defined fermionic excitations -- a prominent example are the quasiparticles at the Fermi surface of a Fermi liquid. In addition to poles, zeros of the Green's function have recently attracted considerable attention~\cite{Dzyaloshinskii2003,Fabrizio2022} as this can, amongst others, account for the missing Luttinger volume~\cite{Luttinger1960,Oshikawa_2000} in the pseudogap phase of the cuprates and certain heavy fermion materials~\cite{ProustTaillefer2019,ColemanRech2005}. In equilibrium systems, zeros of the Green's function (more precisely: zero eigenvalues) can only occur in the presence of sufficiently strong interactions and free fermion Green's functions can only display poles (this is particularly manifest when using the Lehmann representation). Zeros often appear in the context of (orbital) selective Mott phases and may indicate the presence of emergent, potentially fractionalized excitations~\cite{SenthilVojta2003,Fabrizio2023,RoyKoenig2024} or of symmetric mass generation (SMG)~\cite{FidkowskiKitaev2010,YouXu2014,YouVishwanath2018,WangYou2022}.

Just like poles, zeros may form bands in energy vs. momentum space. It is a question of recent interest to ask whether such bands can have topological properties and, if so, whether these imply a bulk-boundary correspondence and thereby the appearance of topological edge zeros~\cite{WagnerSangiovanni2023,ZhaoPhillips2023,SettySi2024,BollmannKoenig2024,WagnerSangiovanni2024,SuMartin2024,StepanovSangiovanni2024,PangburnBanerjee2024,PangburnBanerjee2025,SimonMorice2025,ChenHosur2025,LehmannBudich2025,FloresHooley2025}. Moreover, as Green's-function-based topological invariants~\cite{IshikawaMatsuyama1986,VolovikYakovenko1989,Gurarie2011,ManmanaGurarie2012,WangZhang2012,YouXu2014} can be used to classify interacting topological quantum materials, it is an important observation that topological bands of Green's function zeros contribute to such invariants in an analogous fashion to topological bands of Green's function poles~\cite{Gurarie2011,GavenskyGoldman2023,BlasonFabrizio2023,ZhaoPhillips2023}. As such, topological Green's function zeros may be a means to practically diagnose interacting symmetry protected topological phases (SPT) as well as intrinsically topologically ordered states. Crucially, these observations have so far been restricted to the case of equilibrium systems.

Motivated by recent experimental realizations on Floquet topology in programmable quantum devices~\cite{MiAbanin2022,ZhangWang2022,samanta2024isolatedzeromodequantum,JinDeng2025}, we here study topological Green's function zeros in the Floquet non-equilibrium setting~\cite{Eckardt_2015,OkaRev,Mori2023}.  
We concentrate on one-dimensional Floquet-Kitaev-like models~\cite{Jiang11,ThakurathiDutta2013,Roy16,
YatesMitra2018,
Yates19,SchmidvonOppen2024} and introduce Floquet Green's-function-based topological bulk invariants. Building on the general classification of interacting Floquet SPTs~\cite{PotterVishwanath2016,Khemani16,Sondhi16-I}, we further concentrate on three phases of Floquet SMG of Majorana edge zero and $\pi$ modes. For a Floquet-version of the Fidkowski-Kitaev (FK) interaction~\cite{FidkowskiKitaev2010,WangYou2022} leading to SMG we calculate both edge and bulk Green's function and determine the influence on the newly defined topological invariant. As the experimental observation of Green's function zeros may be challenging in the solid state setup, we conclude our study with a discussion for a gate-based quantum simulator displaying topological Green's function boundary zeros.

This manuscript is structured as follows: In Sec.~\ref{sec:GFTopoInv} we introduce a Green's-function-based topological invariant for Floquet systems of class BDI.  To illustrate it, in Sec.~\ref{sec:FreeFloquet} we consider 
the well-studied periodically driven Kitaev chain 
and present results for its free fermion Green's function. Sec.~\ref{sec:Interactions} contains the main results on topological Green's function zeros in context of Floquet SMG, both in regards to bulk and edge Green's function. Finally, Sec.~\ref{sec:Circuit} contains a discussion about the implementation of this physics in gate-based quantum simulators. We conclude with a summary and outlook. Technical details are relegated to multiple appendices.

\section{Green's-function-based topological invariants}
\label{sec:GFTopoInv}

In this section we introduce Green's-function-based topological invariants for interacting Floquet systems which we connect to known results for non-interacting Floquet systems and, separately, to analogous invariants of (potentially interacting) equilibrium systems. To make the paper self-contained we first review the symmetry properties of 1D systems with parity and time reversal symmetry.

\subsection{Basic definitions}
\begin{figure}
    \centering
\includegraphics[scale=1]{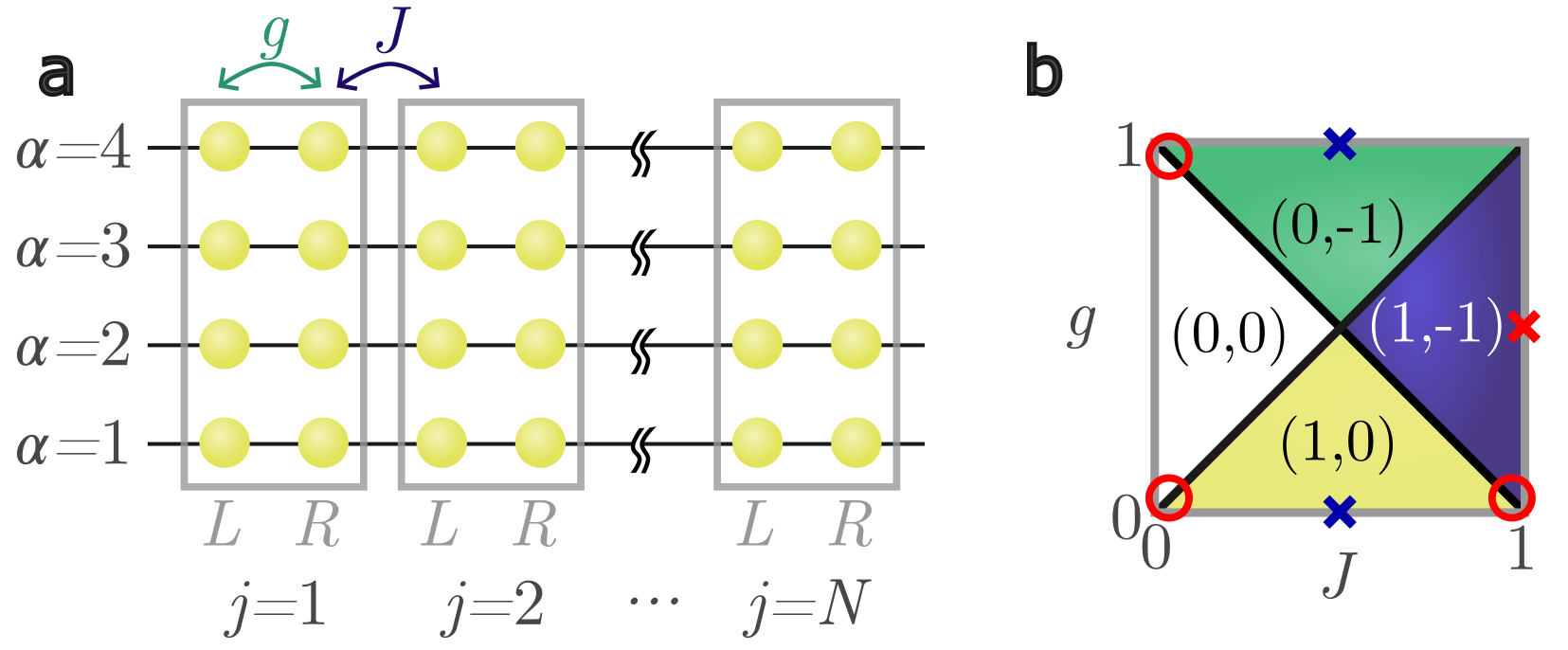}
    \caption{a) Illustration of the system and notational convention: in particular $g$ and  $J$  describe the free fermion Floquet evolution (matchgate circuit) while $w$ adds interactions. b) Free fermion Floquet phase diagram. The labelling $( \nu_0 ,  \nu_\pi)$ corresponds to the topological free fermion invariant Eq.~\eqref{eq:TopoFree} and sets the number $\vert \nu_0\vert $ and $\vert \nu_\pi\vert $ of Majorana zero modes and Majorana $\pi$-modes at a given edge, respectively~\cite{ThakurathiDutta2013,CardosoMitra2025}. Crosses (circles) indicate the points (regimes) in parameter space for calculations of Sec.~\ref{sec:FKFloquetEdge} (\ref{sec:FKFloquetBulk}).}
    \label{fig:Model}
\end{figure}

We consider 1D Floquet systems composed of Majorana operators $\gamma_A=\gamma_{j,D }^\alpha$ with $\{\gamma_A, \gamma_B \} = 2 \delta_{AB}$.  
In this sense, the multiindex $A = (j,D,\alpha)$ of Eq.~\eqref{eq:GreenDef} represent the site $j = 1, \dots, N$, the Majorana sublattice orbital $D = L/R$ and the flavor $\alpha = 1, \dots, N_f$, see Fig.~\ref{fig:Model} a) for an illustration. We further define the Fourier transformed fields in a system with periodic boundary conditions 
\begin{equation}
    \gamma_{\mathcal A
    }(p) = \frac{1}{\sqrt{2N}} \sum_j e^{- i p j} \gamma_{j,
     \mathcal A}, \quad p \in [-\pi, \pi), \label{eq:FT}
\end{equation}
and use the multiindex $\mathcal A=(D, \alpha)$ for the remaining internal degrees of freedom.
In view of the self-adjointness of Majorana operators it follows that $\gamma_{\mathcal A} (p) = \gamma_{\mathcal A}^\dagger(-p)$ 
with canonical commutation relations $\lbrace \gamma_{\mathcal A}^\dagger(p), \gamma_{\mathcal A'}(p') \rbrace = \delta_{\mathcal A, \mathcal A'} \delta_{p,p'}= \lbrace \gamma_{\mathcal A}(-p), \gamma_{\mathcal A'}(p') \rbrace $. Finally, we consider the Fourier transform of the Green's function, Eq.~\eqref{eq:GreenDef} 

\begin{equation}
     G_{\rm R/A}^{\mathcal A \mathcal B}(t; p) = \frac{1 }{2N} \sum_{jj'} e^{- i p(j- j')}  G_{\rm R/A}^{(j,\mathcal A) (j',\mathcal B)}(t).\label{eq:MomspaceGF}
\end{equation}

\subsection{Symmetries}
In addition to fermionic parity symmetry generated by $\hat P = \prod_{A} \gamma_A$,
we consider systems with
time reversal symmetry 
\begin{align}
    \hat T^2 & = 1, &
    \hat T i \hat T  &= -i, &
    \hat T \gamma_A \hat T &= {\Gamma_{AB}}\gamma_B. \label{eq:Symmetries}
\end{align}
Here, 
\begin{equation}
\Gamma = \text{diag}(1,\dots,1, -1, \dots, -1)
\end{equation}
is the $2N N_f \times 2N N_f$ dimensional diagonal matrix with positive (negative) entries for $D = L$ $(D =R)$.
The symmetries imply that the time dependent Hamiltonian may only contain even numbers of fermion operators and restricts bilinears to contain one $L$ and one $R$ Majorana. 
{As outlined in detail in Sec.~\ref{sec:FreeFloq}, App.~\ref{app:BDI}, these two symmetries further ensure time reversal and particle-hole symmetric free fermion unitaries, placing the system in class BDI.}

The symmetries {also}~\cite{Delplace14}
have implications on the micromotion of a time reversal invariant, and time periodic Hamiltonian with period $T_F$, $ \hat H(t) = \hat H(t + T_F) = \hat T \hat H (-t) \hat T$.
Here, we chose the center of time inversion to be $t = 0$. 
These symmetries imply the following relations and a special role of $t = T_F/2$ for the micromotion of the system. In particular the half-period time evolution
\begin{subequations}
\begin{equation}
    \hat F = \hat{\mathcal T} e^{- i \int_0^{T_F/2} \hat H(t) dt}
\label{eq:FOp}
\end{equation}
is related to the the micromotion in the second half of the Floquet period as follows 
\begin{equation}
    \hat{\mathcal T} e^{- i \int_{T_F/2}^{T_F} \hat H(t) dt} = \hat{\mathcal T} e^{- i \int_{-T_F/2}^{0} \hat H(t) dt} = \hat  T \hat F^\dagger \hat T.
\end{equation}
\end{subequations} 
This means that there are two Floquet unitaries, one starting at $t = 0 \mod (T_F)$ \cite{Delplace14}
\begin{subequations}
    \begin{equation}
        \hat U_F = {\hat{\mathcal T} e^{- i \int_0^{T_F} \hat H(t) dt}}= \hat T \hat F^\dagger \hat T \hat F, \label{eq:UF}
    \end{equation}
    and the other one starting at $t = T_F/2 \mod(T_F)$
     \begin{equation}
        \hat{\tilde{U}}_F { =\hat{\mathcal T} e^{- i \int_{T_F/2}^{3 T_F/2} \hat H(t) dt} =}  \hat F \hat T \hat F^\dagger \hat T = \hat F\hat U_F \hat F^{\dagger}. 
    \end{equation}
    \label{eq:TwoCyclesGeneral}
\end{subequations} \noindent 

The special role of $t = T_F/2$ motivates the consideration of the following two discrete-time, matrix-valued Green's functions 
\begin{subequations}
\begin{align}
    \mathbf G_{\rm R/A}
    ^{AB}(n) &=  {\mp i \theta(\pm n) \langle \{\hat U_F^{-n} \gamma_A  \hat U_F^n, \gamma_B \} \rangle,} \\
    \tilde {\mathbf G}_{\rm R/A}
    ^{AB}(n) &= {\mp i \theta(\pm n) \langle \{\hat{\tilde U}_F^{-n} \gamma_A  \hat{\tilde U}_F^n, \gamma_B \} \rangle}.
\end{align}
\label{eq:FloquetGFs}
\end{subequations} \noindent 
Here, $n \in \mathbb Z$ and $\theta(0) = 1/2$ is used for the Heaviside function $\theta(n)$ entering the definition of {the Green's function (}cf. $G_{\rm R/A}(t)$, Eq.~\eqref{eq:GreenDef}). 

For the following symmetry discussion it will be important that the quantum average is taken with respect to a density matrix which is time reversal invariant. For all concrete calculations we will use an appropriate equilibrium density matrix. 

\subsection{Symmetry constraints on the Green's function}

The Fourier transforms of the discrete time Green's functions {are}
\begin{equation}
    \ptwiddle{\mathbf G}_{\rm R}(\Omega) = \sum_n e^{i \Omega^+ n} \ptwiddle{\mathbf G}_{\rm R}(n), \quad \Omega \in [-\pi, \pi),  \label{eq:FTOmegaGF}
\end{equation}
with $\Omega^\pm = \Omega \pm i 0$
{and where the tilda in brackets indicates that both Eqs.~\eqref{eq:FloquetGFs} are treated simultaneously. Given that time reversal symmetry imposes $\hat U_F = \hat T {\hat U}_F^{\dagger} \hat T$ and analogously for $\hat{\tilde U}_F$}, the Green's functions have the property, cf.~Appendix~\ref{app:SymmetryConstraintsGF} 
\begin{subequations}
\begin{align}
    \ptwiddle{\mathbf G}_{\rm R}(\Omega,p) &= -    \ptwiddle{\mathbf G}_{\rm R}^*(-\Omega,-p) = -\Gamma     \ptwiddle{\mathbf G}_{\rm A}(-\Omega,p) \Gamma.
\end{align}
\label{eq:FTConstraints}
\end{subequations}  
\noindent 
{Both the Green's function $\mathbf G_{\rm R/A}(\Omega, p)$ and $\Gamma$ are $2N_f \times 2N_f$ matrices, the latter} being the diagonal matrix with positive (negative) {unit} entries for $D = L$ $(D =R)$.

\subsection{Green's-function-based topological invariants}

Green's-function-based topological invariants have the advantage 
that they can be generalized to interacting systems (i.e.~can be applied beyond matchgate circuits viz. free fermion Floquet cycles).  

A Green's-function-based topological invariant for systems of the present symmetry class BDI and with \textit{continuous} time evolution has been presented in Refs.~\cite{Gurarie2011,ManmanaGurarie2012}. Here, we generalize it to periodically driven systems, where Green's-function-based approaches are less common (see, however, Ref.~\cite{GavenskyGoldman2024}). We introduce 
\begin{subequations}
    \begin{align}
        N_1 &=  { \int_{- \pi}^\pi \frac{dp}{4 \pi i} \tr[\Gamma \mathbf G_{\rm R}^{-1} \partial_p \mathbf G_{\rm R}]_{\Omega = 0},} \\ 
      \tilde N_1 & = { \int_{- \pi}^\pi \frac{dp}{4 \pi i}  \tr[\Gamma \tilde{\mathbf G}_{\rm R}^{-1} \partial_p \tilde{\mathbf G}_{\rm R}]_{\Omega = 0}.} 
    \end{align}
    \label{eq:GFINvariants}
\end{subequations} \noindent  \noindent 
{Here}, we tacitly took the thermodynamic $N \rightarrow \infty$ limit so {to define} integrals over the $p \in [-\pi,\pi)$ interval.
These integrals are the first central result of this work. They are real and quantized (see App.~\ref{app:TopoInvProperties}) under the assumption that there is no spectral weight at $\Omega = 0,\pi$ (i.e.~the system is gapped) and that there are no Green's function zeros crossing $\Omega = 0,\pi$. Using these assumptions we also remark that the $\Omega = 0$ contribution to $N_1$ fully corresponds the equilibrium Green's-function-based invariant introduced in Ref.~\cite{ManmanaGurarie2012} (which is technically written using Matsubara formalism) {and we follow their notation, including the subscript $_1$.}

\subsection{Free fermion Floquet systems {and relationship to free fermion topological invariants}}
\label{sec:FreeFloq}

Using the spinor convention $\gamma_\alpha(p) = (\gamma_L^\alpha(p),\gamma_R^\alpha(p))^T$ 
the second quantized {continuum} time evolution operator of a free fermion system is
\begin{align}
    \hat U (t)  = \mathcal T e^{- i \int_0^t dt' \sum_{p > 0} \gamma^\dagger(p) h(t',p) \gamma (p)},
\end{align}
where particle-hole and time-reversal-symmetry impose {symmetries} on the first quantized matrix Hamiltonian 
\begin{equation}
h(t, p) = - h^T(t,-p) = -\Gamma h(-t, p) \Gamma, \label{eq:FirstQuantSymmetriesHam} 
\end{equation}
{where $^T$ denotes the standard matrix transpose}
and we further assumed spatial translational symmetry. The spinor fermion operator at $p>0$ thus evolves as
\begin{equation}
    \hat U^\dagger (t) \gamma(p) \hat U(t) = U(t,p) \gamma(p),\,\,  \hat T \gamma(p) \hat T=\Gamma \gamma(-p) \label{eq:FirstQuantTEvol}
\end{equation}
with matrix valued first quantized time evolution operator 
\begin{align}
U(t,p) &= \mathcal T e^{-i \int dt' h(t',p)}\\
    U(t,p) &= \Gamma U^\dagger(t,p) \Gamma = U(t,-p)^*. \label{eq:ChiralSymmetry}
\end{align} 

{For the time-periodic, Floquet time evolution,} analogously to the second quantized expression Eq.~\eqref{eq:UF}, the first quantized Floquet operator $U_F = U(T_F)$ 
can {then} be decomposed as
\begin{equation}
    U_F = \Gamma F^\dagger \Gamma F,\,\, \tilde{U}_F = FU_F F^{\dagger}.
\end{equation}
where
\begin{equation}
    F = \mathcal T e^{- i \int_0^{T_F/2} h(t,p) dt}
\end{equation}
is the first quantized version of the half-period time evolution operator introduced in Eq.~\eqref{eq:FOp}. We suppress the $p$ dependence of $U_F, F$ for notational convenience.

The Fourier transformed discrete time Green's functions, Eq.~\eqref{eq:FloquetGFs} in the non-interacting case are given by (cf. Appendix~\ref{app:BulkGF})

\begin{subequations}
\begin{align}
     \mathbf G_{\rm R/A}(\Omega, p)  &= -  \frac{i}{2} \frac{\mathbf 1+e^{i \Omega^{\pm}} U_F}{\mathbf 1-e^{i \Omega^{\pm}} U_F},\\    
  { \tilde{\mathbf G}_{\rm R/A}(\Omega, p)}  &= {-  \frac{i}{2} \frac{\mathbf 1+e^{i \Omega^{\pm}} \tilde U_F}{\mathbf 1-e^{i \Omega^{\pm}} \tilde U_F}}. 
\end{align}
\label{eq:FreeFermionGFs}
\end{subequations} \noindent  
It is a well-known feature of free fermion evolution that this result is independent of the state defining the average $\langle \dots \rangle$ in the defining Eq.~\eqref{eq:GreenDef}. The free fermions Floquet Green's functions, Eqs.~\eqref{eq:FreeFermionGFs}, manifestly fulfill the symmetry constraints, Eqs.~\eqref{eq:FTConstraints}. 

Using the definition of Green's-function-based invariants, Eq.~\eqref{eq:GFINvariants}, as well as the results Eq.~\eqref{eq:FreeFermionGFs} we find (see Appendix \ref{app:BulkGF} for details)

\begin{subequations}
\begin{align}
    N_1 & = {\nu_0 + \nu_\pi} \\
    \tilde N_1 & = {\nu_0 - \nu_\pi},
\end{align}
where we recover free fermion Floquet invariants introduced in Ref.~\cite{CardosoMitra2025} 
\begin{align}
    \nu_{0} & = \frac{1}{2\pi i} \int_{-\pi}^\pi dp \; \tr \left( [F,\Gamma]^{-1} \partial_p F \right ),\\
    \nu_{\pi} & = \frac{1}{2\pi i} \int_{-\pi}^\pi dp \; \tr \left  ( \{F,\Gamma \}^{-1} \partial_p F \right ).
\end{align}
\label{eq:TopoFree}
\end{subequations} \noindent 
In summary, in this section we introduced two Floquet Green's-function-based topological invariants, Eq.~\eqref{eq:GFINvariants}, which -- in the non-interacting limit -- reduce to linear combinations of established Floquet-cycle based invariants, Eq.~\eqref{eq:TopoFree}.

\section{Free Ising Floquet circuit}
\label{sec:FreeFloquet}

As a concrete application of the above formulae, we will concentrate on Kitaev-Floquet (or Ising-Floquet) time evolution with interactions, see Sec.~\ref{sec:Interactions}. To set the stage, we briefly review the corresponding free fermion Floquet unitary $\hat U_F$. By means of the Hamiltonians
\begin{subequations}
\begin{align}
    \hat H_g & = i {g}\sum_{\alpha}  \sum_{j}  \gamma_{j,L}^\alpha \gamma_{j,R}^\alpha, \\
    \hat H_J  & =  i {J}\sum_{\alpha}  \sum_{j} \gamma_{j, R}^\alpha \gamma_{j + 1, L}^\alpha.
\end{align}
\label{eq:FreeFloquetHam}
\end{subequations} 
we introduce
\begin{subequations}
\begin{align}
    \hat  U_g &= e^{- i \frac{\pi}{2} \hat H_g},\\
    \hat U_J  &= e^{- i \frac{\pi}{2} \hat H_J},\\
    \hat F_0 &= \hat  U_{J/2}  \hat  U_{g/2},
\end{align}
so that
\begin{align}
\hat U_F &=  \hat  U_{g/2}  \hat  U_J  \hat  U_{g/2} = \hat T\hat F_0^\dagger \hat T \hat F_0,\\
\hat{\tilde U}_F &=  \hat  U_{J/2}  \hat  U_g  \hat  U_{J/2}= \hat F_0\hat T\hat F_0^\dagger \hat T.
\end{align}
\label{eq:FreeFloquet}
\end{subequations}

The parameters of the Floquet unitary (see Fig.~\ref{fig:Model} a) for illustration) have the following physical meaning: the parameter $g$ corresponds to the on-site chemical potential of the  Floquet-Kitaev chain or equivalently to the transverse field in the Floquet-Ising chain. In contrast,  $J$ is the hopping and $p$-wave pairing of the Floquet-Kitaev chain or, equivalently, the exchange interaction of the Floquet-Ising chain. The factor of $\pi/2$ in front of the Hamiltonian in Eqs.~\eqref{eq:FreeFloquet} is a customary convenience to restrict the phase diagram to $g,J \in [0, 1)$ (see below).

The free fermion Floquet evolution is characterized by $(\nu_0, \nu_\pi) \in \mathbb Z \times \mathbb Z$ topological quantum numbers, where $\vert 
\nu_0 \vert$ corresponds to the number of Majorana zero modes (MZMs) and $\vert \nu_\pi \vert $ to the number of Majorana $\pi$ modes (M$\pi$Ms) at a given edge. As we will shortly see, for $N_f = 1$, the phases with topological quantum numbers $(0,0)$,$(1,0)$,$(0,-1)$,$(1,-1)$ can be accessed by tuning $g,J$ according to the phase diagram Fig.~\ref{fig:Model} b), while additional tuning of $N_f$ allows to access arbitrary number of MZMs and M$\pi$Ms at the edge.

\subsection{Bulk physics}

\label{sec:FreeFloquetBulk}

\subsubsection{Floquet unitary and its topological winding}

Using the spinor convention $\gamma_\alpha(p) = (\gamma_L^\alpha(p),\gamma_R^\alpha(p))^T$ 
the second quantized Floquet unitary, Eq.~\eqref{eq:FreeFloquet}, can be written as (cf. Appendix~\ref{app:FreeIsingFloquet})

\begin{align}
    \hat U_F 
    & = \prod_{p , \alpha} e^{- i \frac{\pi}{2} \gamma_\alpha^\dagger (p) h_{F}(p) \gamma_{\alpha}(p)} = \prod_{p >0, \alpha} e^{- i \pi \gamma_\alpha^\dagger (p) h_{F}(p) \gamma_{\alpha}(p)} \label{eq:UFMainText}
\end{align}
where the effective Floquet Hamiltonian $h_F$ defines the first quantized Floquet unitary~\cite{YehMitra2023}
\begin{subequations}
     \label{eq:Ueff}
     \begin{align}
     U_F &\equiv e^{- i \pi h_F}
     = \vec n(p)\cdot (i \sigma_x, i \sigma_y, \mathbf 1)^T, \\
    \vec n(p)  &= \left (\begin{array}{c}
           \sin(p) \sin(\pi J)\\
           \cos(\pi J) \sin(\pi g) - \cos(p)  \sin(\pi J)\cos(\pi g) \\
           \cos(\pi J) \cos(\pi g) + \cos(p)  \sin(\pi J)\sin(\pi g)
    \end{array}\right ),
\end{align} 
and $\sigma_{x,y,z}$ are Pauli matrices in $L/R$ space. 
The other unitary is
\begin{widetext}
     \label{eq:Ueff}
     \begin{align}
     \tilde{U}_F &\equiv e^{- i \pi \tilde{h}_F}
     = \vec{\tilde{n}}(p)\cdot (i \sigma_x, i \sigma_y, \mathbf 1)^T, \\
    \vec {\tilde{n}}(p)  &= \left (\begin{array}{c}
          \sin(p) \sin(\pi J)\cos(\pi g)+\sin(2p)\sin^2(\pi J/2)\sin(\pi g)\\
         \sin (\pi  g) \{\cos (\pi  J) \cos ^2(p)+\sin ^2(p)\}- \cos (p)\sin (\pi  J)\cos (\pi  g)  \\
          \cos(\pi J) \cos(\pi g){+} \cos(p) \sin(\pi J)\sin(\pi g)
    \end{array}\right ).
\end{align} 
\end{widetext}
 \label{eq:n}
\end{subequations} \noindent

The eigenvalues of $U_F\ket{u_{\pm}(p)} = e^{\mp i \epsilon(p)} \ket{u_{\pm}(p)}$ come in pairs in view of chiral symmetry, Eq.~\eqref{eq:ChiralSymmetry}, where $\Gamma = \sigma_z$ in this concrete model. For most of the parameter space, the Floquet unitary is gapped both at quasienergy zero and $\pi$.
{The same is true for $\tilde U_F$ whose eigenvalues are the same.}
As we will explicitly review in Sec.~\ref{sec:Edge}, the topological gap closing at $g = J$ ($g = 1-J$) illustrated by a diagonal line in Fig.~\ref{fig:Model} b) implies the presence of MZMs (M$\pi$Ms) modes for $g<J$ ($g>1-J$). At the same time, the in-plane components $(n_x(p), n_y(p))$ acquire a non-zero winding about the origin only in the phases denoted $(1,0)$, $(0,-1)$ in Fig.~\ref{fig:Model} b), while in-plane components $(\tilde n_x(p), \tilde n_y(p))$ acquire a non-zero winding about the origin in all phases but the one denoted $(0,0)$, $(0,0)$

\subsubsection{Bulk Green's function}

\begin{figure}
    \centering
    \includegraphics[scale = 1]{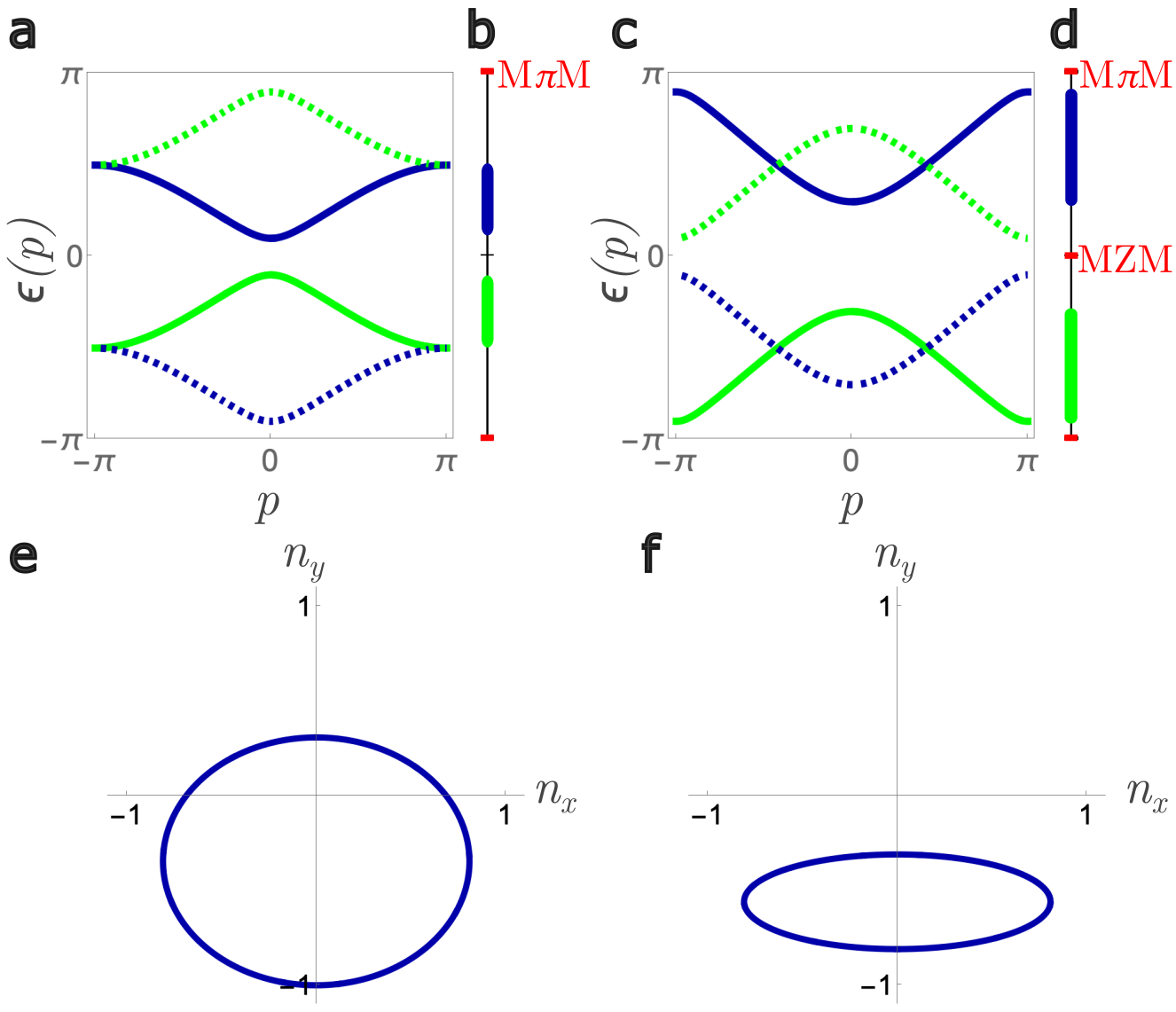}
    \caption{Poles, zeros and winding of the free fermion Green's function Eqs.~\eqref{eq:FreeFermionGFs}, \eqref{eq:Ueff}. a), c) Poles (solid) and zeros (dotted) of $\mathbf G_{\rm R}(\Omega, p)$ for periodic boundary conditions. Note the vicinity to the gap closing at $\Omega = 0$ and $\Omega = \pi$ in panels a) and c) respectively. 
    b), d) Schematic spectrum for open boundary conditions. 
    e), f) Parametric plot of the in-plane vector $(n_x(p),n_y(p))$, Eq.~\eqref{eq:n} entering the Green's function and its winding,~Eq.~\eqref{eq:WindingFree}. 
    Figures (a,b,e) are plotted for $(J,g) = (0.7,0.8)$ and (c,d,f) at $(J,g) = (0.7,0.4)$.
}
    \label{fig:PolesZerosWinding}
\end{figure}

In the case of $N_f = 1$ Kitaev chains, the Green's functions in the bulk in Fourier space are given by Eq.~\eqref{eq:FreeFermionGFs} evaluated using \eqref{eq:Ueff}.
We see from the eigenvalues of $U_F$ that, at given $p$, the Green's function have poles at $\Omega = \pm \epsilon(p)$ and, simultaneously, zeros at $\Omega = \pi \pm \epsilon$, for illustration see Fig.~\ref{fig:PolesZerosWinding}.  

This is a crucial difference between systems with continuum and Floquet time evolution: As noted in the introduction, in the case of continuum evolution, the determinant of the Green's function of free fermions cannot contain any zeros. In contrast, the Floquet Green's function with an infinite number of equally spaced poles in the extended Floquet zone scheme implies interposed zeros of the determinant of the Green's function even in the absence of interactions. As we will see shortly, this is crucial in the characterization of the Floquet topology using Green's function techniques. 

\subsubsection{Topological invariant}

We
now evaluate the topological invariants $N_1,\tilde N_1$ Eqs.~\eqref{eq:GFINvariants} for the free Ising Floquet circuit.
Using Eqs.~\eqref{eq:FreeFermionGFs}, \eqref{eq:Ueff} we obtain

\begin{subequations}
\begin{align} 
    N_1 & = \int_{- \pi}^\pi \frac{dp}{2\pi} \hat n_x \partial_p \hat n_y - \hat n_y \partial_p \hat n_x, \label{eq:WindingFree}\\
      \tilde  N_1 & = \int_{- \pi}^\pi \frac{dp}{2\pi} \hat {\tilde n}_x \partial_p \hat{\tilde n}_y - \hat{\tilde n}_y \partial_p \hat {\tilde n}_x, \label{eq:WindingFree2}
\end{align}
\end{subequations}
with $\hat n_{x,y} = n_{x,y}/\sqrt{n_x^2 + n_y^2}$, $\hat {\tilde n}_{x,y} = {\tilde n}_{x,y}/\sqrt{{\tilde n}_x^2 + {\tilde n}_y^2}$ is the in-plane unit vector. Thus $N_1,\tilde N_1$ count the winding of the two respective Floquet unitaries. This leads to non-zero $\nu_0 = (N_1 + \tilde N_1)/2$ and/or $\nu_\pi= (N_1 - \tilde N_1)/2$ in all colored phases in Fig.~\ref{fig:Model} b). 

It is an interesting observation that $N_1$, which is fully determined by the topology of the Green's function $\mathbf G_{R}(\Omega, p)$ at frequency $\Omega = 0$, changes from nontrivial to trivial at the transition between phases denoted $(1,0)$ in $(1,-1)$ Fig.~\ref{fig:Model} b), even though the gap does not close at $\Omega = 0$. The reason is that $\mathbf G_{R}(0, p)$ has a gap closing of the Green's function \textit{zero}, cf. Fig.~\ref{fig:PolesZerosWinding} b) which illustrates the vicinity of the transition, and, as alluded to in the introduction, gap closings of zeros can change topological invariants similarly to gap closings of poles. Analogous arguments hold for the transition between the phases denoted $(0,-1)$ and $(1,-1)$. Thus, $N_1$ alone is capable to distinguish topological phases of equilibrium systems, but not sufficient for the Floquet case. This is the reason why there is a second topological invariant $\tilde N_1$ associated to the Green's function $\tilde{\mathbf G}_{R}(0, p)$ associated to the Floquet cycle shifted by a half-period. This allows to discriminate all four phases with arbitrary number of MZM and M$\pi$M and leads to the labels in the phase diagram Fig.~\ref{fig:Model} b) (see appendix~\ref{app:FreeIsingFloquet}, in particular Tab.~\ref{tab:Windings} for a review on $F(p)$ and the determination of topological quantum numbers $(\nu_0, \nu_\pi)$ for the present model).

\subsection{Edge Physics}
\label{sec:Edge}

We now switch to edge physics and review certain fine tuned limits in which the problem allows for particularly simple solutions with perfectly localized (i.e.~within a unit cell) edge MZMs or M$\pi$Ms.

For notational simplicity we consider the edge physics for the slightly different Floquet unitary $\hat U_F = \hat U_g \hat U_J$, or, equivalently, we use a basis change $\hat \gamma_{j,D}^\alpha \rightarrow \hat U_{g=1/2} \hat \gamma_{j,D}^\alpha \hat U^\dagger_{g=1/2}$ in the evaluation of the correlators Eq.~\eqref{eq:GreenDef}, see Appendix~\ref{app:Translation} for details. We note that this basis is the one employed in experimental realizations~\cite{MiAbanin2022}, but it is disadvantageous as compared to the basis used in all previous sections in regards to the representation of symmetries of the system, which are less manifest.

First we consider $g = 0,1$. The unitary $\hat U_g$ simplifies substantially to be either $U_{g =0} = 1$ or $U_{g =1}  = \hat P$, where $\hat P = \prod_{j, \alpha} \gamma_{j,L}^\alpha \gamma_{j,R}^\alpha$ is the total parity (up to a phase). 

For $g = 0$, the Majorana edge operators $\gamma_{1,L}^\alpha$ and $\gamma_{N,R}^\alpha$ thereby completely decouple from the rest of the system and commute with $\hat U_F$ -- they are zero modes.
For $g = 1$, odd parity states acquire a fluctuating sign upon Floquet evolution. This particularly concerns the Majorana edge operators $\gamma_{1,L}^\alpha$ and $\gamma_{N,R}^\alpha$ which do not enter $\hat U_J$ and anticommute with $\hat U_g \vert_{g = 1}$ and thus, they are pinned at quasienergy $\pi$.

Next we consider $J = 1$ and $g = 1/2$.
In this case 
\begin{subequations}
\begin{align}
    \hat U_{J = 1} & = \prod_{\alpha = 1}^{N_f}  \prod_{j = 1}^{N-1} \gamma_{j,R}^\alpha\gamma_{j+1,L}^\alpha = \hat P (-1)^{N_f}\prod_{\alpha = 1}^{N_f} \gamma_{1,L}^\alpha \gamma_{N,R}^\alpha \\
    \hat U_{g = 1/2} & = \prod_{\alpha = 1}^{N_f}  \prod_{j = 1}^N \frac{1 + \gamma_{j, L}^\alpha \gamma_{j,R}^\alpha}{\sqrt{2}}
\end{align}
\end{subequations} \noindent  
so that 
\begin{align}
    \gamma_{1,+}^\alpha &= \frac{\gamma_{1L}^\alpha + \gamma_{1R}^\alpha }{\sqrt{2}}, &\gamma_{1,-}^\alpha &= \frac{\gamma_{1L}^\alpha - \gamma_{1R}^\alpha }{\sqrt{2}}, \label{eq:MZMMPM}
\end{align}
are {M$\pi$Ms} and {MZMs, respectively}.

\section{Interaction effects}
\label{sec:Interactions}

\begin{figure}
    \centering
    \includegraphics[scale = 1]{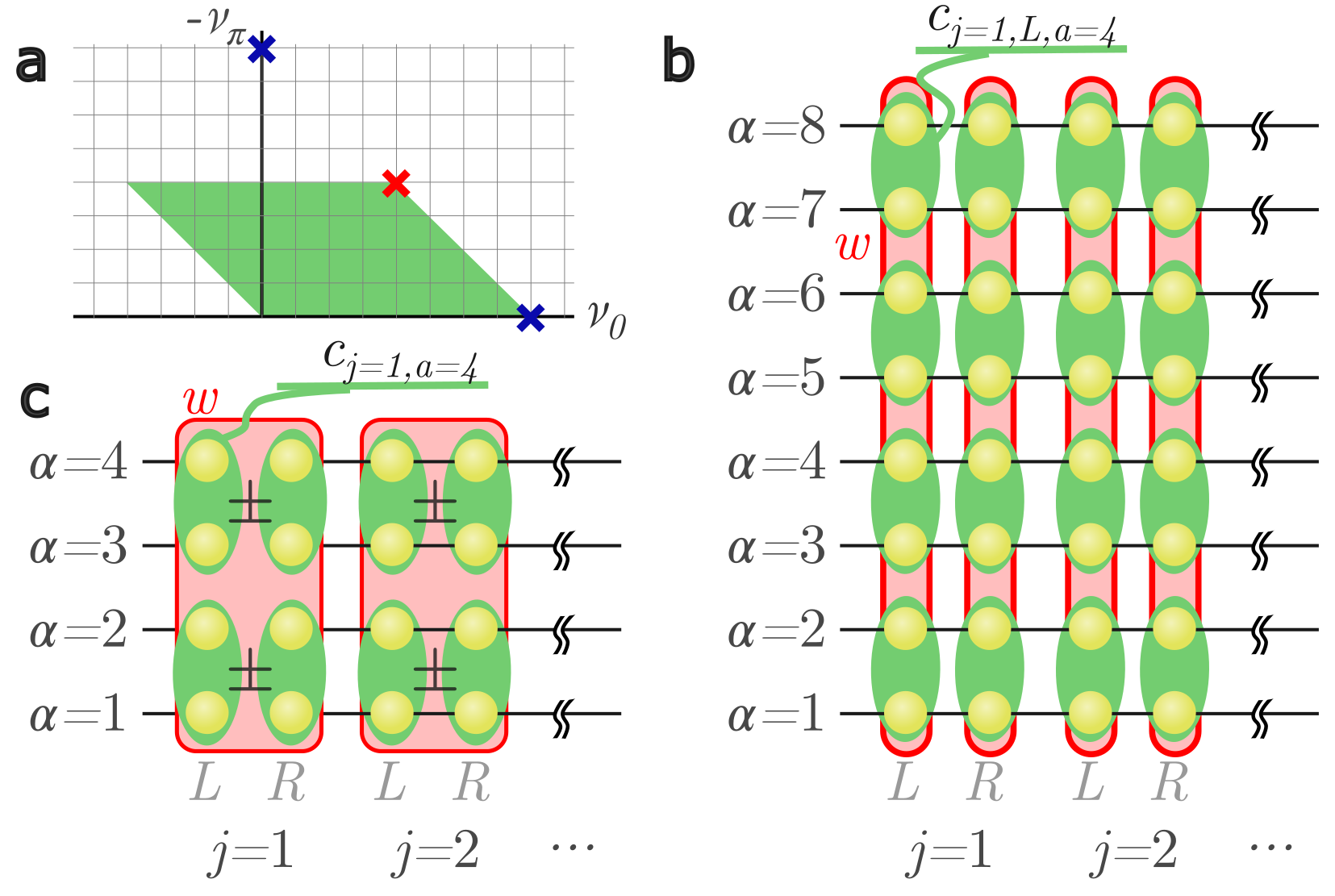}
    \caption{ a) Classification of Floquet SPTs (reproduced from Ref.~\cite{PotterVishwanath2016}): Symmetry preserving interactions fold the non-interacting $\mathbb Z \times \mathbb Z$ classification down to $\mathbb Z_8 \times \mathbb Z_4$ (shaded green). b)Illustration of the interaction for models 1 and 2 corresponding to $N_f = 8$ Kitaev chains (blue crosses in a). The complex $c_{\v j, a}$ fermions entering the interaction term, Eq.~\eqref{eq:FK} are illustrated as green ellipses and composed of Majorana fermions on adjacent wires. c) Illustration of the interaction model 3 (red cross in a) corresponding to $N_f = 4$ wires. The complex $c_{\v j, a}$ fermions entering the interaction term, Eq.~\eqref{eq:FK} are composed of linear combinations of Majorana Eq.~\eqref{eq:MZMMPM} from adjacent wires.   
}
    \label{fig:FKInteractions}
\end{figure}

In the previous section we reviewed a matchgate circuit (i.e. free fermion Floquet system) and its topological properties. Generally, such systems in class BDI are classified by a $\mathbb Z \times \mathbb Z$ topological invariants encoded in $\nu_0, \nu_\pi$, Eq.~\eqref{eq:TopoFree}. We reiterate that, while continuum time (equilibrium) free fermion systems do not display zero-eigenvalues of the Green's function, free driven (Floquet) systems do generally display zeros. {In contrast, the} major focus of this work is the study of interaction induced additional Green's function zeros. 

\subsection{Green's function zeros, classification of SPTs and symmetric mass generation}

{While the possible existence of Green's function zeros follows from general considerations of interacting many-body systems~\cite{Fabrizio2022}, to our knowledge, there are three analytically controlled methods to microscopically derive Green's function zeros in equilibrium systems\footnote{Bosonization in 1D also allows one to controllably derive Green's function zeros, yet only in the vicinity of the Fermi points of the free system.}. The first one assumes ultra-long-range interactions ("Hatsugai-Kohmoto" models and variants thereof)~\cite{HatsugaiKohmoto1992,HuangPhillips2022}. The second is based on models displaying fractionalization~\cite{SenthilVojta2003,BollmannKoenig2024}. The third is the symmetry preserving opening of a many-body gap in the absence of any anomalies, i.e.~SMG ~\cite{FidkowskiKitaev2010,YouVishwanath2018, WangYou2022}. Here, we will focus on the latter.}

{We remind the reader that SMG of boundary states of non-interacting topological systems leads to the reduction of allowed SPT phases as compared to their free fermion analogues. Moreover, while SMG precludes a zero-energy spectral weight at the edge, the edge Green's function may display edge zeros at zero energy (see Eq.~\eqref{eq:GFEquilibrium} and, for a generalization of this statement to the Floquet case, Eq.~\eqref{eq:GreensfunctionsEdgeMaintext}, below).}

{By means of the bulk-boundary correspondence~\cite{Gurarie2011}, SMG at the boundary also trivializes the bulk. In such interacting phases, it has been explicitly shown that bulk Green's functions do not feature topological bands of poles, yet do display topological Green's function zeros~\cite{YouXu2014} (we generalize this statement to the Floquet case, see Eq.~\eqref{eq:BulkGFWithIA}, below).}

{Of course, perturbations leading to SMG at the edge are not always (symmetry) allowed. For example,  equilibrium Kitaev-like Majorana chains (one spatial dimension) with time reversal symmetry, class BDI, have a free fermion topological index in the integers $\mathbb Z$. Interactions reduce this to $\mathbb Z_8$ as 8 edge states can be gapped symmetrically~\cite{FidkowskiKitaev2010}. }

{Returning to driven systems and}
the present symmetry class, the classification of interacting Floquet SPTs~\cite{PotterVishwanath2016} reduces from $\mathbb Z \times \mathbb Z$ to $\mathbb Z_8 \times \mathbb Z_4$, see Fig.~\ref{fig:FKInteractions} a). This is related to the Floquet variant of SMG, i.e.~the opening of symmetry conserving many-body gaps of topological edge states {(at quasienergy zero and/or $\pi$)} upon inclusion of certain interactions. Here we study precisely such interactions. We mention that, while generically in higher dimensions and at equilibrium, a critical interaction strength has to be overcome for the SMG to set in~\cite{YouXu2014,MartinGrover2025}, the present model for the edge of a one-dimensional system displays SMG at infinitesimal interaction strength.

Specifically, we study the model, Eq.~\eqref{eq:FreeFloquet}, starting from three different free fermion phases marked by crosses in Fig.~\ref{fig:FKInteractions} a). 
First, we concentrate on 
the non-interacting topological phase $(\nu_0, \nu_\pi) = (8,0)$ (phase denoted $(1,0)$ in Fig.~\ref{fig:Model} b) but with $N_f = 8$ chains). Second, we still consider $(\nu_0, \nu_\pi) = (0,-8)$ as a non-interacting parent phase (phase denoted $(0,-1)$ in Fig.~\ref{fig:Model} b) with $N_f =8$). Third, we consider $(\nu_0, \nu_\pi) = (4,-4)$ (phase denoted $(1,-1)$ in Fig.~\ref{fig:Model} b) but with $N_f = 4$).

\begin{figure*}
    \centering
    \includegraphics[width = \textwidth]{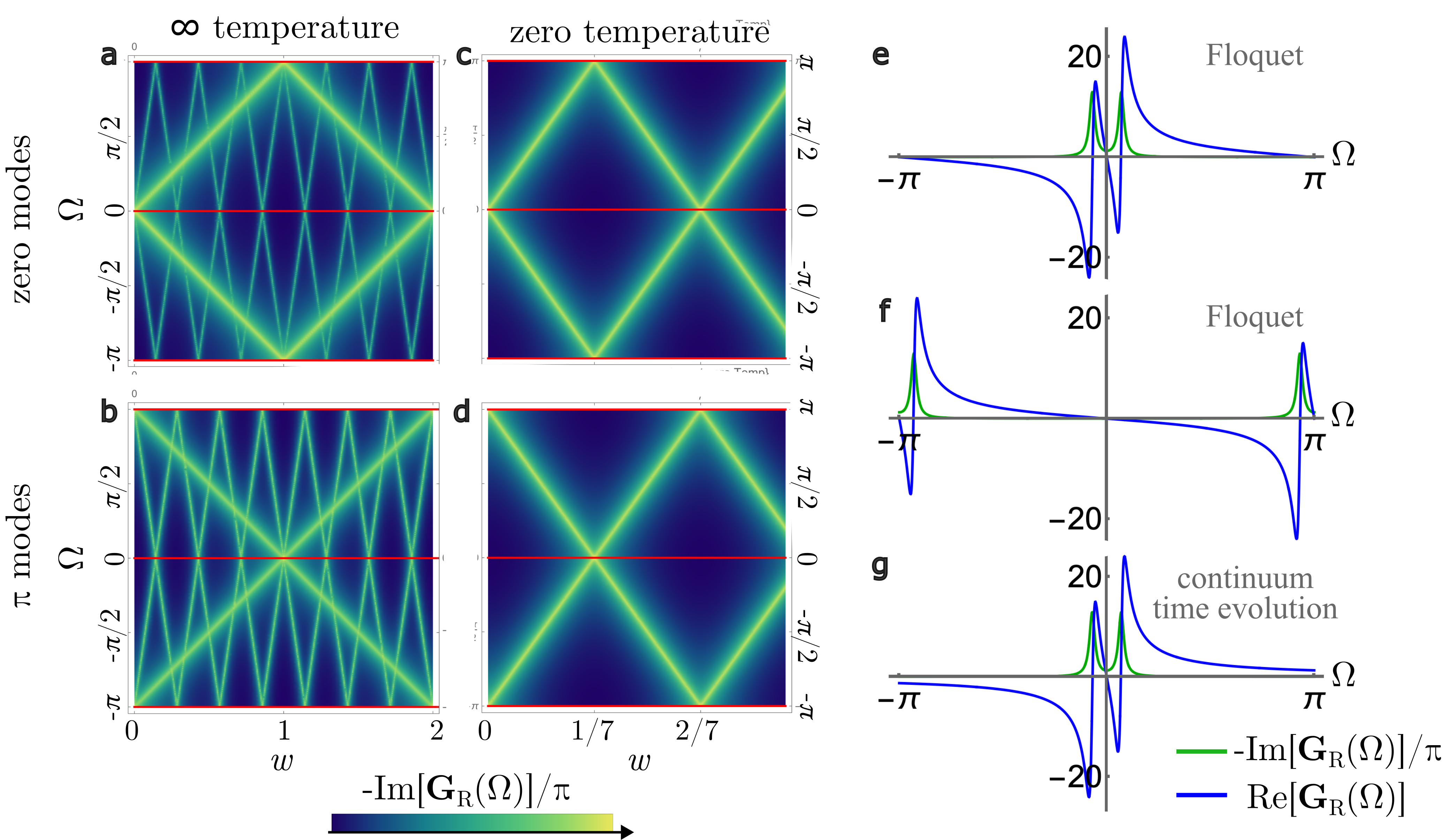}
    \caption{
    Infinite temperature [panels a), b)] and zero temperature [c)-d)]. Green's functions of edge zero modes [a), c), e), g)] and $\pi$ modes [b), d), f)]. a)-f) correspond to Floquet Green's functions while g) is the Fourier transformed continuous time Green's function for the same parameters as e). e)[f)] is a cross-cut of c)[d)] at $w = 0.01$. Red lines in panels a)-d) delineate vanishing Re$[\mathbf G_{\rm R}(\Omega)]$, while the color plot is the spectral function $-{\rm Im}\left[G_R(\Omega)\right]/\pi$. All figures are plotted for finite broadening $\Omega \rightarrow \Omega + i \Gamma$, $\Gamma = 0.05, \lambda = 1$. 
}
    \label{fig:EdgeSpectralWeight}
\end{figure*}

\subsection{Fidkowski-Kitaev unitary} 
\label{sec:FK}
 
We introduce an additional unitary operator $\hat U_w =e^{- i \frac{\pi}{2} \hat H_w}$ to the Floquet cycle,
\begin{align}
\hat H_{w} & = - 8 w \sum_{{\mathbf j}} \biggl[(c_{{\mathbf j},1} c_{{\mathbf j},2} c_{{\mathbf j},3} c_{{\mathbf j},4} + \text{H.c.}) + \frac{\lambda}{8} \sum_{b<a} \hat P_{{\mathbf j},b} \hat P_{{\mathbf j},a}\biggr], \label{eq:FK}
\end{align}
i.e.~the FK interaction term~\cite{FidkowskiKitaev2010}. Here $c_{{\mathbf j},a}$, $a \in \{1,2,3,4 \}$ are complex fermionic operators $\{ c_{{\mathbf j},a} ,c_{{\mathbf j}',a'}^\dagger\} = \delta_{{\mathbf j}{\mathbf j}'} \delta_{aa'}$ and $\hat P_{{\mathbf j},a} = 1- 2 [c_{{\mathbf j},a}]^\dagger c_{{\mathbf j},a}$. These fermionic operators are linear combinations of $\gamma_{j,D}^\alpha$: for details, e.g. on the index $\mathbf j$ (it accounts for the position of the particle), see below and Fig.~\ref{fig:FKInteractions}.  
We note that there are multiple versions of FK-like interaction terms leading to SMG and our model corresponds to the SU(4) invariant case (as opposed to, e.g., the more general SO(7) invariant form).

We remind the reader about the equilibrium FK-Hamiltonian $H_w$~\cite{FidkowskiKitaev2010,WangYou2022}: For $w>0, \lambda>0$ its ground state is the Greenberger-Horne-Zeilinger (GHZ) state in occupation number basis of $c$ fermions $[\ket{0000} + \ket{1111}]/\sqrt{2}$ and has energy $E_1 = - [8 + 6 \lambda]w$. The other GHZ state $[\ket{0000} - \ket{1111}]/\sqrt{2}$ has eigenvalue $E_2 = [8 - 6 \lambda]w$ and the remaining 6 even parity states have energy $E_3 = 2 \lambda w$, while all 8 odd parity states have vanishing energy. {The zero temperature Green's function in equilibrium systems is thus dominated by the transition from the ground state to the odd parity manifold~\cite{YouXu2014}
\begin{equation}
    G_{\rm R}(\Omega) = \frac{2\Omega}{(\Omega^+)^2-E_{1}^2}. \label{eq:GFEquilibrium}
\end{equation}
Note the zero of the Green's function at zero energy.
}

\subsection{Models of interactions}
\label{sec:ModelsIA}

{In this work we concentrate on three instances of SMG induced Green's function zeros: First, $N_f = 8$ Kitaev chains in a parameter regime (potentially) supporting eight copies of MZMs at the edge. This corresponds to the blue cross at $(\nu_0, -\nu_\pi) = (8, 0)$ in Fig.~\ref{fig:FKInteractions} a). Second, $N_f = 8$ Kitaev chains in a parameter regime (potentially) supporting eight copies of M$\pi$Ms at the edge. This corresponds to the blue cross at $(\nu_0, -\nu_\pi) = (0, 8)$ in Fig.~\ref{fig:FKInteractions} a). Third, $N_f = 4$ Kitaev chains in a parameter regime (potentially) supporting four copies of MZMs and four copies of M$\pi$Ms at the edge. This corresponds to the red cross at $(\nu_0, -\nu_\pi) = (4, 4)$ in Fig.~\ref{fig:FKInteractions} a).}

{Depending on these different instances we introduce interactions differently:}
To study Floquet SMG in systems with $N_f =8$ Kitaev chains (blue crosses in Fig.~\ref{fig:FKInteractions}) 
 we group $c_{{\mathbf j},a} = [\gamma_{j,D}^{2a-1} + i\gamma_{j,D}^{2a}]/2$, $a = 1,2,3,4$, and in this case the index ${\mathbf j}$ in Eq.~\eqref{eq:FK} is short for ${\mathbf j} = (j, D)$. In contrast, for the model with $N_f = 4$ Kitaev chains we use Eq.~\eqref{eq:MZMMPM} so that $c_{{\mathbf j},1} =[\gamma_{j,{+}}^{1} + i\gamma_{j,{+}}^{2}]/2$, $c_{{\mathbf j},2} =[\gamma_{j,{-}}^{1} + i\gamma_{j,
{-}}^{2}]/2$ $c_{{\mathbf j},3} =[\gamma_{j,{+}}^{3} + i\gamma_{j,{+}}^{4}]/2$, and  $c_{{\mathbf j},4} =[\gamma_{j,{-}}^{3} + i\gamma_{j,
{-}}^{4}]/2$,  
and we can identify ${\mathbf j} = j$. At the edge {and for $(J,g) = (1,1/2)$}, $c_{{\mathbf j},a} \rightarrow (-1)^a c_{{\mathbf j},a}$ under one Floquet cycle.

{As will be clear shortly, the form of these interactions is chosen to allow for analytically exact results at special points in parameter space.}

\subsection{Edge Green's function}

\label{sec:FKFloquetEdge}

As in Sec.~\ref{sec:Edge}, we consider a Floquet unitary 
\begin{equation}
\hat U_F = \hat U_g \hat U_J \hat U_w
\end{equation}
with less manifest symmetries but which is experimentally more realistic (see App.~\ref{app:Translation} for details about the basis conversion).

We further concentrate on three fine tuned cases: Model 1 with $N_f =8, (J,g) = (1/2,0)$; model 2 with $N_f =8, (J,g) = (1/2,1)$; and model 3 with $N_f =4, (J, g) =(1,1/2)$. In all three cases, the edge states are perfectly localized and the edge Green's function is essentially entirely determined by the few-body physics of Eq.~\eqref{eq:FK} in the 16-dimensional Hilbert space of edge zero/$\pi$ modes. Depending on the model under consideration (eight or four Kitaev chains) we implement the FK interaction differently, see Sec.~\ref{sec:ModelsIA}.

{We evaluate the edge Green's functions with respect to the thermal density matrix 
\begin{equation}
    \hat \rho= \frac{e^{- \beta \hat H_w}}{\tr[e^{- \beta \hat H_w}]}
\end{equation}
in the two limits $\beta \rightarrow \infty$ (zero temperature) and $\beta \rightarrow 0$ (infinite temperature). Specifically, we find that the Green's function for zero modes is diagonal, i.e.~for model 1 describing physics at the lower blue cross in Fig.~\ref{fig:Model} b)
\begin{subequations}
\begin{align}
     \mathbf G^{\alpha, \alpha'}_{\rm R}(n)  \equiv - i {\theta(n)} \langle \{\gamma_{1,L}^\alpha(n),\gamma_{1,L}^{\alpha'}(0)\} \rangle = \delta^{\alpha {\alpha'}}\mathbf G^{(0)}_{\rm R}(n) 
\end{align}
for model 2 describing physics at the upper blue cross in Fig.~\ref{fig:Model} b)
\begin{align}
     \mathbf G^{\alpha, \alpha'}_{\rm R}(n)  \equiv - i {\theta(n)} \langle \{\gamma_{1,L}^\alpha(n),\gamma_{1,L}^{\alpha'}(0)\} \rangle = \delta^{\alpha {\alpha'}}\mathbf G^{(\pi)}_{\rm R}(n), 
\end{align}
while for model 3 describing physics at the red cross in Fig.~\ref{fig:Model} b) ($\xi, \zeta = \pm$)
\begin{align}
     \mathbf G^{\alpha, \alpha'}_{\rm R}(n)  \equiv &- i {\theta(n)} \langle \{\gamma_{1,\xi}^\alpha(n),\gamma_{1,\zeta}^{\alpha'}(0)\} \rangle \notag\\
     &= \delta^{\alpha \alpha'} \delta_{\xi, \zeta} 
      [\delta_{\xi, -}\mathbf G^{(0)}_{\rm R}(n) +\delta_{\xi, +}\mathbf G^{(\pi)}_{\rm R}(n)].
\end{align}
\end{subequations}
}

We find that the Green's function of MZMs is 
(see Appendix \ref{app:EdgeGF} for derivations and $\Delta_{1,2,3} = E_{1,2,3} \pi/2$)
\begin{widetext}
\begin{equation}
   \mathbf G^{(0)}_{\rm R}(n) = - i \frac{\theta(n)}{4}\begin{cases}  8 \cos \left(\Delta_1 n \right ), & \beta \rightarrow \infty, \\
    \cos \left(\Delta_1 n \right ) + \cos \left(\Delta_2 n  \right ) +  6 \cos \left(\Delta_3 n  \right ), & \beta \rightarrow 0, 
    \end{cases}
    \label{eq:DiscreteTimeGF}
\end{equation}
or in Fourier space
    \begin{align}
    \mathbf G^{(0)}_{\rm R}(\Omega) 
    & =  \begin{cases}  \frac{\sin(\Omega)}{\cos(\Delta_1 ) - \cos(\Omega^+)}, & \beta \rightarrow \infty, \\
    \frac{1}{8}\frac{\sin(\Omega)}{\cos(\Delta_1) - \cos(\Omega^+)} + \frac{1}{8}\frac{\sin(\Omega)}{\cos(\Delta_2) - \cos(\Omega^+)}+\frac{6}{8}\frac{\sin(\Omega)}{\cos(\Delta_3) - \cos(\Omega^+)}, & \beta \rightarrow 0. 
    \end{cases}
    \label{eq:GreensfunctionsEdgeMaintext}
\end{align}
\end{widetext}
The Green's function $\mathbf G^{(\pi)}_{\rm R}(\Omega)$ for M$\pi$Ms 
is identical to \eqref{eq:DiscreteTimeGF},\eqref{eq:GreensfunctionsEdgeMaintext} except that $\Delta_{1,2,3}  \rightarrow \Delta_{1,2,3} + \pi$.\footnote{The attentive reader might have noticed that the numerator of Eq.~\eqref{eq:GreensfunctionsEdgeMaintext} at $\beta \rightarrow \infty$ contains a resonances at $\Delta_1 = \pm \pi E_1/2$ while in the case of continuum time evolution, Eq.~\eqref{eq:GFEquilibrium}, the resonance occurs at $\pm E_1$. This is entirely due to our convention $\hat U_w = e^{- i \pi \hat H_w/2}$ of the Floquet unitary and is not of further physical relevance.  } 

 These results are central to this paper and illustrated in Fig.~\ref{fig:EdgeSpectralWeight} and also Fig.~\ref{fig:JW} c) below. First, it is worthwhile to recapitulate that edge zeros and edge poles of the Green's function both appear already in the non-interacting limit $w = 0$. The Green's function zeros pinned at zero and $\pi$ persist at finite $w$ and should be contrasted to the case of continuum time evolution, {cf. 4 g)}. Second, we highlight that infinitesimal $w$ opens up a many-body gap so that the spectral weight is shifted away from zero/$\pi$. We highlight that a single pole splits up into two poles and one zero as $w$ is increased. Third, as apparent from Eq.~\eqref{eq:GreensfunctionsEdgeMaintext}, the zero temperature spectral weight contains only one typical energy difference $\Delta_1$, while the infinite temperature Green's function contains information about several transition frequencies $\Delta_{1,2,3}$. 
 
 Finally, one may wonder if the non-interacting Green's function in the topologically trivial regime (denoted (0,0) in Fig.~\ref{fig:Model} b) may display similar phenomenology as Eq.~\eqref{eq:GreensfunctionsEdgeMaintext}, Fig.~\ref{fig:EdgeSpectralWeight}. The answer is negative as particularly manifest in the limit $J = 0$. In this case the real space Green's function is local and given by a matrix in $L/R$ space (see Eq.~\eqref{eq:FreeFermionGFs} at $J = 0$)
 \begin{equation}
     G_{\rm R}^{jj'}(\Omega) = \tilde G^{jj'}_{\rm R}(\Omega) = - i  \delta_{jj'}\frac{1 + e^{i \Omega^+} e^{i \pi g \sigma_y}}{1 - e^{i \Omega^+}e^{i \pi g \sigma_y}}.
 \end{equation}

 We readily see that so long $g \notin \{0,1\}$, {$\det [G_{\rm R}(\Omega)] \neq 0$} at $\Omega = 0, \pi$, {i.e.~there are no Green's function zeros and, instead, the Green's function contains finite off-diagonal matrix elements in $R/L$ space}. In contrast the determinant of the many-body Green's function (Eq.~\eqref{eq:GreensfunctionsEdgeMaintext} at the edge) does contain true zeros.

\begin{figure}
    \centering
\includegraphics[width = 1 \linewidth]{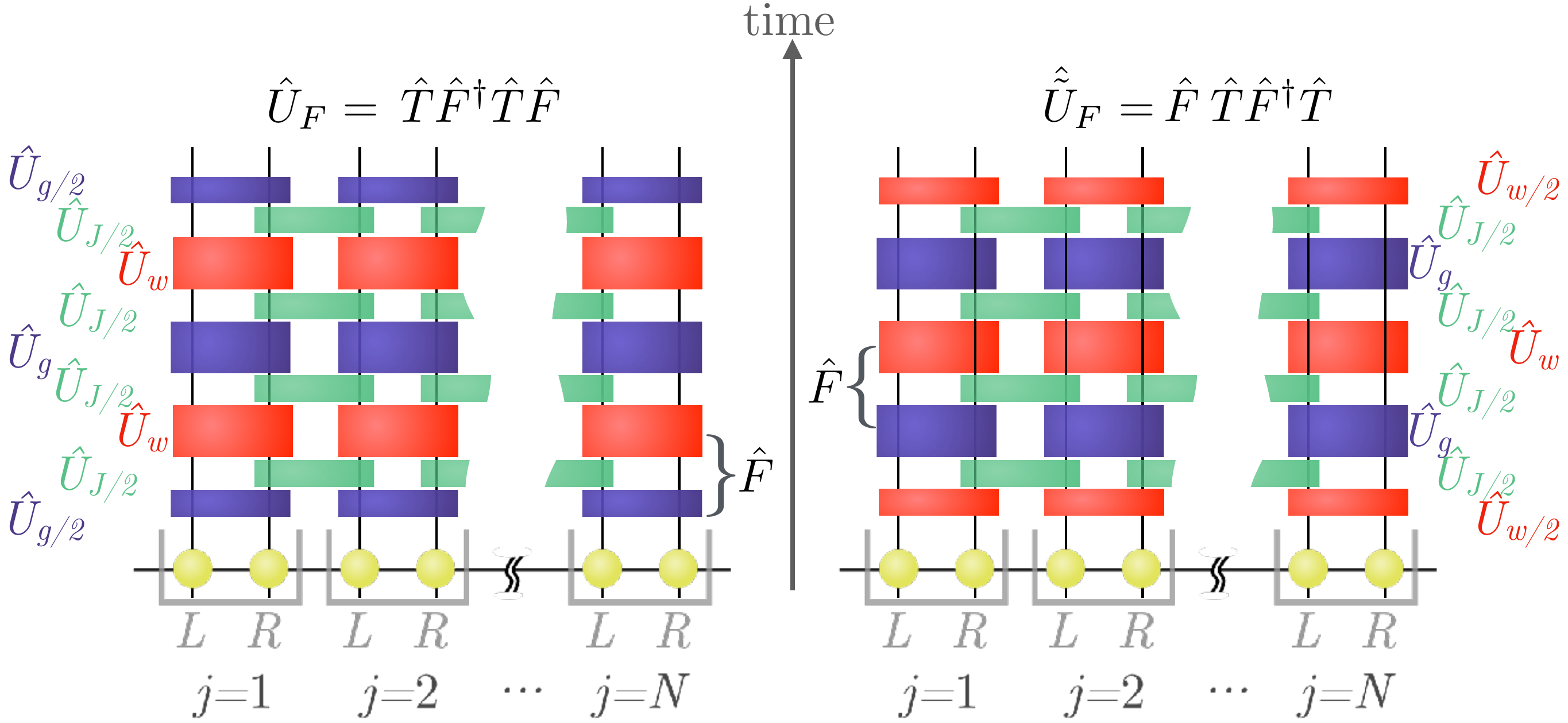}
    \caption{ Graphical illustration of the two interacting Floquet cycles, Eqs.~\eqref{eq:FloquetCycles}
}
    \label{fig:FloquetCycles}
\end{figure}

\subsection{Bulk Green's function}

\label{sec:FKFloquetBulk}

\begin{figure}
    \centering
    \includegraphics[scale = 1]{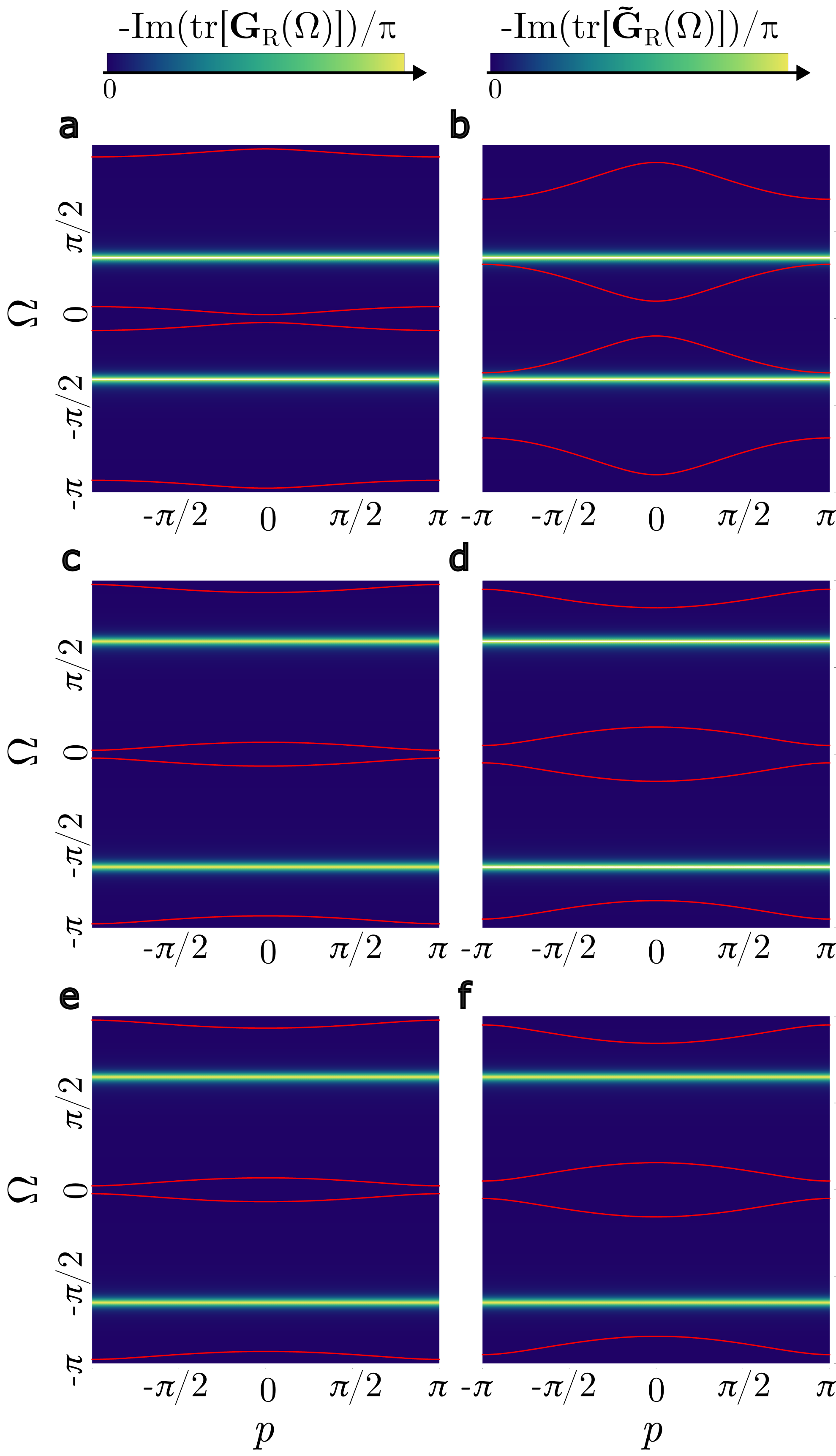}
    \caption{ Illustration of zeros of the bulk Green's functions Eq.~\eqref{eq:BulkPertGFModel1} (panel a), Eq.~\eqref{eq:GtildaModel1} (b), Eq.~\eqref{eq:BulkPertGFModel2} (c),
    Eq.~\eqref{eq:GtildaModel2} (d), 
    Eq.~\eqref{eq:BulkPertGFModel3} (e),
    Eq.~\eqref{eq:GtildaModel3} (f). Color plots illustrate the spectral weight, red lines Green's function zeros. Panels a),b) are plotted for $(J,g) = (0.2,0.1)$, c),d) are plotted for $(J,g) = (0.2,0.9)$, e),f) are plotted for $(J,g) = (0.8,0.1)$. In all figures we used $w = 0.05, \lambda = 1$ and a finite broadening $\Omega^+ \rightarrow \Omega + i \Gamma$ with $\Gamma = 0.05$.
}
    \label{fig:BulkSpectralWeight}
\end{figure}
{The previous section demonstrates that, by means of Floquet-SMG, edge MZM and M$\pi$M of the non-interacting cycle may be gapped out, leaving edge Green's function zeros at $\Omega = 0$ and $\Omega = \pi$ behind. With the perspective of a bulk-boundary correspondence for Floquet Green's function zeros, in this section we study the interacting Green's functions $\mathbf G_R(\Omega,p), \tilde{\mathbf G}_R(\Omega,p)$ defining bulk invariants $N_1,\tilde N_1$ introduced in Eq.~\eqref{eq:GFINvariants}. To this end}
it is convenient to return to {two related} Floquet cycles as in Eq.~\eqref{eq:TwoCyclesGeneral} 
which ensures all the symmetries of the Green's function discussed in Sec.~\ref{sec:GFTopoInv}. We here concentrate on, cf. Fig.~\ref{fig:FloquetCycles} 
\begin{subequations}
\begin{align}
\hat U_F &= \hat U_{g/2}\hat U_{J/2}\hat U_w \hat U_{J/2}\hat U_{g/2}
\equiv \hat T \hat F^\dagger_0 \hat T \hat U_w \hat F_0, \\
\hat{\tilde U}_F & = \hat U_{w/2} \hat U_{J/2} \hat U_g \hat U_{J/2} \hat U_{w/2}.
\end{align}
\label{eq:FloquetCycles}
\end{subequations}

We consider a weak perturbation about limits at which the interacting bulk system decays into a collection of FK-quantum dots (essentially behaving like the edge states discussed in the previous section). The main evolution is thus due to $\hat U_w$ and we employ Floquet perturbation theory at zero temperature, at $\lambda = 1$ and near trivial limits of free evolution marked by circles in Fig.~\ref{fig:Model} b), see App.~\ref{app:BulkGFFull} for a derivation.

\begin{subequations}
The perturbative solutions for the Green's functions of the strongly interacting system are conveniently expressed using the single-particle Hamiltonian of the Kitaev chain and the extended Kitaev chain (i.e., of the cluster-Ising model~\cite{DohertyBartlett2009, SonVedral2011})
\begin{align}
h_{(J,g)}(p) &= (J\cos(p) - g) \sigma_y - J \sin(p) \sigma_x,\\
\tilde h_{(J,g)}(p) &= (J\cos(p) - g\cos(2p)) \sigma_y \notag \\
&- (J \sin(p) - g \sin(2p)) \sigma_x],
\end{align}
which, crucially, have the same eigenvalues.

For $N_f =8$ chains near $(J,g) = (0,0)$ (cf. Fig.~\ref{fig:Model} b), bottom left circle) we find 
\begin{align}
    \mathbf G_{\rm R}(\Omega,p)  &\simeq  \frac{1}{2}\frac{\sin(\Omega) +\frac{\pi}{2}\cos(7 \pi w) h_{(J,g)}(p)}{\cos(7 \pi w) - \cos(\Omega^+)},  \label{eq:BulkPertGFModel1}
    \\
    \tilde{\mathbf G}_{\rm R}(\Omega,p) &\simeq  \frac{1}{2}\frac{\sin(\Omega) +\frac{\pi}{2} h_{(J,g)}(p)}{\cos(7 \pi w) - \cos(\Omega^+)}.  \label{eq:GtildaModel1}
\end{align}

For $N_f = 8$ near $(J,g) \simeq (0,1)$ (which is expected to adiabatically connect to model 2 if $J>1-g$ and corresponds to Fig.~\ref{fig:Model} top left circle)
\begin{align}
    \mathbf G_{\rm R}(\Omega,p)  &\simeq  -\frac{1}{2}\frac{\sin(\Omega) -\frac{\pi}{2}\cos(7 \pi w) h_{(J,g-1)}(-p)}{\cos(7 \pi w) + \cos(\Omega^+)},  \label{eq:BulkPertGFModel2}
    \\
\tilde{\mathbf G}_{\rm R}(\Omega,p) & = -\frac{1}{2}\frac{\sin(\Omega) -\frac{\pi}{2} h_{(J,g-1)}(p)}{\cos(7 \pi w) + \cos(\Omega^+)}.  \label{eq:GtildaModel2}
\end{align}

Finally to make connection to model 3 featuring both MZMs and M$\pi$Ms at the edge of the free system, we consider $N_f = 4$ chains driven near $(J,g) \simeq (1,0)$ (cf. bottom right circle in Fig.~\ref{fig:Model} b). 
We find
\begin{align}
    \mathbf G_{\rm R}(\Omega,p)  &\simeq  -\frac{1}{2}\frac{\sin(\Omega) -\frac{\pi}{2}\cos(7 \pi w) h_{(J-1,g)}(p)}{\cos(7 \pi w) + \cos(\Omega^+)},  \label{eq:BulkPertGFModel3}
    \\
   \tilde{\mathbf G}_{\rm R}(\Omega, p) &\simeq -\frac{1}{2}\frac{\sin(\Omega) -\frac{\pi}{2} \tilde{h}_{(J-1,g)}(p)}{\cos(7 \pi w) + \cos(\Omega^+)} .\label{eq:GtildaModel3}
\end{align}
\label{eq:BulkGFWithIA}
\end{subequations}

These analytic expressions constitute the third major result of this work and are illustrated in Fig.~\ref{fig:BulkSpectralWeight} (there we push $(J,g)$ to the limit of perturbation theory for a better visualization of the dispersion). Several comments are in order: 
First, it is worthwhile to highlight that the presented results for the momentum space Green's function $\mathbf{G}(\Omega, p)$ at zeroth order in perturbation theory ($h,\tilde{h}=0$) and the real space edge Green's function Eq.~\eqref{eq:GreensfunctionsEdgeMaintext} are consistent with each other (the factor 1/2 stems from the definition of the Fourier transform Eq.~\eqref{eq:FT}). Both $\mathbf G_{\rm R}(\Omega,p)$ and $\tilde{\mathbf G}_{\rm R}(\Omega,p)$ manifestly fulfill all symmetries Eq.~\eqref{eq:FTConstraints}. Note that we suppressed flavor indices $\alpha$ in Eq.~\eqref{eq:BulkGFWithIA} as the Green's functions are trivial in flavor space.

Second, the bands of poles in $\mathbf{G}(\Omega, p), \tilde{
\mathbf G}(\Omega,p)$ are completely flat in the present first order perturbation theory. They do not contribute to the topological winding numbers $N_1, \tilde N_1$, Eq.~\eqref{eq:GFINvariants}. 

Third, while these results are perturbative in $(J,g)$, $(J, 1-g)$,  $(J-1, g)$, respectively, they are not perturbative in $\Omega, w$. However, when additionally expanded in the small $\Omega, w$ limit,  Eqs.~\eqref{eq:BulkPertGFModel1},\eqref{eq:GtildaModel1} are consistent with results reported in Ref.~\cite{YouXu2014} for analogous equilibrium models. 

Fourth, there are bands of zeros in $\mathbf G_{\rm R}(\Omega ,p)$, which in turn can or cannot be topological (depending on the properties of the Hamiltonians $h_{(J,g)}(p),\tilde{h}_{(J,g)}(p)$). On the one hand there are zeros near $\Omega = 0$. As in the equilibrium counterpart~\cite{YouXu2014}, the "Hamiltonian" of these zeros is opposite to the Hamiltonian of poles in the corresponding non-interacting system (see App.~\ref{app:BulkGFPertTheory} for details). On the other hand, 
additional bands of zeros near $\Omega = \pi$ with opposite "Hamiltonian" appear. 

{Fifth, it is manifest that within perturbation theory, the Green's-function-based topological invariants Eq.~\eqref{eq:GFINvariants} are entirely determined by $h_{(J,g)}(p), \tilde h_{(J,g)}(p)$ describing the topology of Green's function zeros. Crucially, the topological invariants $N_1, \tilde N_1$ (or equivalently $\nu_0, \nu_\pi$) obtained from Green's functions of the strongly coupled systems Eq.~\eqref{eq:BulkGFWithIA} are the same as the topological invariants for free systems at the same parameter range in the $J-g$ phase diagram, Fig.~\ref{fig:Model} b). 

Finally, when taken together with the appearance of boundary Green's function zeros at $\Omega = 0$ and $\Omega =\pi$ in the topological regime, Fig.~\ref{fig:EdgeSpectralWeight}, this can be interpreted as a generalized bulk-boundary correspondence for topological Floquet Green's function zeros.}

\section{Circuit implementation}
\label{sec:Circuit}

In view of the recent progress in the digital emulation of Floquet dynamics, in particular also in view of reports of noise-resilient Majorana $\pi$-modes in such experiments~\cite{MiAbanin2022,SchmidvonOppen2024} we here discuss the implementation of the above physics in a qubit array
using Jordan-Wigner (JW) transformation. To be concrete we present such a JW transformation for the most interesting case of $N_f = 4$ chains (which can host the nontrivial Floquet-boundary zeros discussed in Sec.~\ref{sec:FKFloquetEdge}) .

\subsection{Jordan-Wigner transformation}

\begin{figure}
    \centering
\includegraphics[scale=1]{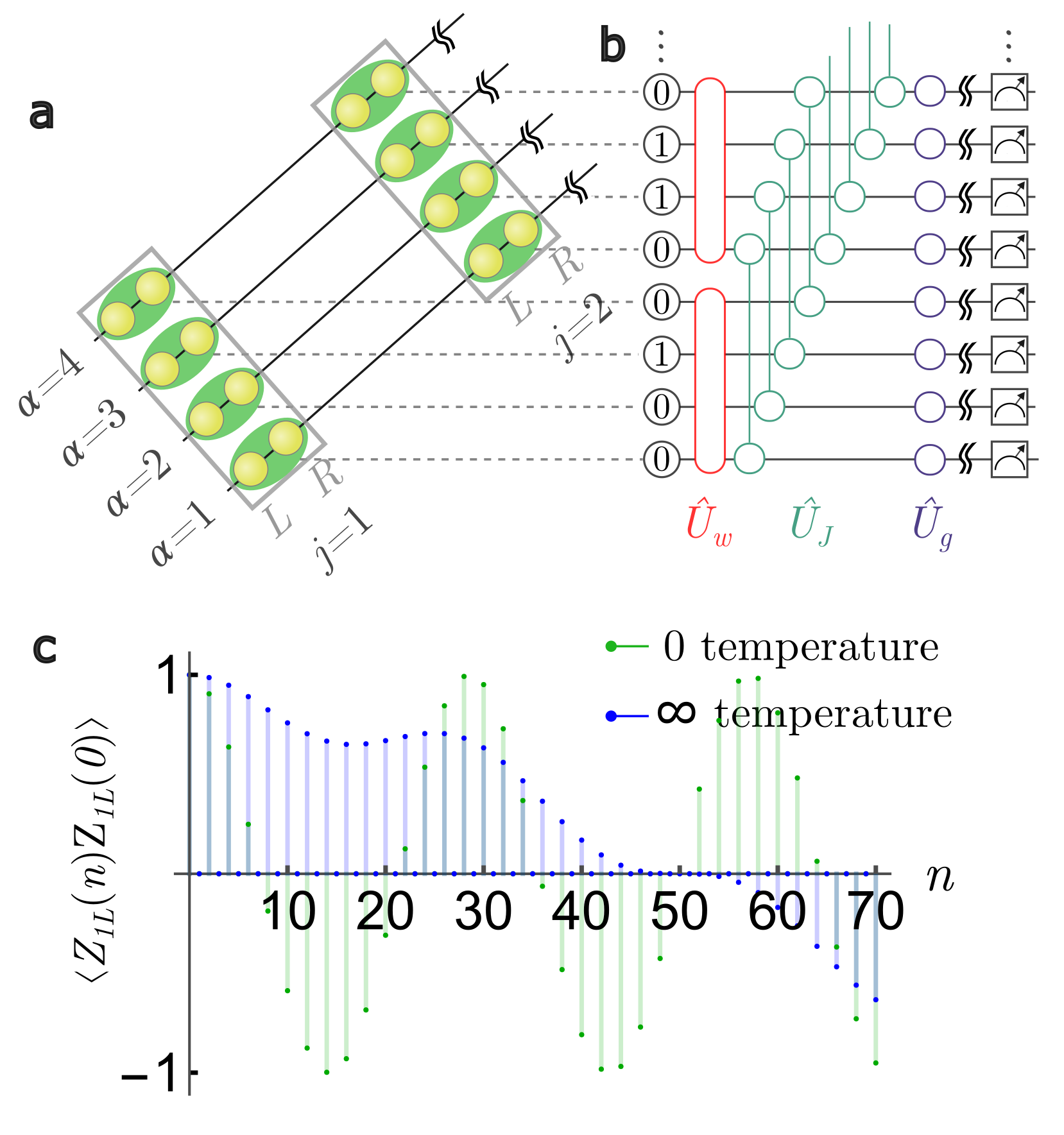}
    \caption{a) Illustration of the system with $N_f = 4$ Kitaev chains which is Jordan-Wigner mapped to a quantum chip (the string runs from bottom to top along the array of qubits), panel b). The qubits are initialized as a random product state. The Floquet unitary $\hat U_F$ is decomposed into subsequent application of $\hat U_w$, cf. Sec.\ref{sec:FKFloquetEdge} and $\hat U_J $, $\hat U_g$ (previously implemented on quantum chips). The Floquet unitary $\hat U_F$ acts $n$-times (indicated by wavy lines) before the final bitstring is measured. c) The edge autocorrelation function is direcetly related to the retarded edge Green's function.}
    \label{fig:JW}
\end{figure}

For the implementation of the fermionic model on a qubit system we use the Pauli-X operator to define the local fermionic parity $X_{j, \alpha} = i \gamma_{j,L}^\alpha \gamma_{j,R}^\alpha$ and $Z_{j,\alpha} = S_{j, \alpha} \gamma_{j,L}^\alpha, Y_{j,\alpha} =  S_{j,\alpha} \gamma_{j, R}^\alpha$ where $S_{j, \alpha}$ is the product of all Majoranas on sites before $j,D $ ordered along a JW string. Reversely, $\gamma_{j,L}^\alpha = \tilde S_{j,\alpha} Z_{j,\alpha}, \gamma_{j,R}^\alpha = \tilde S_{j,\alpha} X_{j,\alpha} Z_{j,\alpha}$ where $\tilde S_{j,\alpha}$ is a product over all $X$ operators at sites before $j,\alpha$. Our choice of the JW string corresponds to the ordering of qubits (from bottom to top) in Fig.~\ref{fig:JW} a).

\subsection{Initial state and observables}

We next assume an initial state which is a random bitstring, e.g. $\ket{\psi_0} =\ket{0111010100111\dots}$ and measure $\braket{Z_{j,D }^\alpha(n)} \equiv \braket{\psi_0 \vert [\hat U_F^\dagger ]^n Z_{j,D }^\alpha \hat U_F^n\vert \psi_0}$ and repeat this measurement for various $t \in \mathbb N_0$. Since the initial state is an eigenstate of $Z_{j,D }^\alpha$ this expectation value contains all information about the autocorrelator $\braket{Z_{j,D }^\alpha(n)Z_{j,D }^\alpha(0)} = \braket{Z_{j,D }^\alpha(n)}\braket{Z_{j,D }^\alpha(0)}$. Moreover, by Jordan-Wigner transformation the spin autocorrelation function at the edge determines the Majorana Green's function. 
Averaging this result over equally weighted initial states results in the infinite temperature retarded Green's function. We can thus summarize for the special points in parameter space

\begin{equation}
  \braket{Z_{1,L }^\alpha(n)Z_{1,L }^\alpha} = \begin{cases}
  i \frac{\mathbf G_{\rm R}^{(0)}(n)}{2}, (N_f,J,g) = (8,\frac{1}{2},0), \\
  i \frac{ \mathbf G_{\rm R}^{(\pi)}(n)}{2}, (N_f,J,g) = (8,\frac{1}{2},1), \\
  i \frac{\mathbf G_{\rm R}^{(0)}(n)+ \mathbf G_{\rm R}^{(\pi)}(n)}{4}, (N_f,J,g) = (4,1,\frac{1}{2}),
  \end{cases}
  \label{eq:Autocorrel}
\end{equation}
where $\mathbf G_{\rm R}^{(0,\pi)}$ are defined in Eq.~\eqref{eq:DiscreteTimeGF} and the limit $ \beta \rightarrow 0$ is assumed everywhere. This expression is valid for all positive $n$. At $n =0$ the autocorrelation function is unity while $i \mathbf G_{\rm R}(0)/2 = 1/2$.

The third expressions of Eq.~\eqref{eq:Autocorrel} is illustrated in  Fig.~\ref{fig:JW} c). Note the alternation of the edge Green's function between finite and vanishing values in view of the superposition Eq.~\eqref{eq:MZMMPM}. A cosine of the edge autocorrelation in discrete real time starting at unity corresponds, once Fourier transformed, to a Green's function with Green's function with edge zeros.

\subsection{Implementation of circuit}

With this choice of the Jordan-Wigner transformation, $\hat U_g$ corresponds to a single qubit gate (rotation about the $X$ axis), $\hat U_J $ is a two-qubit gate (typically decomposed into two $\sqrt{Z Z}$ gates), see Fig.~\ref{fig:JW} b) and the experimental implementation~\cite{MiAbanin2022}. 
The missing ingredient for gate-based simulation of topological Floquet is a circuit generating the Fidkowski-Kitaev interaction. 

We concentrate on the case $N_f = 4$ chains. Using the Jordan-Wigner transformation discussed above, Fig.~\ref{fig:JW} a) we consider a set of 4 qubits interacting with each other by means of $\hat U_w$. As mentioned $\hat H_w$ has a GHZ ground state and by now, fast algorithms to create such a state from a product state are available.~\cite{Bao2024} 
We further observe that $\hat U_w$ at special values $w = 1/3, 3/8, 5/8$ essentially corresponds to generalized Toffoli gates (see Appendix~\ref{app:Toffoli}) which attracted increased interest in recent years~\cite{YouZhang2022,NieSun2024}.

\section{Summary and Outlook}

In summary, in this work we studied topological Green's function zeros in Floquet systems of symmetry class BDI and discussed their possible experimental implementation in NISQ devices. We first introduced Green's-function-based topological invariants, Eq.~\eqref{eq:GFINvariants}. While designed for interacting systems, these two invariants reproduce the free fermion topological Floquet invariants~\cite{CardosoMitra2025} when interactions are switched off. One of these invariants relates to the the continuum time Green's function invariant~\cite{Gurarie2011, ManmanaGurarie2012} in the Trotter limit of vanishing Floquet time.
A major focus of our work are Green's function zero eigenvalues for Floquet systems. While such Green's function zeros generally do not appear in free continuum time systems, we highlight that they do generally appear in free Floquet systems. Along with the poles, these Green's function zeros have an impact on the quantized value of Floquet Green's-function-based topological invariants. Furthermore, we consider interaction induced Green's function zeros. We take inspiration from previous works~\cite{FidkowskiKitaev2010,YouXu2014} on the symmetric mass generation in topological superconductors and transcribe the mechanism from equilibrium continuum time evolution to the Floquet setting. We concentrate on three instances in parameter space which are known to illustrate interaction induced trivialization according to the reduced classification $\mathbb Z \times \mathbb Z \rightarrow \mathbb Z_8 \times \mathbb Z_4$ of interacting Floquet SPTs~\cite{PotterVishwanath2016} of the present symmetry class BDI. Indeed, the interacting Floquet system loses all of its topological boundary excitations at quasienergy zero and $\pi$, but always sustains Green's function zeros at the corresponding frequencies, Eqs.~\eqref{eq:GreensfunctionsEdgeMaintext} and Fig.~\ref{fig:EdgeSpectralWeight}. Similarly the bulk Green's functions, Eq.~\eqref{eq:BulkGFWithIA}, which we evaluate perturbatively in the limit of predominant interactions, ceases to display topological bands of excitations, but does display topological bands of zeros. {Somewhat analogously to the equilibrium case, the Floquet Green's-function-based topological invariants $N_1, \tilde N_1$ are now determined entirely by Green's function zeros, and turn out to be the same as in corresponding non-interacting systems.}
{They thereby account for the newly appeared Green's function boundary zeros at $\Omega = 0, \pi$ which is consistent with} a generalized bulk-boundary correspondence in "trivialized" Floquet SPT phases {(in the sense of symmetric mass generation~\cite{PotterVishwanath2016})}. As explicitly illustrated in Figs.~\ref{fig:EdgeSpectralWeight}, \ref{fig:JW}, Green's function zeros have distinct experimentally observable consequences.

We conclude with a short outlook: On the one hand, the experimental implementation in terms of a quantum circuit strongly depends on the feasibility of the implementation of $\hat U_w$ using reasonably shallow circuits.  
On the theoretical end, a generalization of the concept of topological Floquet Green's function zeros to different symmetry classes and dimensions is desirable. Moreover, as mentioned in the introduction, apart from symmetric mass generation there are other mechanisms, including fractionalization, which are of particular interest in this context.

\section{Acknowledgements}

 Support for this research was provided by the Office of the Vice Chancellor
for Research and Graduate Education at the University
of Wisconsin–Madison with funding from the Wisconsin
Alumni Research Foundation. This research was supported in part by grants NSF PHY-1748958 and PHY-
2309135 to the Kavli Institute for Theoretical Physics
(KITP) (EJK) and by the US Department of Energy, Office of Science, Basic Energy Sciences, under Award No. DE-SC0010821 (AM). EJK acknowledges hospitality
by the KITP.

\appendix 

\section{Bulk topological invariant and symmetry constraints}

This appendix contains details on the symmetry constraints of the Green's functions and the Green's-function-based topological invariants.

\subsection{Review of symmetries in first and second quantization.}
\label{app:BDI}

Time reversal symmetry, as defined for second quantized operators Eq.~\eqref{eq:Symmetries}, immediately implies the standard time reversal symmetry of first quantized single particle Hamiltonians $h(t,p) = \Gamma h^T(-t, -p) \Gamma$, Eq.~\eqref{eq:FirstQuantSymmetriesHam}.

It is a little more subtle to highlight that the ubiquituos parity symmetry implies particle-hole symmetry of non-interacting Hamiltonians and corresponding Floquet cycles. Parity symmetry enforces terms with even number of Majorana operators in the Hamiltonian, specifically in the non-interacting case
\begin{equation}
    \hat H = i \gamma_A h_{AB} \gamma_B.
\end{equation}
Hermiticity of $\hat H$ and the basic properties of Majorana fermions ensure that that the first quantized matrix Hamiltonian fulfills $h = h^\dagger = -h^T$. The second equality sign implies particle-hole symmetry, i.e. $h(p) = -h^T(-p)$ for translationally invariant systems and in momentum space, cf.~Eq.~\eqref{eq:FirstQuantSymmetriesHam}.

If only parity symmetry is needed to impose $h = - h^T$, why are then not all free fermion Hamiltonians particle-hole symmetric? To answer this question, we follow the standard prescription of the Altland-Zirnbauer classification to first block-diagonalize a given $h$ with respect to all available unitary symmetries and only afterwards discuss the symmetry classes block-by-block. In the presence of particle-number conservation $U(1) \simeq O(2)$, the $2N \times 2N$ matrix $h$ can be decomposed as $h = iA \mathbf 1_\sigma + S \sigma_y$, where $i\sigma_y$ here is the generator of $O(2)$ and $A = -A^T$, $S = S^T$ are $N \times N$ matrices. Using a unitary transformation, $h$ can be brought to standard Nambu form (as in superconductors, but without pairing)
\begin{equation}
    h = \left (\begin{array}{cc}
        \mathfrak{h} & 0  \\
        0 & -\mathfrak{h}^T 
    \end{array} \right ),
\end{equation}
where each block $\mathfrak{h} = i A +S$ may or may not be particle-hole symmetric.

\subsection{Symmetry constraints on the Green's function}
\label{app:SymmetryConstraintsGF}

In this section we use Eq.~\eqref{eq:UF} to deduce properties of the discrete time Green's functions, Eq.~\eqref{eq:FloquetGFs}. 
\paragraph*{Particle-hole symmetry.} The particle-hole symmetry inherent to the Majorana nature of the correlator implies 
\begin{align}
     [{\mathbf G}^{AB}_{\rm R}(\Omega)]^* & = i \sum_n \theta(n) e^{- i \Omega^- n} \langle \{ \hat{{U}}_F^{-n} \gamma_A \hat{{U}}_F^n, \gamma_B  \} \rangle^* \notag\\
     &= - {\mathbf G}^{AB}_{\rm R}(-\Omega),
\end{align}
and analogously for $\tilde{\mathbf G}_{\rm R}(\Omega)$.

\paragraph*{Time-reversal symmetry.} 

We find 
\begin{align}
[\Gamma {\mathbf G}_{\rm R} (\Omega) \Gamma ]^{AB} & = - i \sum_{n = 0}^\infty e^{i \Omega^+ n} \langle \{ \hat U_F^{-n}  \hat T \gamma_A \hat T \hat U_F^n, \hat T \gamma_B \hat T \} \rangle \notag  \\
    & = - i \sum_{n = 0}^\infty e^{i \Omega^+ n}  \langle \{ \underbrace{\hat T \hat U_F^{-n} \hat T}_{\hat U_F^{n}} \gamma_A  \underbrace{\hat T \hat U_F^{n} \hat T}_{\hat U_F^{-n }}, \gamma_B \} \rangle\notag  \\
     & = - i \sum_{n = -\infty}^{0} e^{-i \Omega^+n}  \langle \{\hat U_F^{-n} \gamma_A \hat U_F^{n}, \gamma_B \} \rangle \notag \\
    &  = - {\mathbf G}_{\rm A}^{AB}(- \Omega).
\end{align}
At the second equality sign we used the time reversal invariance of the density matrix defining the average. {The analogous equation for $\tilde{\mathbf G}_{\rm R}(\Omega)$ follows by replacing $\hat U_F \rightarrow \hat{\tilde U}_F$.}
Fourier transformation of the previous equations yields Eqs.~\eqref{eq:FTConstraints} of the main text. 

\subsection{Green's-function-based topological invariant}
\label{app:TopoInvProperties}

We here demonstrate the topological invariants, Eqs.~\eqref{eq:GFINvariants}, are quantized and real. 

For the reality of $N_1$, we use the first equality sign in Eqs.~\eqref{eq:FTConstraints} a):

\begin{align}
    N_1^* &=  \left (\int_{- \pi}^\pi \frac{dp}{4 \pi i} \tr[\Gamma \mathbf G_{\rm R}^{-1} (\Omega, p)\partial_p \mathbf G_{\rm R}(\Omega, p)]_{\Omega = 0} \right )^* \notag \\
    & = -\int_{- \pi}^\pi \frac{dp}{4 \pi i} \tr[\Gamma \mathbf G_{\rm R}^{-1,*} (\Omega, p)\partial_p \mathbf G^*_{\rm R}(\Omega, p)]_{\Omega = 0} \notag \\
    & = -\int_{- \pi}^\pi \frac{dp}{4 \pi i} \tr[\Gamma \mathbf G_{\rm R}^{-1} (-\Omega, -p)\partial_p \mathbf G_{\rm R}(-\Omega,- p)]_{\Omega = 0} \notag \\
    & = \int_{- \pi}^\pi \frac{dp}{4 \pi i} \tr[\Gamma \mathbf G_{\rm R}^{-1} (\Omega, p)(\partial_p) \mathbf G_{\rm R}(\Omega, p)]_{\Omega = 0} \notag \\
    & = N_1.
\end{align}
The reality of $\tilde N_1$ follows analogously from the first equality sign in Eq.~\eqref{eq:FTConstraints} b). We also use that $ \mathbf G_{\rm R} (-\Omega,p) =\mathbf G_{\rm R} (\Omega,p)$ for $\Omega = 0$.

Next, we consider the change in $\ptwiddle{N}_1$ under a small change $\ptwiddle{\mathbf G} \rightarrow \ptwiddle{\mathbf G} + \delta \ptwiddle{\mathbf G}$. We find that the variation of the argument of the $p$ integral is
\begin{align}
   & \delta \tr[\Gamma \ptwiddle{\mathbf G}_{\rm R}^{-1} \partial_p \ptwiddle{\mathbf G}_{\rm R}]_{\Omega = 0} \notag \\
   & = \tr[\Gamma (\ptwiddle{\mathbf G}_{\rm R}^{-1} \partial_p \delta \ptwiddle{\mathbf G}_{\rm R} - \ptwiddle{\mathbf G}_{\rm R}^{-1} \delta \ptwiddle{\mathbf G}_{\rm R} \ptwiddle{\mathbf G}_{\rm R}^{-1}\partial_p \ptwiddle{\mathbf G}_{\rm R})]_{\Omega = 0} \notag \\
    & = \tr[\Gamma (\ptwiddle{\mathbf G}_{\rm R}^{-1} \partial_p \delta \ptwiddle{\mathbf G}_{\rm R} + \partial_p \ptwiddle{\mathbf G}_{\rm R}^{-1}  \delta \ptwiddle{\mathbf G}_{\rm R} )]_{\Omega = 0},
\end{align}
which is a total derivative and thus vanishes under the integral. We used time reversal symmetry and  
our assumption of gapped systems at $\Omega = 0,\pi$, implies, per definition, vanishing spectral weight
\begin{equation}
    0 = \mathbf G_{\rm R}(\Omega, p) - \mathbf G_{\rm A}(\Omega,p) , \quad \Omega = 0,\pi,
\end{equation}
i.e.~the 
retarded and advanced Green's function become the same.

\section{Free bulk Green's function}
\label{app:BulkGF}

This appendix contains details about the matrix valued Green's function of the free fermion Floquet evolution. 

\subsection{Time evolution and Green's function}
Let us note that,
\begin{align}
    U_F =\Gamma F^{\dagger} \Gamma F\,, \tilde{U}_F = FU_F F^{\dagger}.
\end{align}
The time evolved Majorana operators, Eq.~\eqref{eq:FirstQuantTEvol}, imply
the following Green's function in momentum space (we use ${\mathcal A} = (\alpha, D)$)

\begin{subequations}
\begin{align}
    \mathbf G_{\rm R/A}^{{\mathcal A}, {\mathcal B}}(n; p) &= \mp i \theta(\pm n)
    \left [U_F^n \right ]_{{\mathcal A} {\mathcal B}}, \\
    {\tilde{\mathbf G}_{\rm R/A}^{{\mathcal A}, {\mathcal B}}(n; p)}&={{ \mp i \theta(\pm n) \left [\underbrace{\tilde U_F^n}_{= (F U_F F^{\dagger})^n} \right ]_{{\mathcal A} {\mathcal B}}}}.
\end{align}
\end{subequations}

Eqs.~\eqref{eq:FreeFermionGFs} of the main text follow by straightforward application of the Fourier transform, Eq.~\eqref{eq:FTOmegaGF}, using the matrix valued geometric series, for example
\begin{subequations}
\begin{align}
    \mathbf G_{A}(\Omega) & = \frac{i}{2} + i \sum_{n = - \infty}^{-1} e^{i \Omega^-n}U_F^n \notag \\
    & = - \frac{i}{2} \frac{1 + e^{i \Omega^-} U_F}{1 - e^{i \Omega^-} U_F}, \\
    \tilde{\mathbf G}_{A}(\Omega) &= - \frac{i}{2} \frac{1 + e^{i \Omega^-} F U_F F^{\dagger}}{1 - e^{i \Omega^-} F U_F F^{\dagger}}, 
\end{align}
\end{subequations}
Analogously, we find
\begin{subequations}
\begin{align}
    \mathbf G_{R}(\Omega) & =- \frac{i}{2} \frac{1 + e^{i \Omega^+} U_F}{1 - e^{i \Omega^+} U_F},\label{gr1} \\
  \mathbf{  \tilde{G}}_{\rm R}(\Omega) &= 
    - \frac{i}{2} 
    \frac{1 + e^{i \Omega^+} F U_F F^{\dagger}}{1 - e^{i \Omega^+} F U_F F^{\dagger}}\label{gr2}.
\end{align}
\end{subequations}
Also note that
\begin{align}
    \mathbf{G}_{\rm R}(0) = \frac{-i}{2}\{F^{\dagger},\Gamma\}\left[F^{\dagger},\Gamma\right]^{-1}, \label{gr3}\\ 
     \mathbf{\tilde{G}}_{\rm R}(0)=\frac{-i}{2}\{\Gamma, F\}\left[\Gamma,F\right]^{-1}.
     \label{gr4}
\end{align}

\subsection{Free fermion topological invariant}
\label{app:FreeFermionTopoInv}

Employing \eqref{gr3},\eqref{gr4}, we find for Eq.~\eqref{eq:GFINvariants} 
\begin{align}
    N_1 &= \int_{-\pi}^{\pi}\frac{dp}{2\pi i}\tr\biggl[\biggl(\left[F^{\dagger},\Gamma\right]^{-1} - \{F^{\dagger},\Gamma\}^{-1}\biggr)\partial_p F^{\dagger}\biggr] \notag \\
    & = \left ( \int_{-\pi}^{\pi}\frac{dp}{2\pi i}\tr\biggl[\biggl( \underbrace{- \left[F^{\dagger},\Gamma\right]^{-\dagger}}_{=[F,\Gamma]^{-1}} + \underbrace{\{F^{\dagger},\Gamma\}^{-\dagger}}_{=\{F, \Gamma \}^{-1}}\biggr)\partial_p F\biggr] \right )^*,\\
    \tilde N_1&=  \int_{-\pi}^{\pi}\frac{dp}{2\pi i}\tr\biggl[\biggl(\left[F,\Gamma\right]^{-1} - \{F,\Gamma\}^{-1}\biggr)\partial_p F\biggr].
\end{align}

Since we had proven previously that $N_1, \tilde N_1$ are real, one may directly calculate $N_1^*$ and use this final result and obtain 
Eq.~\eqref{eq:TopoFree} of the main text.

\subsection{Floquet unitary for Ising/Kitaev Floquet cycle}
\label{app:FreeIsingFloquet}

This section contains notational details about the bulk Green's function in the absence of interactions. Using the notation of the main text, we can rewrite Eq.~\eqref{eq:FreeFloquet} as $\hat H_{g,J} =\sum_\alpha \sum \gamma_\alpha^\dagger (p) h_{g,J}(p) \gamma_{\alpha}(p) $
\begin{subequations}
\begin{align}
    h_g (p) &= -  g \sigma_y, \\
    h_J (p) & = J [\cos(p) \sigma_y - \sin(p) \sigma_x].
\end{align}
\label{eq:FirstQuantFloquet}
\end{subequations} \noindent 
The second quantized Floquet unitary is then

\begin{align}
\hat U_F &= \prod_{p >0, \alpha} e^{- i \pi \sum_\alpha \sum_{p} \Gamma_\alpha^\dagger (p) h_{g/2}(p) \Gamma_{\alpha}(p)} \notag \\
    &\times e^{- i \pi \sum_\alpha  \Gamma_\alpha^\dagger (p) h_{J}(p) \Gamma_{\alpha}(p)} \notag \\
    &\times  e^{- i \pi \sum_\alpha  \Gamma_\alpha^\dagger (p) h_{g/2}(p) \Gamma_{\alpha}(p)} \notag \\
& = \prod_{p >0, \alpha} e^{- i \pi \sum_\alpha\Gamma_\alpha^\dagger (p) h_{F}(p) \Gamma_{\alpha}(p)} ,
\end{align}
where the effective Floquet Hamiltonian $h_F$ defines the first quantized Floquet unitary. In particular, denoting $F$ as the first-quantized half-period evolution, and $\Gamma$ as the first quantized chiral symmetry,
\begin{align}
    F(p) =  e^{-i \pi h_{J/2}} e^{- i \pi h_{g/2}}, \, \Gamma = \sigma_z,
\end{align}
the first-quantized Floquet unitary 
\begin{align}
    U_F &=    \Gamma F^{\dagger} \Gamma F, \\
       \tilde U_F &= F   \Gamma F^{\dagger} \Gamma,
\end{align}
specifically
\begin{align}
    U_F = e^{-i\pi h_F} =e^{- i \pi h_{g/2}} e^{-i \pi h_{J}} e^{- i \pi h_{g/2}},\\ 
  \tilde  U_F = e^{-i\pi \tilde h_F} =e^{- i \pi h_{J/2}} e^{-i \pi h_{g}} e^{- i \pi h_{J/2}}.
\end{align}
The explicit expressions for $U_F, \tilde U_F$ are given in Eq.~\eqref{eq:Ueff} of the main text. 
We can parameterize $F(p)$ as 
\begin{subequations}
\begin{equation}
     F = \left ( \begin{array}{cc}
       \zeta   & - \eta^* \\
        \eta & \zeta^*
    \end{array} \right ), \,\, |\zeta|^2 + |\eta|^2 =1.
\end{equation}
where 
\begin{align}
    \zeta &= \cos \left(\frac{\pi  g}{2}\right) \cos \left(\frac{\pi  J}{2}\right)+e^{-i p} \sin \left(\frac{\pi  g}{2}\right) \sin \left(\frac{\pi  J}{2}\right),  \\
    \eta & =-\sin \left(\frac{\pi  g}{2}\right) \cos \left(\frac{\pi  J}{2}\right)+e^{i p} \cos \left(\frac{\pi  g}{2}\right) \sin \left(\frac{\pi  J}{2}\right).
\end{align}
\end{subequations}

{ The above implies
\begin{align}
    U_F = \begin{pmatrix}|\zeta|^2 -|\eta|^2 & -2 \zeta^* \eta^*\\ 2\zeta  \eta & |\zeta|^2 -|\eta|^2 \end{pmatrix}.
\end{align}
while, the unitary whose starting point is shifted by  half a drive cycle is given by
\begin{align}
    &\tilde{U}_F = F U_F F^{\dagger} 
    =\begin{pmatrix} |\zeta|^2 -|\eta|^2& -2 \zeta \eta^*\\ 2 \eta \zeta^*& |\zeta|^2 -|\eta|^2\end{pmatrix}
\end{align}
}

Using this parametrization~\cite{CardosoMitra2025}

\begin{subequations}
\begin{align}
    \nu_0 & = \int \frac{dp}{4\pi i}[ \partial_p {\ln [\eta]} -\partial_p {\ln [\eta^*]}], \\
    \nu_\pi & = \int \frac{dp}{4\pi i}[\partial_p {\ln [\zeta]} -\partial_p {\ln [\zeta^*]}], 
\end{align}
\end{subequations}
we find the topological quantum numbers indicated in Fig.~\ref{fig:Model} b). 

It is a helpful illustration to list $\zeta, \eta$ at the special points of the phase diagram and express these results in terms of the winding of $\vec{n}$, i.e. $N_1$, and the winding of $\vec{\tilde{n}}$, i.e. $\tilde N_1$. The winding number associated to $0$ and $\pi$ modes are given by \cite{Obuse13,Delplace14,YatesMitra2018} $\nu_0=(N_1+\tilde N_1)/2$, $\nu_\pi = (N_1-\tilde N_1)/2$ respectively, see Tab.~\ref{tab:Windings}. 

\begin{table}
\begin{center}
    \begin{tabular}{c|c|c|c|c|c|c|c|c}
         $\left [\begin{array}{c} J\\ g\end{array} \right
         ]$ & $\eta $& $\zeta$  & $\left (\begin{array}{c}
              n_x  \\
              n_y
         \end{array} \right )$ & $\left (\begin{array}{c}
              \tilde{n}_x  \\
              \tilde{n}_y
         \end{array}  \right )$ & $N_1$ & ${\tilde{N}}_1$ &$\nu_0$ & $\nu_\pi $ \\
         \hline \hline
         $\left [\begin{array}{c}\frac{1}{2}\\0\end{array} \right]$& $\frac{e^{i p}}{\sqrt{2}}$& $\frac{1}{\sqrt{2}}$ & $\left ( \hspace{-.2cm}\begin{array}{c} \sin(p) \\ -\cos(p) \end{array} \hspace{-.2cm}\right)$& $\left ( \hspace{-.2cm}\begin{array}{c} \sin(p) \\ -\cos(p) \end{array} \hspace{-.2cm}\right)$ & 1& 1 &1 &0 \\ 
       \hline
         $\left [\begin{array}{c}1\\\frac{1}{2}\end{array} \right]$& $\frac{e^{i p}}{\sqrt{2}} $& $\frac{e^{-i p}}{\sqrt{2}}$&  $\left ( \begin{array}{c} 0 \\ -1\end{array} \right)$& $\left (\hspace{-.2cm} \begin{array}{c} \sin(2p) \\ -\cos(2p) \end{array}\hspace{-.2cm} \right)$ & 0& 2 &1 &-1\\
         \hline
         $\left [\begin{array}{c}\frac{1}{2}\\1\end{array} \right]$& $\frac{-1}{\sqrt{2}} $& $\frac{e^{-i p}}{\sqrt{2}}$& $\left ( \hspace{-.2cm}\begin{array}{c} \sin(p) \\ \cos(p) \end{array} \hspace{-.2cm}\right)$ & $\left ( \hspace{-.2cm}\begin{array}{c}- \sin(p) \\ \cos(p) \end{array}\hspace{-.2cm} \right)$&-1 & 1&0 &-1\\
         \hline
         $\left [\begin{array}{c}0\\\frac{1}{2} \end{array} \right]$& $\frac{-1}{\sqrt{2}} $& $\frac{1}{\sqrt{2}}$ & $\left ( \begin{array}{c} 0 \\ 1 \end{array} \right)$ &$\left ( \begin{array}{c}0 \\ {1} \end{array} \right)$& 0& 0 &0 &0\\
    \end{tabular}
\end{center}
\caption{Windings at special points in the phase diagram Fig.~\ref{fig:Model} b)}
\label{tab:Windings}
\end{table}

\section{Evaluation of edge Green's function}
\label{app:EdgeGF}

In this section of the appendix we provide details about the edge Floquet Green's function for four flavors of Dirac fermions $a = 1,2,3,4$ coupled by FK-like interactions Eq.~\eqref{eq:FK}. We concentrate on the fine-tuned point illustrated as crosses in Fig.~\ref{fig:Model} b) and use the expression of Majorana fermions $\gamma$ in terms of complex fermions, cf. Sec.~\eqref{sec:ModelsIA}.

The Green's function of complex fermions is

\begin{align}
\mathcal G_{\rm R}^{ab} (t) &\equiv - i  \theta(t) 
    \langle \{ c_a(t), c^\dagger_b(0) \} \rangle \\
    \Rightarrow \mathcal G_{\rm R}^{ab} (t)^* &=  i  \theta(t) 
    \langle \{ c_a^\dagger(t), c_b(0) \} \rangle
\end{align}
and we find $ \mathcal G_{\rm R}^{ab} (t)= \delta_{ab} \mathcal G_{\rm R} (t)$. At the fine tuned points, the complex fermions of interest are simply evolved using the Heisenberg picture $c_a(t) = \hat U_w^{\dagger} c_a \hat U_w$ and a time evolution operator associated to the Fidkowski-Kitaev interaction. In part of the calculations we assume a ground state with respect to the very same FK Hamiltonian, so the equilibrium calculation follow.  

Note that the anomalous Green's functions of the type $\langle c_a(t) c_b(0)$ vanish due to $\mathbb Z_4$ symmetry $c_a \rightarrow i c_a, \forall a$ of Eq.~\eqref{eq:FK}.

\subsection{Connection of Green's function of Majorana edge fermions and complex fermions}

Generally, there is a connection $c_{\mathbf j, a} = (\gamma_{\mathbf j}^{2a-1} + i\gamma_{\mathbf j}^{2a})/2$, $a = 1,2,3,4$ with different meaning of $a$, $\mathbf j$ depending on which physical system we are considering, see Sec.~\ref{sec:FKFloquetEdge}. In this basis the FK-Green's function for Majorana fermions is 
\begin{align}
    G_{\rm R}^{AB}(t) = \delta_{A B}[\mathcal G_{\rm R}(t) - \mathcal G_{\rm R}(t)^*] =  2\delta_{A B} \mathcal G_{\rm R}(t)
\end{align}
 with $A,B \in \{ 2a, 2a-1 \vert a = 1,2,3,4\}$. As we will shortly see $\mathcal G_{\rm R}(t)$ is purely imaginary leading to the final equation. In the case of M$\pi$Ms the discrete time evolution Green's function acquires an additional oscillating sign. 

\subsection{Continuum time evolution}

We first consider continuum time evolution assuming a thermal 
density matrix
\begin{equation}
    \hat \rho= \frac{e^{- \beta \hat H_w}}{\tr[e^{- \beta \hat H_w}]}
\end{equation}
and reinstating the Floquet time $T_F$ and $W = \pi w/ 2 T_F$ and $\beta = 1/(k_B T)$ the inverse temperature

\begin{widetext}
\begin{subequations}
\begin{align}
&    \mathcal G_{\rm R} (t)  = -i \theta(t) \times \notag \\
    &\frac{(e^{6 \beta \lambda W} \cosh(8 \beta W) +1) \cos(8Wt) \cos(6 Wt\lambda W) - e^{6 \beta \lambda W}\sinh(8 \beta W) \sin(8 Wt)\sin(6 Wt\lambda) + 3(1+e^{- 2 \beta \lambda W}) \cos(2 Wt\lambda)}{e^{6 \beta \lambda W} \cosh(8 \beta W) + 4 + 3e^{-2 \beta \lambda W}} .
 \end{align} 

It is instructive to highlight the following limits

 \begin{align}
   \text{1.:} \mathcal G_{\rm R} (t) & \stackrel{\beta \rightarrow \infty}{\simeq} - i \theta(t) \cos((8 + 6 \lambda)Wt),\\
   \text{2.:} \mathcal G_{\rm R} (t) & \stackrel{\beta \rightarrow 0}{\simeq} - i \theta(t)\frac{1}{8} (\cos (Wt  (6 \lambda +8))+\cos (Wt  (8-6 \lambda ))+6 \cos (2 Wt  \lambda )),\\
   \text{3.:} \mathcal G_{\rm R} (t) & \stackrel{\lambda = 1}{=} - i \theta(t) \left [ \frac{ 1+ e^{-14 \beta} }{ 1+ 8e^{-14 \beta} + 7 e^{-16 \beta}} \cos(14 Wt ) + \frac{ 7e^{-14 \beta} + 7 e^{-16 \beta} }{ 1+ 8e^{-14 \beta} + 7 e^{-16 \beta}} \cos(2 Wt) \right ].\label{eq:RetardedGF}
\end{align}
\label{eq:ConttimeretardedGF}
\end{subequations} \noindent

In Fourier space this is
\begin{subequations}
\begin{align}
    \mathcal G_{\rm R}(\Omega) & =  \frac{e^{6 \beta \lambda W} \cosh(8 \beta W) + 1}{e^{6 \beta \lambda W} \cosh(8 \beta W) + 4 + 3e^{-2 \beta \lambda W}} \frac{1}{2}\left [    \frac{\Omega}{\Omega_+^2 - (8 + 6 \lambda)^2 W^2} + \frac{\Omega}{\Omega_+^2 - (8 - 6 \lambda)^2 W^2} \right] \notag\\
    & + \frac{e^{6 \beta \lambda W} \sinh(8 \beta W)}{e^{6 \beta \lambda W} \cosh(8 \beta W) + 4 + 3e^{-2 \beta \lambda W}} \frac{1}{2} \left [    \frac{\Omega}{\Omega_+^2 - (8 + 6 \lambda)^2 W^2} - \frac{\Omega}{\Omega_+^2 - (8 - 6 \lambda)^2 W^2} \right] \notag\\
    &+ \frac{3 (1 + e^{- 2\beta \lambda W})}{e^{6 \beta \lambda W} \cosh(8 \beta W) + 4 + 3e^{-2 \beta \lambda W}} \frac{\Omega}{\Omega_+^2 - 4 \lambda^2 W^2} ,\\
    \mathcal G_{\rm R}(\Omega) \vert_{\lambda = 1} &{=}  \frac{ 1+ e^{-14 \beta W} }{ 1+ 8e^{-14 \beta W} + 7 e^{-16 \beta W}} \frac{\Omega}{\Omega_+^2 - (14W)^2} + \frac{ 7e^{-14 \beta W} + 7 e^{-16 \beta W} }{ 1+ 8e^{-14 \beta W} + 7 e^{-16 \beta W}} \frac{\Omega}{\Omega_+^2 - (2W)^2} .
\end{align}
\end{subequations}
\end{widetext}
A couple of physical comments on this result, in particular the last line, follow.

The first pair of delta functions has strength 1/2 (1/16) at zero (infinite) temperature, while the second one has strength 0 (7/16) in the corresponding limits. This coincides with the degeneracy of corresponding transitions.  
The Green's function zero is pinned to zero remains as the system has particle-hole symmetry and the Green's function has to change sign at some point. 
Zero energy Green's function zeros without superimposing poles exist for any $\lambda \neq 0, \pm 4/3$

\subsection{Discrete time evolution}

For discrete time evolution under $H_w$, only, we may simply evaluate Eq.~\eqref{eq:ConttimeretardedGF} at discrete times $t = n T_F$. Note that the fermionic parity operator commutes with $H_w$.

The retarded Green's function at the boundary is 
\begin{equation}
    \mathbf G_{\rm R}^{AB}(n) = 2e^{-i \epsilon_A n} \mathcal G_{\rm R}(n T_F) \delta_{AB} ,
\end{equation}
where 
$\mathcal G_{\rm R}(t)$ is the Green's function under continuous time evolution given by Eq.~\eqref{eq:ConttimeretardedGF} evaluated at discrete times $n$ and $\epsilon_A \in \{0, \pi \}$ is the quasi energy of the $A$th mode.
Note that this result holds for multiple limits
\begin{itemize}
    \item In the case $g = 0, J = 1/2$, all modes are zero modes, $\epsilon_A =0 \forall A$.
    \item In the case $g = 1, J = 1/2$, all modes are $\pi$ modes, $\epsilon_A =\pi \forall A$.
    \item In the case $g = 1/2, J = 1$,  $e^{-i \epsilon_A n} = (-1)^{A}$.
\end{itemize}

For the Fourier transform we generally use that
\begin{align}
    -i \sum_{n} \theta(n) e^{i \Omega^+ n} \cos(\epsilon n) 
    & = -i\frac{1}{2} \left [1 + \sum_\pm \frac{e^{i \Omega \pm i\epsilon}}{1-e^{i \Omega^+ \pm i\epsilon}}  \right ] \notag \\
    & = \sum_\pm \frac{1}{4}\cot\left ( \frac{\Omega^+ \pm \epsilon}{2} \right ) \notag \\
    & = \frac{1}{2} \frac{\sin(\Omega^+)}{\cos(\epsilon) - \cos(\Omega^+)},
\end{align}
so that Eq.~\eqref{eq:GreensfunctionsEdgeMaintext} immediately follows from Eq.~\eqref{eq:ConttimeretardedGF} (note the factor of $2$ in the conversion from $c-$Green's function $\mathcal G_{\rm R}$ to Majorana Green's function $G_{\rm R}$).

\section{Interacting bulk Green's function}
\label{app:BulkGFFull}

In this appendix we provide extensive details for the derivation of Eqs.~\eqref{eq:BulkGFWithIA}.

\subsection{Basics of Floquet perturbation theory}
\label{sec:PertTheoryBasics}

Consider a Floquet operator 
\begin{subequations}
\begin{align}
    \hat U &= \hat U_1^{1/2} \hat U_0\hat U_1^{1/2}, \label{eq:FloquetBasic}\\
    \hat U_1^{1/2} &= e^{- i \hat V/2}, \label{eq:Vhalf}\\
    \hat U_0 &= e^{- i \hat H_0}, \label{eq:H0}\\
\end{align}
where we will use the eigenbasis
\begin{equation}
    \hat U_0 \ket{n^{(0)}}  = e^{- i E_n^{(0)}} \ket{n^{(0)}}.
\end{equation}\label{eq:FloqPert}
\end{subequations}

Under the assumption that matrix elements $V_{nm} \equiv \braket{n^{(0)} \vert \hat V \vert m^{(0)}}$ vanish for degenerate states
\begin{equation}
    V_{nm} = 0, \quad \forall n,m \text{ s.th. } E_n^{(0)} = E_m^{(0)}
\end{equation}
the perturbatively corrected eigenstates and frequencies are (see, e.g., Ref.~\cite{SchmidvonOppen2024} and $\alpha$ an unimportant phase)
\begin{subequations}
\begin{align}
    \ket{n} & \simeq e^{i \alpha} \ket{n^{(0)}} + \frac{1}{2} \sum_{m \neq n} \cot\left (\frac{E_n^{(0)}-E_m^{(0)}}{2}\right) V_{mn}\ket{m^{(0)}}, \\
    E_{n} & \simeq E_n^{(0)}  + \frac{1}{2} \sum_{m \neq n} \cot\left (\frac{E_n^{(0)}-E_m^{(0)}}{2}\right) \vert V_{mn} \vert^2 .
\end{align}
\label{eq:FloqPert}
\end{subequations}
These equations indicate that the smallness of $\vert \cot\left (\frac{E_n^{(0)}-E_m^{(0)}}{2}\right) V_{mn} \vert$ controls the perturbation theory.

We also need the Floquet perturbation theory for the reverse order of operators
\begin{subequations}
\begin{align}
    \hat U &= \hat U_0^{1/2} \hat U_1\hat U_0^{1/2}, \label{eq:FloquetBasicReverse}\\
    \hat U_1 &= e^{- i \hat V}, \label{eq:V}\\
    \hat U_0^{1/2} &= e^{- i \hat H_0/2}, \label{eq:H0half}\\
\end{align}
where we will use the eigenbasis
\begin{equation}
    \hat U_0^{1/2} \ket{n^{(0)}}  = e^{- i E_n^{(0)}/2} \ket{n^{(0)}}.
\end{equation}\label{eq:FloqPert2}
\end{subequations}

Again, we assume that matrix elements $V_{nm} \equiv \braket{n^{(0)} \vert \hat V \vert m^{(0)}}$ vanish for degenerate states
\begin{equation}
    V_{nm} = 0, \quad \forall n,m \text{ s.th. } E_n^{(0)} = E_m^{(0)}.
\end{equation}
Then, perturbatively corrected eigenstates and frequencies are (again, $\alpha$ is an unimportant phase)
\begin{subequations}
\begin{align}
    \ket{n} & \simeq e^{i \alpha} \ket{n^{(0)}} + \frac{1}{2} \sum_{m \neq n} \frac{1}{\sin \left (\frac{E_n^{(0)}-E_m^{(0)}}{2}\right)} V_{mn}\ket{m^{(0)}}, \\
    E_{n} & \simeq E_n^{(0)}  + \frac{1}{2} \sum_{m \neq n} \cot\left (\frac{E_n^{(0)}-E_m^{(0)}}{2}\right) \vert V_{mn} \vert^2 .
\end{align}
\label{eq:FloqPert2}
\end{subequations}
For this order of operators $\hat U_0, \hat U_1$, smallness of $V_{mn}/\vert \sin\left (\frac{E_n^{(0)}-E_m^{(0)}}{2}\right)  \vert$ controls the perturbation theory.

\subsection{Lehman representation of Green's functions}
\label{app:BulkGFPertTheory}

We will calculate the Green's functions for the following two many-body Floquet operators
\begin{align}
    \hat U_F &= \hat U_{g/2} \hat U_{J/2} \hat U_{w} \hat U_{J/2} \hat U_{g/2}. \\
    \hat{\tilde U}_F &= \hat U_{w/2} \hat U_{J/2} \hat U_{g} \hat U_{J/2} \hat U_{w/2}.
\end{align}
We introduce the corresponding exact many-body eigenstates $\hat U_F \ket{m} = e^{i E_m} \ket{m}$, $\hat{\tilde U}_F \ket{\tilde m} = e^{i \tilde E_m} \ket{\tilde m}$.

We will perturb in small $g,J$ or near special points in parameter space where $g$ and/or $J$ are close to unity. In these cases, one may adiabatically connect the spectrum of eigenvalues of the purely interacting Hamiltonian $\hat H_w$, Eq.~\eqref{eq:FK}, to the eigenphases $\{E_m\}$, $\{\tilde{E}_m\}$ of the Floquet unitary. Using the correspondence to the spectrum of the Hamiltonian, we can order $\{E_m\}$, $\{\tilde{E}_m\}$ on the real axis and, e.g., define a ground state. This also allows to define a minimal energy and a thermal density matrix (see next section).

\subsubsection{Lehmann representation for the Green's function to $\hat U_F$}
For simplicity, consider the Floquet cycle $\hat U_F$. We can use its eigenbasis  
to find the retarded Green's function for a thermal density matrix 
\begin{equation}
\hat \rho = \frac{\sum_m \ket{m}e^{- \beta E_m} \bra{m}}{\sum_k e^{- \beta E_k} }
\end{equation}
(energy units of $1/T_F$) the Green's function may be generally expressed as
\begin{widetext}
\begin{align}
    \mathbf G_{\rm R}^{AB}(n) & = - i \theta(n) \sum_{ml} \braket{m \vert \gamma_A \vert l} \braket{l \vert \gamma_B \vert m} e^{i (E_m-E_l) n} \frac{e^{- \beta E_m} + e^{- \beta E_l}}{\sum_k e^{- \beta E_k}}, \\
    \mathbf G_{\rm R}^{AB}(\Omega) & = \sum_{ml} \frac{e^{- \beta E_m} + e^{- \beta E_l}}{2\sum_k e^{- \beta E_k}} \braket{m \vert \gamma_A \vert l} \braket{l \vert \gamma_B \vert m}  \cot \left ( \frac{\Omega^+ - E_{l-m}}{2}\right ) \notag \\
    & = \sum_{ml} \frac{e^{- \beta E_m}}{2\sum_k e^{- \beta E_k}} \Big [  \frac{\sin(\Omega)}{\cos (E_{l-m}) - \cos(\Omega^+)}\left (\braket{m \vert \gamma_A \vert l} \braket{l \vert \gamma_B \vert m}  + A \leftrightarrow B \right ) \notag \\
  & \hspace{2.8cm}
    + \frac{\sin(E_{l-m})}{\cos (E_{l-m}) - \cos(\Omega^+)}\left (\braket{m \vert \gamma_A \vert l} \braket{l \vert \gamma_B \vert m}  - A \leftrightarrow B \right )\Big]. \label{eq:LehmannLike}
\end{align}
\end{widetext}

The advanced Green's function follows from $\Omega^+ \rightarrow \Omega^-$.

\subsubsection{Lehmann representation for the Green's function to $\hat{\tilde U}_F$.}

Using the exact same steps and defining quantum averages with respect to the density matrix 
\begin{equation}
\hat \rho = \frac{\sum_m \ket{\tilde{m}}e^{- \beta \tilde{E}_m} \bra{\tilde{m}}}{\sum_k e^{- \beta \tilde{E}_k} }
\end{equation}
one may obtain the Green's function for the cycle $\hat{\tilde U}_F$
\begin{widetext}
\begin{align}
    \tilde{\mathbf G}_{\rm R}^{AB}(\Omega) 
    & = \sum_{ml} \frac{e^{- \beta  \tilde{E}_m}}{2\sum_k e^{- \beta  \tilde{E}_k}} \Big [  \frac{\sin(\Omega)}{\cos ( \tilde{E}_{l-m}) - \cos(\Omega^+)}\left (\braket{ \tilde{m} \vert \gamma_A \vert \tilde{l}} \braket{\tilde{l} \vert \gamma_B \vert  \tilde{m}}  + A \leftrightarrow B \right ) \notag \\
  & \hspace{2.8cm}
    + \frac{\sin( \tilde{E}_{l-m})}{\cos ( \tilde{E}_{l-m}) - \cos(\Omega^+)}\left (\braket{\tilde{m} \vert \gamma_A \vert \tilde{l}} \braket{\tilde{l} \vert \gamma_B \vert  \tilde{m}}  - A \leftrightarrow B \right )\Big]. \label{eq:LehmannLikeTilde}
\end{align}
\end{widetext}

\subsection{Perturbative Green's functions near $(J,g) = (0,0)$ for $N_f = 8$}

\subsubsection{Green's function to $\hat U_F$.}

We now evaluate Eq.~\eqref{eq:LehmannLike} using the general results of perturbation theory Eq.~\eqref{eq:FloqPert} where $\hat V = \pi(\hat H_g + \hat H_J)/2$ and $\hat U_0 = \hat U_w$. We restrict ourselves to the first order in $g,J$. We may thus omit all corrections to the eigenfrequencies $E_m = E_m^{(0)} + \mathcal O((g,J)^2)$. 

We consider zero temperature correlators defined with respect to the eigenstate of $\hat U_F$ which is adiabatically connected to the unperturbed ground state of the Fidkowski Kitaev model. The latter is a product state of the local groundstate of $H_w$ on each "FK quantum dot" $\mathbf j = (j,D)$. Any $\gamma_A$ connects to the local odd-parity manifold $\{ \ket{o^{(0)}} \}$, so that $E_l-E_m = 7 \pi w + \mathcal O((g,J)^2)$. Then Eq.~\eqref{eq:LehmannLike} simplifies to 
\begin{align}
    \mathbf G_{\rm R}^{AB}(\Omega) & = \frac{1}{2} \frac{2 \delta_{AB}\sin(\Omega) + \sin(7 \pi w) \braket{g \vert [\gamma_A, \gamma_B]
    \vert g}}{\cos(7 \pi w) - \cos(\Omega^+)}. \label{eq:SimplePertTheory00}
\end{align}
In the unperturbed limit, the ground state 
\begin{equation} \label{eq:GSUnperturbed}
    \ket{g^{(0)}} = \otimes \ket{g_{\rm FK}}, \quad \ket{g_{\rm FK}} = \frac{\ket{0000} + \ket{1111}}{\sqrt{2}}
\end{equation}
is a direct product of local groundstates of the FK model (in the $c-$basis introduced in Sec.~\ref{sec:FK}). We further use that $\hat V$ acts on two different FK-quantum dots, so that each term generates a many body state with excitation energy $14 \pi w$ above the ground state. The wave function corrections hence take the form 
\begin{align}
    \ket{g} &=   \ket{g^{(0)}} - \frac{1}{2} \cot\left (7 \pi w \right ) \sum_o \ket{o^{(0)}}\braket{o^{(0)} \vert \hat V \vert g^{(0)}} \notag \\
    &=\ket{g^{(0)}} - \frac{1}{2} \cot\left (7 \pi w \right )   \hat V \ket{g^{(0)}} . \label{eq:gcorrect}
\end{align}
In the second line we used that, since $\hat V$ only connects to the odd sector, we may replace $\ket{o}$ by a summation over all virtual states, i.e. an insertion of identity.

We next use that in the $c-$basis of Sec.~\ref{sec:FK}
\begin{subequations}
\begin{align}
    \braket{g^{(0)} \vert [c_{\mathbf j, a},c_{\mathbf j, b}]\vert g^{(0)} }&= 0,\\
    \braket{g^{(0)} \vert [c_{\mathbf j, a}^\dagger,c_{\mathbf j, b}^\dagger]\vert g^{(0)} }&= 0 ,\\
    \braket{g^{(0)} \vert [c_{\mathbf j, a}^\dagger,c_{\mathbf j, b}]\vert g^{(0)}} &= 0.
\end{align}
\label{eq:localaverages}
\end{subequations}
The first two follow from the fact that the GHZ ground state has well-defined $\mathbb Z_4$ parity. The last equation vanishes obviously for $a \neq b$, while for $a = b$ we use that the average parity on a given orbital vanishes. Using that all $c_{\mathbf j, a}$ are local linear combinatinos of $\gamma_A$ we thus conclude
\begin{equation}
    \braket{g^{(0)} \vert [\gamma_A, \gamma_B]
    \vert g^{(0)}} = 0,
\end{equation}
where we also used that intersite averages vanish trivially. This leads to 
\begin{align}
    \mathbf G_{\rm R}^{AB}(\Omega) & = \frac{ \delta_{AB}\sin(\Omega) - \frac{\cos(7 \pi w)}{4}  \braket{ \{ [\gamma_A, \gamma_B], \hat V \}}_{g^{(0)}}}{\cos(7 \pi w) - \cos(\Omega^+)}. \label{eq:GintermediateStep00}
\end{align}

We next use the notation 
\begin{equation}
    \hat V = \frac{\pi}{2} [\hat H_{g}+\hat H_{ J}] = i v_{CD}\gamma_{C} \gamma_{D}, \label{eq:Littlev}
\end{equation}
where the matrix $v$
is real and antisymmetric and vanishes if $C,D$ are on the same FK quantum dot.

Thus, 
\begin{align}
    \braket{ \{ [\gamma_{A}, \gamma_B], \hat V \}}_{g^{(0)}}
    &  
    =- 4 i v_{AB}.
\end{align}
We used that $\braket{\gamma_{A} \gamma_B}_{g^{(0)}} = \delta_{AB}$ on a given FK quantum dot.

In summary we thus find

\begin{align}
    \mathbf G_{\rm R}^{AB}(\Omega) & = \frac{ \delta_{AB} \sin(\Omega) + i v_{AB} \cos(7 \pi w)}{\cos(7 \pi w) - \cos(\Omega^+)}. \label{eq:GWithLittlev}
\end{align}
We mention that the first term should and is consistent with Eq.~\eqref{eq:GreensfunctionsEdgeMaintext} which was obtained by other means. Moreover, in the limit $7 \pi \omega, \Omega \rightarrow 0$ this is consistent with the continuum time evolution result of Ref.~\cite{YouXu2014} (recall that $7 \pi w = 14 W$).

We use the definition of the Fourier transform, Eq.~\eqref{eq:MomspaceGF} as well as
\begin{subequations}
    \begin{align}
        \frac{1}{2N}\sum_{jj'} e^{- i p(j- j')} \delta_{AB} &\doteq \frac{1}{2}  \mathbf 1 , \\
        \frac{1}{2N}\sum_{jj'} e^{- i p(j- j')} i v_{AB} &\doteq \frac{\pi}{4}  h_{(J,g)}(p),
    \end{align}
\end{subequations}
to find
\begin{align}
    \mathbf G_{\rm R}(\Omega,p) & = \frac{1}{2}\frac{\sin(\Omega) +\frac{\pi}{2}\cos(7 \pi w) h_{(J,g)}(p)}{\cos(7 \pi w) - \cos(\Omega^+)}. \label{eq:AppGFWithPert}
\end{align}
This concludes the derivation of Eq.~\eqref{eq:BulkPertGFModel1} of the main text in the limit of small $J,g$.

\begin{figure}
    \centering
\includegraphics[scale=1]{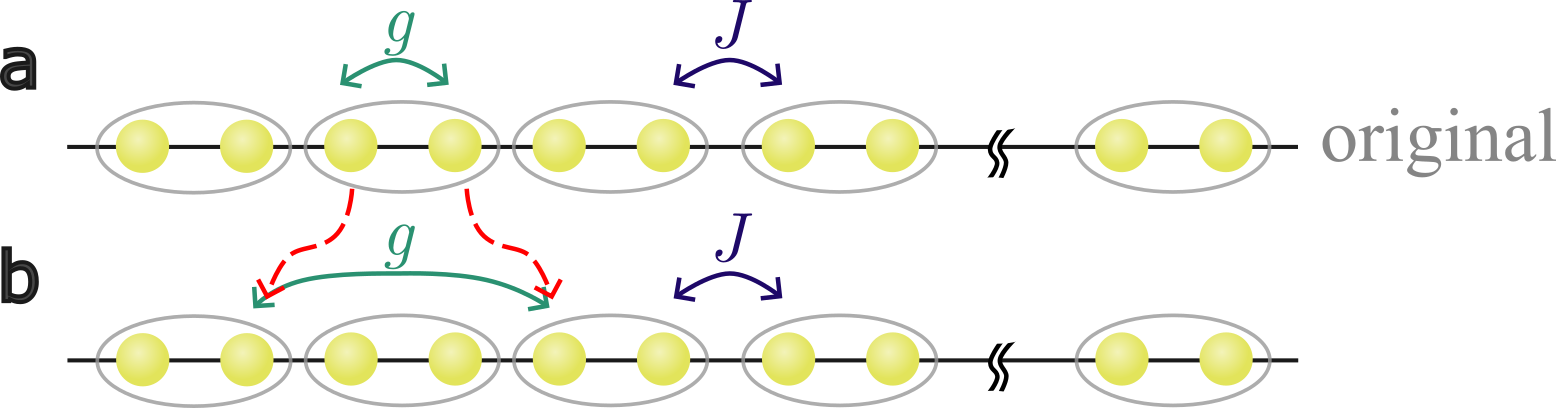}
    \caption{Basis transformations of relevance for App.~\ref{app:BulkGFFull}. a)     original basis b) basis of relevance near $(J,g) \simeq (1,0)$. }
    \label{fig:NewBasis}
\end{figure}

\subsubsection{Green's function to $\hat{\tilde U}_F$.}

In view of the similarity of Eqs.~\eqref{eq:LehmannLike}, \eqref{eq:LehmannLikeTilde} we readily find the analogue to Eq.~\eqref{eq:SimplePertTheory00}, i.e.
\begin{align}
    \tilde{\mathbf G}_{\rm R}^{AB}(\Omega) & = \frac{1}{2} \frac{2 \delta_{AB}\sin(\Omega) + \sin(7 \pi w) \braket{\tilde{g} \vert [\gamma_A, \gamma_B]
    \vert \tilde{g}}}{\cos(7 \pi w) - \cos(\Omega^+)}.
\end{align}
The only difference in its evaluation is the replacing $\cot(7 \pi w)$ by $1/\sin(7 \pi w)$ in Eq.~\eqref{eq:gcorrect}, cf.~Sec.~\ref{sec:PertTheoryBasics}, so that 

\begin{align}
    \tilde{\mathbf G}_{\rm R}(\Omega,p) & = \frac{1}{2}\frac{\sin(\Omega) +\frac{\pi}{2} h_{(J,g)}(p)}{\cos(7 \pi w) - \cos(\Omega^+)}. \label{eq:AppGFTildeWithPert}
\end{align}
This concludes the derivation of Eq.~\eqref{eq:GtildaModel1} of the main text.

\subsection{Perturbation theory near $(J,g) = (0,1)$ for $N_f = 8$}
\subsubsection{Green's function to $\hat U_F$}
\label{sec:gminus1} 

Next we consider the vicinity of the top left corner of the phase diagram, Fig.~\ref{fig:Model} b). 
In this case, it is useful to introduce the perturbation
\begin{subequations}
\begin{equation}
\hat V = \frac{\pi}{2} [\hat H_{J}  + \hat H_{g \rightarrow (g-1)}] \label{eq:Vgminus1}
\end{equation}
and the basis transformation 
\begin{equation}
    \check{A} = \hat U_{g = 1/2} \hat A \hat U_{g = 1/2}^\dagger
\end{equation}
for arbitrary operator $\hat A$ 
to express the Floquet unitary as
\begin{align}
    \hat U_F & = \hat U_{g = 1/2} \underbrace{e^{- i \hat V/2} \hat U_w e^{- i \hat V/2}} \underbrace{\hat U_{g = 1/2}}_{ = \hat P \hat U_{g = -1/2}} \\
    & = \underbrace{e^{- i \check V/2} \check U_w e^{- i \check V/2}} \hat P.
\end{align}
\label{eq:UeffApp}
\end{subequations}

Here, $\hat P$ is the total fermionic parity. 
It is important to highlight that the unperturbed ground state of relevance to $\hat U_F$ is -- differently from Eq.~\eqref{eq:GSUnperturbed} --
\begin{equation}
    \ket{\check g^{(0)}} = \hat U_{g = 1/2} \ket{g^{(0)}} = \hat U_{g = 1/2}\otimes \ket{g_{\rm FK}}
\end{equation}
and its leading order correction is -- differently from Eq.~\eqref{eq:gcorrect} --
\begin{align}
    \ket{\check g} &=   \ket{\check g^{(0)}} - \frac{1}{2} \cot\left (7 \pi w \right ) \sum_o \ket{o^{(0)}}\braket{o^{(0)} \vert \check V \vert  \check g^{(0)}} \notag \\
    &=\ket{\check g^{(0)}} - \frac{1}{2} \cot\left (7 \pi w \right )   \check V \ket{\check g^{(0)}} . \label{eq:gcorrectWithCheck}
\end{align}

We now follow the same steps as from Eq.~\eqref{eq:LehmannLike} to Eq.~\eqref{eq:SimplePertTheory00}. There are three major distinctions as compared to the previous calculation: First, the term $\hat P^n$ leads to a factor $(-1)^n$ in $\mathbf G_{\rm R}(n)$ and thus to $\Omega \rightarrow \Omega + \pi$ in $\mathbf G_{\rm R}(\Omega)$. Second, in the perturbation $\hat V$, we replace $g \rightarrow g-1$. Third, we highlight that all calculations are performed in the checked basis now. 

\begin{align}
    \mathbf G_{\rm R}^{AB}(\Omega) & = \frac{1}{2} \frac{-2 \delta_{AB}\sin(\Omega) + \sin(7 \pi w) \braket{\check{g} \vert [\gamma_A, \gamma_B]
    \vert \check g}}{\cos(7 \pi w) + \cos(\Omega^+)} \notag \\
    & {=  \frac{- \delta_{AB}\sin(\Omega) - \frac{\cos(7 \pi w)}{4} \braket{ \{[\gamma_A, \gamma_B], \check V \}
    }_{{\check g}^{(0)}}}{\cos(7 \pi w) + \cos(\Omega^+)}.}
\end{align}
where we highlight the consistency with Eq.~\eqref{eq:GintermediateStep00}.
We next use
\begin{align}
    \check \gamma_A = \hat U_{g = -1/2} \gamma_A \hat U_{g = 1/2} = (i \sigma_y \gamma)_A, \label{eq:rotatedSpinor}
\end{align}
so that
\begin{align}
    \braket{ \{[\gamma_A, \gamma_B], \check V \}
    }_{{\check g}^{(0)}} &= (\sigma_y)_{AA'}  \braket{ \{[\gamma_{A'}, \gamma_{B'}], \hat V \}
    }_{{g}^{(0)}} ( \sigma_y)_{B'B}.
\end{align}

We can then apply the results of the previous section and find 
\begin{align}
    \mathbf G_{\rm R}(\Omega,p) & =  \sigma_y \frac{1}{2}\frac{-\sin(\Omega) +\frac{\pi}{2}\cos(7 \pi w) h_{(J,g-1)}(p)}{\cos(7 \pi w) + \cos(\Omega^+)} \sigma_y \notag \\
    & = \frac{1}{2}\frac{-\sin(\Omega) +\frac{\pi}{2}\cos(7 \pi w) h_{(J,g-1)}(-p)}{\cos(7 \pi w) + \cos(\Omega^+)}.
\end{align}
This concludes the derivation of Eq.~\eqref{eq:BulkPertGFModel2} of the main text in the limit of small $J,1-g$.

\subsubsection{Green's function to $\hat{\tilde U}_F$.}

While we use the same $\hat V$, Eq.~\eqref{eq:Vgminus1} as in the previous section, however there is no need to introduce the checked basis
\begin{align}
    \hat {\tilde U}_F = \hat U_{w/2} \hat U_{J/2} \hat U_{g} \hat U_{J/2} \hat U_{w/2} \simeq \hat P\hat U_{w/2} e^{- i \hat V} \hat U_{w/2}.
\end{align}

Therefore, as compared to the steps from Eq.~\eqref{eq:LehmannLikeTilde} to Eq.~\eqref{eq:AppGFTildeWithPert}, the differences are that the term $\hat P^n$ leads to a factor $(-1)^n$ in $\mathbf G_{\rm R}(n)$ and thus to $\Omega \rightarrow \Omega + \pi$ in $\mathbf G_{\rm R}(\Omega)$, and of course that, in the perturbation $\hat V$, we replace $g \rightarrow g-1$. We hence find

\begin{align}
    \tilde{\mathbf G}_{\rm R}(\Omega,p) & = \frac{1}{2}\frac{-\sin(\Omega) +\frac{\pi}{2} h_{(J,g-1)}(p)}{\cos(7 \pi w) + \cos(\Omega^+)}. 
\end{align}

This concludes the derivation of Eq.~\eqref{eq:GtildaModel2}. of the main text in the limit of small $J,1-g$.

\subsection{Perturbation theory near $(J, g) = (1,0)$ for $N_f = 4$}
\subsubsection{Green's function to $\hat U_F$.}
We highlight that in this case, the interaction unitary is slightly differently defined, see Fig.~\ref{fig:FKInteractions}. We can however follow similar steps as in the previous sections. 

In particular, it is useful to introduce the perturbation
\begin{subequations}
\begin{equation}
\hat V = \frac{\pi}{2} [\hat H_{J \rightarrow J-1}  + \hat H_{g }] \label{eq:VJminus1}
\end{equation}
and the basis transformation 
\begin{equation}
    \check{A} = \hat U_{J = 1/2} \hat A \hat U_{J = 1/2}^\dagger \label{eq:JTrafo}
\end{equation}
for arbitrary operator $\hat A$ 
to express the Floquet unitary as
\begin{align}
    \hat U_F & = \hat U_{g/2}\hat U_{J/2}\hat U_{w}\underbrace{\hat U_{J/2}}_{= \hat P \hat U_{J = -1/2} \hat U_{J/2 \rightarrow J/2-1/2}}\hat U_{g/2} \\
    & = \underbrace{e^{- i \hat V/2} \check U_w e^{- i \hat V/2}} \hat P.
\end{align}
\label{eq:UeffJApp}
\end{subequations}
As in Sec.~\ref{sec:gminus1}, it is important to highlight that the unperturbed ground state of relevance to $\hat U_F$ is -- differently from Eq.~\eqref{eq:GSUnperturbed} --
\begin{equation}
    \ket{\check g^{(0)}} = \hat U_{ J = 1/2} \ket{g^{(0)}} = \hat U_{ J = 1/2} \otimes \ket{g_{\rm FK}}
\end{equation}
and its leading order correction is -- differently from Eq.~\eqref{eq:gcorrect} and Eq.~\eqref{eq:gcorrectWithCheck} --
\begin{align}
    \ket{\check g} &=   \ket{\check g^{(0)}} - \frac{1}{2} \cot\left (7 \pi w \right ) \sum_o \ket{o^{(0)}}\braket{o^{(0)} \vert \hat V \vert  \check g^{(0)}} \notag \\
    &=\ket{\check g^{(0)}} - \frac{1}{2} \cot\left (7 \pi w \right )   \hat V \ket{\check g^{(0)}} . 
\end{align}

We now follow the same steps as from Eq.~\eqref{eq:LehmannLike} to Eq.~\eqref{eq:SimplePertTheory00}. There are three major distinctions as compared to that calculation: First, the term $\hat P^n$ leads to a factor $(-1)^n$ in $\mathbf G_{\rm R}(n)$ and thus to $\Omega \rightarrow \Omega + \pi$ in $\mathbf G_{\rm R}(\Omega)$. Second, in the perturbation $\hat V$ we replace $J \rightarrow J-1$, but keep $g$ unaltered. Third, we highlight that all calculations are performed with respect to checked ground states, but with unchecked $\hat V$

\begin{align}
    \mathbf G_{\rm R}^{AB}(\Omega) & = \frac{1}{2} \frac{-2 \delta_{AB}\sin(\Omega) + \sin(7 \pi w) \braket{\check{g} \vert [\gamma_A, \gamma_B]
    \vert \check g}}{\cos(7 \pi w) + \cos(\Omega^+)} \notag \\
   & {=\frac{- \delta_{AB}\sin(\Omega) - \frac{\cos(7 \pi w)}{4} \braket{ \{[\gamma_A, \gamma_B], \hat V \}
    }_{{\check g}^{(0)}}}{\cos(7 \pi w) + \cos(\Omega^+)}.}
\end{align}
where we highlight the consistency with Eq.~\eqref{eq:GintermediateStep00}.
We next use 
\begin{subequations}\label{eq:rotatedSpinorJ}
\begin{equation}
    \check \gamma_{A} = \hat U_{J = -1/2} \gamma_{A} \hat U_{J = 1/2} = O_{AA'} \gamma_{A'},
\end{equation}
 where the orthogonal matrix $O_{AA'}$ is specified by, see Fig.~\ref{fig:NewBasis} b)
\begin{align}
    \check \gamma_{j,R} &=  \gamma_{j+1,L}, \\
    \check \gamma_{j,L} &= -\gamma_{j-1,R}, 
\end{align}
\end{subequations}
so that for $\hat V = i v_{CD} \gamma_C \gamma_D$, cf. Eq.~\eqref{eq:Littlev}
\begin{align}
    \braket{ \{[\gamma_A, \gamma_B], \hat V \}
    }_{{\check g}^{(0)}} &=   i v_{CD }\braket{ \{[\check \gamma_{A}, \check \gamma_{B}], \check\gamma_C \check \gamma_D \}
    }_{{g}^{(0)}} \notag \\
    & = - i 4 v_{AB},
\end{align}
so that we recover an expression in full analogy to Eq.~\eqref{eq:GWithLittlev}, yet with $\Omega \rightarrow \Omega + \pi$ and $v_{CD}$ representing $h_{(J-1,g)}(p)$. In summary we thus find in the limit $1-J, g \ll 1$
\begin{align}
    \mathbf G_{\rm R}(\Omega, p) &\simeq \frac{1}{2}\frac{-\sin(\Omega) +\frac{\pi}{2}\cos(7 \pi w) h_{(J-1,g)}(p)}{\cos(7 \pi w) + \cos(\Omega^+)} .
\end{align}
This concludes the derivation of Eq.~\eqref{eq:BulkPertGFModel3}.

\subsubsection{Green's function to $\hat{\tilde U}_F$}
We use the same Eqs.~\eqref{eq:VJminus1},\eqref{eq:JTrafo} as above to express
\begin{align}
    \hat{\tilde U}_F & = \hat U_{w/2}\hat U_{J/2}\hat U_{g}\underbrace{\hat U_{J/2}}_{= \hat P \hat U_{J = -1/2} \hat U_{J/2 \rightarrow J/2-1/2}}\hat U_{w/2} \\
    & = \hat U_{w/2}e^{- i \check V} \hat U_{w/2} \hat P.
\end{align}
Note that $\hat U_{w/2}$ is not checked, but the perturbation $\check V$ is. The ground state is thus
\begin{align}
    \ket{\tilde g} &=   \ket{g^{(0)}} - \frac{1}{2} \frac{1}{\sin\left (7 \pi w \right )} \sum_o \ket{o^{(0)}}\braket{o^{(0)} \vert \check V \vert g^{(0)}} \notag \\
    &=\ket{g^{(0)}} - \frac{1}{2} \frac{1}{\sin\left (7 \pi w \right )  } \check V \ket{g^{(0)}} .
\end{align}

We thus find in full analogy to the previous sections
\begin{align}
    \tilde{\mathbf G}_{\rm R}^{AB}(\Omega) & = \frac{1}{2} \frac{-2 \delta_{AB}\sin(\Omega) + \sin(7 \pi w) \braket{\tilde{g} \vert [\gamma_A, \gamma_B]
    \vert \tilde g}}{\cos(7 \pi w) + \cos(\Omega^+)} \notag \\
    &{=  \frac{-\delta_{AB}\sin(\Omega) - \frac{1}{4} \braket{ \{[\gamma_A, \gamma_B], \check V \}
    }_{{g}^{(0)}}}{\cos(7 \pi w) + \cos(\Omega^+)}. }
\end{align}
We use Eq.~\eqref{eq:rotatedSpinorJ} to evaluate
\begin{align}
    \check V &= i v_{CD} \check \gamma_C \check \gamma_D \\
     &= i \underbrace{O^T_{C'C}v_{CD} O_{DD'}}_{\check v_{C' D'}} \gamma_{C'} \gamma_{D'} \\
    & = ig \sum_j (-) \gamma_{j-1,R} \gamma_{j+1,L} + i (J-1) \sum_j \gamma_{j{+} 1, L} (-\gamma_{j,R}) \notag\\
    & = \sum_p \left ( \gamma_L^\dagger, \gamma_R^\dagger \right )_{p} \tilde h_{(J-1,g)}(p) \left ( \begin{array}{c}
         \gamma_L  \\
         \gamma_R 
    \end{array}\right )_p
\end{align}
where
\begin{align}
    \tilde h_{(\delta J,g)}(p) = \left (\begin{array}{cc}
        0 & i g e^{-i 2 p} - i \delta J e^{- i p} \\
        -i g e^{i 2 p} + i \delta J e^{ i p} & 0
    \end{array} \right ).
\end{align}
Thus
\begin{align}
    \braket{ \{[\gamma_A, \gamma_B], \check V \}
    }_{{\check g}^{(0)}} &=  - i 4 \check v
\end{align}
leads to 

\begin{align}
    \tilde{\mathbf G}_{\rm R}(\Omega, p) &\simeq \frac{1}{2}\frac{-\sin(\Omega) +\frac{\pi}{2} \tilde{h}_{(J-1,g)}(p)}{\cos(7 \pi w) + \cos(\Omega^+)} .
\end{align}
This concludes the derivation of Eq.~\eqref{eq:GtildaModel3}.

\section{Basis conversions}
\label{app:Translation}

In this appendix, we provide details about the relationship between the various Floquet cycles considered in this work, i.e.
\begin{align}
    \hat U_F &= \hat U_{g/2} \hat U_{J/2} \hat U_w \hat U_{J/2} \hat U_{g/2}, \quad \text{Secs.~\ref{sec:GFTopoInv}, \ref{sec:FreeFloquetBulk}, \ref{sec:FKFloquetBulk}}, \\
   \hat U_F &\rightarrow \hat{\bar U}_F = \hat U_{g} \hat U_{J} \hat U_w , \quad \text{Secs.~\ref{sec:Edge}, \ref{sec:FKFloquetEdge}, \ref{sec:Circuit}}.
\end{align}
While in the main text we avoid overloading the notation by the addition of a bar, we here clearly distinguish the cycles, which effectively correspond to two different bases. This is readily seen from the Green's function
\begin{align}
    \mathbf G_{\rm R}^{AB}(n) & = - i \theta(n) \langle \lbrace \hat U^{-n}_F\gamma_A \hat U_F^n, \gamma_B\rbrace\rangle  \\
    & =- i \theta(n) \langle \lbrace \hat{\bar U}^{-n}_F \bar \gamma_A \hat {\bar U}_F^n , \bar \gamma_B\rbrace\rangle_{-}
\end{align}
where 
\begin{subequations}
\begin{align}
    \bar \gamma_A & = [\hat U_{g}\hat U_{J}]^{1/2}  \gamma_A [\hat U_{g}\hat U_{J}]^{-1/2}, \\
    \gamma_A & = [\hat U_{g}\hat U_{J}]^{-1/2}  \bar \gamma_A [\hat U_{g}\hat U_{J}]^{1/2},
\end{align}
and the subscript $\langle \dots \rangle_{-}$ denotes an average with respected to the rotated density matrix.
Here, 
\begin{equation}
    [\hat U_{g}\hat U_{J}]^{-1/2} = \hat U_{g/2}\hat U_{J/2}^{-1} \hat U_g^{-1}.
\end{equation}
    \label{eq:RotGamma}
\end{subequations}

Note that in Sec.~\ref{sec:Edge} the transformation only involves the $g$ rotation, i.e. we employ Eq.~\eqref{eq:RotGamma} at $J = 0$.

The bar basis is used in Secs.~\ref{sec:Edge}, \ref{sec:FKFloquetEdge}, \ref{sec:Circuit} to access the edge Green's functions. In this basis, at certain fine tuned points, the edge modes are perfectly localized, e.g.~at ${N = 1, D= L}$ on the very first Majorana site. While perfectly localized on a single site or unit cell in the bar-basis, this simple localization is generally lost upon rotation back to basis without bar (used in the remaining sections of the main text). Here we focus on expressing the perfectly localized edge modes of the bar basis in the basis without bar.

For notational simplicity we drop the flavor index $\alpha$ in this section of the appendix and focus on the left edge.

\paragraph{Case $g = 0$.} In this case the edge modes are perfectly localized in the bar basis, e.g. $\bar \gamma_{0}^{\rm left  \; edge} =  \gamma_{1,L}$. In the original basis, the MZM localized at the left edge
\begin{align}
    \gamma_0^{\rm left \; edge} = \hat U_{J}^{-1/2}  \gamma_{1,L} \hat U_J = \gamma_{1,L}, \label{eq:Transl1}
\end{align}
i.e. it happens to be the same as $\bar \gamma_{0}^{\rm left  \; edge}$.

\paragraph{Case $g = 1$.} Also in this case the edge modes (now they are M$\pi$M) are perfectly localized, $\bar \gamma_\pi^{\rm left \; edge} = \gamma_{1,L}^\alpha$. However, in this case
\begin{align}
    \gamma_\pi^{\rm left \; edge}&= (-1) \hat U_{g}^{1/2}\hat U_{J}^{-1/2}  \gamma_{1,L} \hat U_J \hat U_{g}^{-1/2} \notag \\
    &= \hat U_{g}^{-1/2} \gamma_{1,L} \hat U_{g}^{1/2} \\
    &= [(i \sigma_y) \gamma]_{1,L}  = \gamma_{1,R},  \label{eq:Transl2}
\end{align}
where we used Eq.~\eqref{eq:rotatedSpinor}. Thus, in the basis of without bar the M$\pi$M is localized on the second Majorana site.

\paragraph{Case $J = 1$, $g = 1/2$.} This case is slightly more complicated. The MZM and M$\pi$M in bar basis are
\begin{align}
\bar \gamma_{\pi/0}^{\rm left \; edge} & = \frac{\gamma_{1,L} \pm\gamma_{1,R}}{\sqrt{2}} \equiv \gamma_{1, \pm}.
\end{align}
In the original basis this is
\begin{align}
    \gamma_{\pi/0}^{\rm left \; edge} & = \hat U_{g = 1/4} \hat U_{J =1/2}^\dagger \hat U_{g =1/2}^\dagger \gamma_{1,\pm} \hat U_{g = 1/2} \hat U_{J =1/2} \hat U_{g = 1/4}^\dagger \notag\\
    & = \mp \hat U_{g = 1/4} \hat U_{J =1/2}^\dagger \frac{}{} \gamma_{1,\pm}  \hat U_{J =1/2} \hat U_{g = 1/4}^\dagger \notag \\
    & =\mp \hat U_{g = 1/4} \frac{\gamma_{1,L} + \gamma_{2,L}}{\sqrt{2}}\hat U_{g = 1/4}^\dagger \\
    & = \mp \frac{\gamma_{1,L} - \gamma_{1,R} \pm (\gamma_{2,L}- \gamma_{2,R})}{2}.  \label{eq:Transl3}
\end{align}
Thus, the edge modes are localized over the first 4 Majorana sites (=first two unit cells) in the basis without bar. 

In summary, in the basis without bar, the edge state Green's function of the main text, Eqs.~\eqref{eq:GreensfunctionsEdgeMaintext}, correspond to the correlators of modes $\gamma_{0/\pi}^{\rm left edge}$ derived in Eqs.~\eqref{eq:Transl1},  \eqref{eq:Transl2}, \eqref{eq:Transl3} of this appendix.

\section{Implementation of Floquet-Fidkowski-Kitaev interaction}

\label{app:Toffoli}

In this section we discuss gates to implement Fidkowski-Kitaev interactions, Eq.~\eqref{eq:FK}. Note that this implementation is valid for either case of 8 chains or 4 chains (note however that the notion of what we call $c_{1,2,3,4}$ is not the same in the two cases).

The two terms in the definition of $H_w$ commute

\begin{equation}
    [c_1 c_2 c_3 c_4, P_a P_b] = 0.
\end{equation}

Thus we can write, cf. Eq.~\eqref{eq:FK} using $\pi w/2 = \theta$,

\begin{align}
    \hat U_w = \underbrace{e^{i \frac{w \pi}{2}\lambda \sum_{b<a} P_a P_b}}_{\hat U_1} \underbrace{ e^{i 4 w \pi [c_1 c_2 c_3 c_4 + H.c.]}}_{\hat U_2}
\end{align}

$\hat U_1$ is a product of relatively standard two-qubit gates, see e.g. the fSim gate which is natural on the google machine ~\cite{MorvanRoushan2022,MiAbanin2022}. We further note that, in the basis $\{ \ket{1111}, \dots, \ket{0000} \}$ 

\begin{align}
\hat U_2 = \left ( \begin{array}{ccccc}
     \cos(4 \pi w)& 0 & \dots &0&  - i \sin(4 \pi w) \\
    0& 1 &  &   &0 \\
    \vdots&  & \ddots & & \vdots \\
    0&  0 &  & 1 &0 \\
    - i \sin(4 \pi w)  & 0 & \dots & 0 &\cos(4 \pi w)
\end{array}\right) ,
\end{align}
Particularly at $w = 1/8, 3/8, \dots$ this unitary is a generalized Toffoli gate (CCCNOT) up to a several phase gates and CNOTs swapping the basis states.

\bibliography{FloquetZeros.bib}

@book{AGD,
  title={Methods of Quantum Field Theory in Statistical Physics},
  author={Abrikosov, A.A. and Gorkov, L.P. and Dzyaloshinski, I.E.},
  isbn={9780486140155},
  series={Dover Books on Physics},
  url={https://books.google.com/books?id=JYTCAgAAQBAJ},
  year={1975},
  publisher={Dover Publications}
}

@article{samanta2024isolatedzeromodequantum,
  title={Isolated zero mode in a quantum computer from a duality twist},
  author={Samanta, Sutapa and Wang, Derek S and Rahmani, Armin and Mitra, Aditi},
  journal={Quantum},
  volume={9},
  pages={1957},
  year={2025},
  publisher={Verein zur F{\"o}rderung des Open Access Publizierens in den Quantenwissenschaften}
}

@article{Eckardt_2015,
doi = {10.1088/1367-2630/17/9/093039},
url = {https://dx.doi.org/10.1088/1367-2630/17/9/093039},
year = {2015},
month = {sep},
publisher = {IOP Publishing},
volume = {17},
number = {9},
pages = {093039},
author = {Eckardt, André and Anisimovas, Egidijus},
title = {High-frequency approximation for periodically driven quantum systems from a Floquet-space perspective},
journal = {New Journal of Physics}
}

@article{Bao2024,
  title={Creating and controlling global Greenberger-Horne-Zeilinger entanglement on quantum processors},
  author={Bao, Zehang and Xu, Shibo and Song, Zixuan and Wang, Ke and Xiang, Liang and Zhu, Zitian and Chen, Jiachen and Jin, Feitong and Zhu, Xuhao and Gao, Yu and others},
  journal={Nature Communications},
  volume={15},
  number={1},
  pages={8823},
  year={2024},
  publisher={Nature Publishing Group UK London}
}

@article{Sondhi16-I,
  title = {Phase structure of one-dimensional interacting Floquet systems. I. Abelian symmetry-protected topological phases},
  author = {von Keyserlingk, C. W. and Sondhi, S. L.},
  journal = {Phys. Rev. B},
  volume = {93},
  issue = {24},
  pages = {245145},
  numpages = {18},
  year = {2016},
  month = {Jun},
  publisher = {American Physical Society},
  doi = {10.1103/PhysRevB.93.245145},
  url = {https://link.aps.org/doi/10.1103/PhysRevB.93.245145}
}

@article{BlasonFabrizio2023,
  title = {Unified role of Green's function poles and zeros in correlated topological insulators},
  author = {Blason, Andrea and Fabrizio, Michele},
  journal = {Phys. Rev. B},
  volume = {108},
  issue = {12},
  pages = {125115},
  numpages = {10},
  year = {2023},
  month = {Sep},
  publisher = {American Physical Society},
  doi = {10.1103/PhysRevB.108.125115},
  url = {https://link.aps.org/doi/10.1103/PhysRevB.108.125115}
}

@article{BollmannKoenig2024,
  title = {Topological Green's Function Zeros in an Exactly Solved Model and Beyond},
  author = {Bollmann, Steffen and Setty, Chandan and Seifert, Urban F. P. and K\"onig, Elio J.},
  journal = {Phys. Rev. Lett.},
  volume = {133},
  issue = {13},
  pages = {136504},
  numpages = {7},
  year = {2024},
  month = {Sep},
  publisher = {American Physical Society},
  doi = {10.1103/PhysRevLett.133.136504},
  url = {https://link.aps.org/doi/10.1103/PhysRevLett.133.136504}
}

@article{CardosoMitra2025,
  title = {Gapless Floquet topology},
  author = {Cardoso, Gabriel and Yeh, Hsiu-Chung and Korneev, Leonid and Abanov, Alexander G. and Mitra, Aditi},
  journal = {Phys. Rev. B},
  volume = {111},
  issue = {12},
  pages = {125162},
  numpages = {19},
  year = {2025},
  month = {Mar},
  publisher = {American Physical Society},
  doi = {10.1103/PhysRevB.111.125162},
  url = {https://link.aps.org/doi/10.1103/PhysRevB.111.125162}
}

@article{ChenHosur2025,
  title = {Robust boundary Luttinger surfaces in topological band structures},
  author = {Chen, Kai and Hosur, Pavan},
  journal = {Phys. Rev. B},
  volume = {111},
  issue = {12},
  pages = {125132},
  numpages = {9},
  year = {2025},
  month = {Mar},
  publisher = {American Physical Society},
  doi = {10.1103/PhysRevB.111.125132},
  url = {https://link.aps.org/doi/10.1103/PhysRevB.111.125132}
}

@article{ColemanRech2005,
  title = {Sum rules and Ward identities in the Kondo lattice},
  author = {Coleman, P. and Paul, I. and Rech, J.},
  journal = {Phys. Rev. B},
  volume = {72},
  issue = {9},
  pages = {094430},
  numpages = {14},
  year = {2005},
  month = {Sep},
  publisher = {American Physical Society},
  doi = {10.1103/PhysRevB.72.094430},
  url = {https://link.aps.org/doi/10.1103/PhysRevB.72.094430}
}

@article{DohertyBartlett2009,
  title = {Identifying Phases of Quantum Many-Body Systems That Are Universal for Quantum Computation},
  author = {Doherty, Andrew C. and Bartlett, Stephen D.},
  journal = {Phys. Rev. Lett.},
  volume = {103},
  issue = {2},
  pages = {020506},
  numpages = {4},
  year = {2009},
  month = {Jul},
  publisher = {American Physical Society},
  doi = {10.1103/PhysRevLett.103.020506},
  url = {https://link.aps.org/doi/10.1103/PhysRevLett.103.020506}
}

@article{Dzyaloshinskii2003,
  title = {Some consequences of the Luttinger theorem: The Luttinger surfaces in non-Fermi liquids and Mott insulators},
  author = {Dzyaloshinskii, Igor},
  journal = {Phys. Rev. B},
  volume = {68},
  issue = {8},
  pages = {085113},
  numpages = {6},
  year = {2003},
  month = {Aug},
  publisher = {American Physical Society},
  doi = {10.1103/PhysRevB.68.085113},
  url = {https://link.aps.org/doi/10.1103/PhysRevB.68.085113}
}

@article{EcksteinHolmes2024,
  title={Large-scale simulations of Floquet physics on near-term quantum computers},
  author={Eckstein, Timo and Mansuroglu, Refik and Czarnik, Piotr and Zhu, Jian-Xin and Hartmann, Michael J and Cincio, Lukasz and Sornborger, Andrew T and Holmes, Zo{\"e}},
  journal={npj Quantum Information},
  volume={10},
  number={1},
  pages={84},
  year={2024},
  publisher={Nature Publishing Group UK London}
}

@article{Fabrizio2022,
  title={Emergent quasiparticles at Luttinger surfaces},
  author={Fabrizio, Michele},
  journal={Nature communications},
  volume={13},
  number={1},
  pages={1561},
  year={2022},
  publisher={Nature Publishing Group UK London}
}

@article{Fabrizio2023,
  title = {Spin-Liquid Insulators Can Be Landau's Fermi Liquids},
  author = {Fabrizio, Michele},
  journal = {Phys. Rev. Lett.},
  volume = {130},
  issue = {15},
  pages = {156702},
  numpages = {6},
  year = {2023},
  month = {Apr},
  publisher = {American Physical Society},
  doi = {10.1103/PhysRevLett.130.156702},
  url = {https://link.aps.org/doi/10.1103/PhysRevLett.130.156702}
}

@article{FloresHooley2025,
  title = {Weyl-Mott point: Topological and non-Fermi liquid behavior from an isolated Green's function zero},
  author = {Flores-Calder\'on, R. and Hooley, Chris},
  journal = {Phys. Rev. B},
  volume = {111},
  issue = {23},
  pages = {235139},
  numpages = {14},
  year = {2025},
  month = {Jun},
  publisher = {American Physical Society},
  doi = {10.1103/k4st-77ph},
  url = {https://link.aps.org/doi/10.1103/k4st-77ph}
}

@article{FidkowskiKitaev2010,
  title = {Effects of interactions on the topological classification of free fermion systems},
  author = {Fidkowski, Lukasz and Kitaev, Alexei},
  journal = {Phys. Rev. B},
  volume = {81},
  issue = {13},
  pages = {134509},
  numpages = {9},
  year = {2010},
  month = {Apr},
  publisher = {American Physical Society},
  doi = {10.1103/PhysRevB.81.134509},
  url = {https://link.aps.org/doi/10.1103/PhysRevB.81.134509}
}

@article{GavenskyGoldman2023,
  title = {Connecting the Many-Body Chern Number to Luttinger's Theorem through Streda's Formula},
  author = {Peralta Gavensky, Lucila and Sachdev, Subir and Goldman, Nathan},
  journal = {Phys. Rev. Lett.},
  volume = {131},
  issue = {23},
  pages = {236601},
  numpages = {7},
  year = {2023},
  month = {Dec},
  publisher = {American Physical Society},
  doi = {10.1103/PhysRevLett.131.236601},
  url = {https://link.aps.org/doi/10.1103/PhysRevLett.131.236601}
}

@article{GavenskyGoldman2024,
  title = {Streda Formula for Floquet Systems: Topological Invariants and Quantized Anomalies from Ces\`aro Summation},
  author = {Peralta Gavensky, Lucila and Usaj, Gonzalo and Goldman, Nathan},
  journal = {Phys. Rev. X},
  volume = {15},
  issue = {3},
  pages = {031067},
  numpages = {37},
  year = {2025},
  month = {Sep},
  publisher = {American Physical Society},
  doi = {10.1103/b3pw-my97},
  url = {https://link.aps.org/doi/10.1103/b3pw-my97}
}

@article{GiorgadzeVayrynen2025,
  title={Characterizing maximally many-body entangled fermionic states by using $ M $-body density matrix},
  author={Giorgadze, Irakli and Huang, Haixuan and Gaines, Jordan and K{\"o}nig, Elio J and V{\"a}yrynen, Jukka I},
  journal={Quantum},
  volume={9},
  pages={1778},
  year={2025},
  publisher={Verein zur F{\"o}rderung des Open Access Publizierens in den Quantenwissenschaften}
}

@article{Gurarie2011,
  title = {Single-particle Green's functions and interacting topological insulators},
  author = {Gurarie, V.},
  journal = {Phys. Rev. B},
  volume = {83},
  issue = {8},
  pages = {085426},
  numpages = {15},
  year = {2011},
  month = {Feb},
  publisher = {American Physical Society},
  doi = {10.1103/PhysRevB.83.085426},
  url = {https://link.aps.org/doi/10.1103/PhysRevB.83.085426}
}

@article{HatsugaiKohmoto1992,
  title={Exactly solvable model of correlated lattice electrons in any dimensions},
  author={Hatsugai, Yasuhiro and Kohmoto, Mahito},
  journal={Journal of the Physical Society of Japan},
  volume={61},
  number={6},
  pages={2056--2069},
  year={1992},
  publisher={The Physical Society of Japan}
}

@article{HuangPhillips2022,
  title={Discrete symmetry breaking defines the Mott quartic fixed point},
  author={Huang, Edwin W and Nave, Gabriele La and Phillips, Philip W},
  journal={Nature Physics},
  volume={18},
  number={5},
  pages={511--516},
  year={2022},
  publisher={Nature Publishing Group UK London}
}

@article{IshikawaMatsuyama1986,
  title={Magnetic field induced multi-component {Q}{E}{D} 3 and quantum Hall effect},
  author={Ishikawa, Kenzo and Matsuyama, Toyoki},
  journal={Zeitschrift f{\"u}r Physik C Particles and Fields},
  volume={33},
  pages={41--45},
  year={1986},
  publisher={Springer},
  url={https://link.springer.com/article/10.1007/BF01410451}
}

@article{Landau1957,
  title={The theory of a Fermi liquid},
  author={Landau, Lev Davidovich},
  journal={Soviet Physics Jetp-Ussr},
  volume={3},
  number={6},
  pages={920--925},
  year={1957},
  publisher={AMER INST PHYSICS 1305 WALT WHITMAN RD, STE 300, MELVILLE, NY 11747-4501 USA}
}

@article{LehmannBudich2025,
  title = {Probing Green's Function Zeros by Cotunneling through Mott Insulators},
  author = {Lehmann, Carl and Crippa, Lorenzo and Sangiovanni, Giorgio and Budich, Jan Carl},
  journal = {Phys. Rev. Lett.},
  volume = {135},
  issue = {10},
  pages = {106303},
  numpages = {8},
  year = {2025},
  month = {Sep},
  publisher = {American Physical Society},
  doi = {10.1103/jnq4-sykq},
  url = {https://link.aps.org/doi/10.1103/jnq4-sykq}
}

@article{LiuVerstraete2007,
  title = {Quantum Computational Complexity of the $N$-Representability Problem: QMA Complete},
  author = {Liu, Yi-Kai and Christandl, Matthias and Verstraete, F.},
  journal = {Phys. Rev. Lett.},
  volume = {98},
  issue = {11},
  pages = {110503},
  numpages = {4},
  year = {2007},
  month = {Mar},
  publisher = {American Physical Society},
  doi = {10.1103/PhysRevLett.98.110503},
  url = {https://link.aps.org/doi/10.1103/PhysRevLett.98.110503}
}

@article{Luttinger1960,
  title = {{F}ermi Surface and Some Simple Equilibrium Properties of a System of Interacting Fermions},
  author = {Luttinger, J. M.},
  journal = {Phys. Rev.},
  volume = {119},
  issue = {4},
  pages = {1153--1163},
  numpages = {0},
  year = {1960},
  month = {Aug},
  publisher = {American Physical Society},
  doi = {10.1103/PhysRev.119.1153},
  url = {https://link.aps.org/doi/10.1103/PhysRev.119.1153}
}

@article{ManmanaGurarie2012,
  title = {Topological invariants and interacting one-dimensional fermionic systems},
  author = {Manmana, Salvatore R. and Essin, Andrew M. and Noack, Reinhard M. and Gurarie, Victor},
  journal = {Phys. Rev. B},
  volume = {86},
  issue = {20},
  pages = {205119},
  numpages = {12},
  year = {2012},
  month = {Nov},
  publisher = {American Physical Society},
  doi = {10.1103/PhysRevB.86.205119},
  url = {https://link.aps.org/doi/10.1103/PhysRevB.86.205119}
}

@article{MartinGrover2025,
  title={A Perturbative Approach to Symmetric Mass Generation},
  author={Martin, Simon and Grover, Tarun},
  journal={arXiv preprint arXiv:2507.23032},
  year={2025}
}

@article{MiAbanin2022,
  title={Noise-resilient edge modes on a chain of superconducting qubits},
  author={Mi, Xiao and Sonner, Michael and Niu, M Yuezhen and Lee, Kenneth W and Foxen, Brooks and Acharya, Rajeev and Aleiner, Igor and Andersen, Trond I and Arute, Frank and Arya, Kunal and others},
  journal={Science},
  volume={378},
  number={6621},
  pages={785--790},
  year={2022},
  publisher={American Association for the Advancement of Science}
}

@article{Mori2023,
   author = "Mori, Takashi",
   title = "Floquet States in Open Quantum Systems", 
   journal= "Annual Review of Condensed Matter Physics",
   year = "2023",
   volume = "14",
   number = "Volume 14, 2023",
   pages = "35-56",
   doi = "https://doi.org/10.1146/annurev-conmatphys-040721-015537",
   url = "https://www.annualreviews.org/content/journals/10.1146/annurev-conmatphys-040721-015537",
   publisher = "Annual Reviews",
   issn = "1947-5462",
   type = "Journal Article",
  }

@article{MorvanRoushan2022, 
title={Formation of robust bound states of interacting microwave photons}, volume={612}, 
DOI={10.1038/s41586-022-05348-y}, 
number={7939}, 
journal={Nature}, 
author={Morvan, A. and Andersen, T. I. and Mi, X. and Neill, C. and Petukhov, A. and Kechedzhi, K. and Abanin, D. A. and Michailidis, A. and Acharya, R. and Arute, F. and et al.}, 
year={2022}, 
month={Dec}, 
pages={240–245}
}

@article{NieSun2024,
  title={Quantum circuit for multi-qubit toffoli gate with optimal resource},
  author={Nie, Junhong and Zi, Wei and Sun, Xiaoming},
  journal={arXiv preprint arXiv:2402.05053},
  year={2024}
}

@article{Oshikawa_2000,
    doi = {10.1103/physrevlett.84.3370},
    url = {https://doi.org/10.11032Fphysrevlett.84.3370},
    year = 2000,
    month = {apr},
    publisher = {American Physical Society ({APS})},
    volume = {84},
    number = {15},
    pages = {3370--3373},
    author = {Masaki Oshikawa},
    title = {Topological Approach to {L}uttinger{\textquotesingle}s Theorem and the {F}ermi Surface of a {K}ondo Lattice},
    journal = {Physical Review Letters}
}

@article{PangburnBanerjee2024,
  title = {Topological charge excitations and Green's function zeros in paramagnetic Mott insulators},
  author = {Pangburn, Emile and P\'epin, Catherine and Banerjee, Anurag},
  journal = {Phys. Rev. B},
  volume = {112},
  issue = {8},
  pages = {085105},
  numpages = {18},
  year = {2025},
  month = {Aug},
  publisher = {American Physical Society},
  doi = {10.1103/4k4y-4hj4},
  url = {https://link.aps.org/doi/10.1103/4k4y-4hj4}
}

@article{PangburnBanerjee2025,
  title = {Impurity-induced Mott ring states and Mott zeros ring states in the Hubbard operator formalism},
  author = {Pangburn, Emile and Banerjee, Anurag and P\'epin, Catherine and Bena, Cristina},
  journal = {Phys. Rev. B},
  volume = {112},
  issue = {12},
  pages = {125157},
  numpages = {20},
  year = {2025},
  month = {Sep},
  publisher = {American Physical Society},
  doi = {10.1103/lysb-n9zp},
  url = {https://link.aps.org/doi/10.1103/lysb-n9zp}
}

@article{PotterVishwanath2016,
  title = {Classification of Interacting Topological Floquet Phases in One Dimension},
  author = {Potter, Andrew C. and Morimoto, Takahiro and Vishwanath, Ashvin},
  journal = {Phys. Rev. X},
  volume = {6},
  issue = {4},
  pages = {041001},
  numpages = {19},
  year = {2016},
  month = {Oct},
  publisher = {American Physical Society},
  doi = {10.1103/PhysRevX.6.041001},
  url = {https://link.aps.org/doi/10.1103/PhysRevX.6.041001}
}

@article{Preskill2018,
  doi = {10.22331/q-2018-08-06-79},
  url = {https://doi.org/10.22331/q-2018-08-06-79},
  title = {Quantum {C}omputing in the {NISQ} era and beyond},
  author = {Preskill, John},
  journal = {{Quantum}},
  issn = {2521-327X},
  publisher = {{Verein zur F{\"{o}}rderung des Open Access Publizierens in den Quantenwissenschaften}},
  volume = {2},
  pages = {79},
  month = aug,
  year = {2018}
}

@article{ProustTaillefer2019,
  title={The remarkable underlying ground states of cuprate superconductors},
  author={Proust, Cyril and Taillefer, Louis},
  journal={Annual Review of Condensed Matter Physics},
  volume={10},
  number={1},
  pages={409--429},
  year={2019},
  publisher={Annual Reviews}
}

@article{RosenbergRoushan2024,
  title={Dynamics of magnetization at infinite temperature in a {H}Eisenberg spin chain},
  author={Rosenberg, Eliott and Andersen, TI and Samajdar, Rhine and Petukhov, Andre and Hoke, JC and Abanin, Dmitry and Bengtsson, Andreas and Drozdov, IK and Erickson, Catherine and Klimov, PV and others},
  journal={Science},
  volume={384},
  number={6691},
  pages={48--53},
  year={2024},
  doi={10.1126/science.adi7877},
  publisher={American Association for the Advancement of Science}
}

@article{RoyKoenig2024,
  title = {${\mathbb{Z}}_{N}$ lattice gauge theories with matter fields},
  author = {Roy, Kaustubh and K\"onig, Elio J.},
  journal = {Phys. Rev. B},
  volume = {109},
  issue = {19},
  pages = {195108},
  numpages = {22},
  year = {2024},
  month = {May},
  publisher = {American Physical Society},
  doi = {10.1103/PhysRevB.109.195108},
  url = {https://link.aps.org/doi/10.1103/PhysRevB.109.195108}
}

@article{Jiang11,
  title = {Majorana Fermions in Equilibrium and in Driven Cold-Atom Quantum Wires},
  author = {Jiang, Liang and Kitagawa, Takuya and Alicea, Jason and Akhmerov, A. R. and Pekker, David and Refael, Gil and Cirac, J. Ignacio and Demler, Eugene and Lukin, Mikhail D. and Zoller, Peter},
  journal = {Phys. Rev. Lett.},
  volume = {106},
  issue = {22},
  pages = {220402},
  numpages = {4},
  year = {2011},
  month = {Jun},
  publisher = {American Physical Society},
  doi = {10.1103/PhysRevLett.106.220402},
  url = {https://link.aps.org/doi/10.1103/PhysRevLett.106.220402}
}

@article{Obuse13,
  title = {Bulk-boundary correspondence for chiral symmetric quantum walks},
  author = {Asb\'oth, J\'anos K. and Obuse, Hideaki},
  journal = {Phys. Rev. B},
  volume = {88},
  issue = {12},
  pages = {121406},
  numpages = {5},
  year = {2013},
  month = {Sep},
  publisher = {American Physical Society},
  doi = {10.1103/PhysRevB.88.121406},
  url = {https://link.aps.org/doi/10.1103/PhysRevB.88.121406}
}

@article{JinDeng2025,
  title={Topological prethermal strong zero modes on superconducting processors},
  author={Jin, Feitong and Jiang, Si and Zhu, Xuhao and Bao, Zehang and Shen, Fanhao and Wang, Ke and Zhu, Zitian and Xu, Shibo and Song, Zixuan and Chen, Jiachen and others},
  journal={Nature},
  volume={645},
  number={8081},
  pages={626--632},
  year={2025},
  publisher={Nature Publishing Group UK London}
}

@article{Delplace14,
  title = {Chiral symmetry and bulk-boundary correspondence in periodically driven one-dimensional systems},
  author = {Asb\'oth, J. K. and Tarasinski, B. and Delplace, P.},
  journal = {Phys. Rev. B},
  volume = {90},
  issue = {12},
  pages = {125143},
  numpages = {7},
  year = {2014},
  month = {Sep},
  publisher = {American Physical Society},
  doi = {10.1103/PhysRevB.90.125143},
  url = {https://link.aps.org/doi/10.1103/PhysRevB.90.125143}
}

@article{OkaRev,
title={Floquet Engineering of Quantum Materials},
   volume={10},
   ISSN={1947-5462},
   url={http://dx.doi.org/10.1146/annurev-conmatphys-031218-013423},
   DOI={10.1146/annurev-conmatphys-031218-013423},
   number={1},
   journal={Annual Review of Condensed Matter Physics},
   publisher={Annual Reviews},
   author={Oka, Takashi and Kitamura, Sota},
   year={2019},
   month=mar, pages={387–408} }

@article{Roy16,
  title = {Abelian Floquet symmetry-protected topological phases in one dimension},
  author = {Roy, Rahul and Harper, Fenner},
  journal = {Phys. Rev. B},
  volume = {94},
  issue = {12},
  pages = {125105},
  numpages = {12},
  year = {2016},
  month = {Sep},
  publisher = {American Physical Society},
  doi = {10.1103/PhysRevB.94.125105},
  url = {https://link.aps.org/doi/10.1103/PhysRevB.94.125105}
}

@article{Khemani16,
  title = {Phase Structure of Driven Quantum Systems},
  author = {Khemani, Vedika and Lazarides, Achilleas and Moessner, Roderich and Sondhi, S. L.},
  journal = {Phys. Rev. Lett.},
  volume = {116},
  issue = {25},
  pages = {250401},
  numpages = {6},
  year = {2016},
  month = {Jun},
  publisher = {American Physical Society},
  doi = {10.1103/PhysRevLett.116.250401},
  url = {https://link.aps.org/doi/10.1103/PhysRevLett.116.250401}
}

@article{SchmidvonOppen2024,
  title = {Robust Spectral $\ensuremath{\pi}$ Pairing in the Random-Field Floquet Quantum Ising Model},
  author = {Schmid, Harald and Penner, Alexander-Georg and Yang, Kang and Glazman, Leonid and von Oppen, Felix},
  journal = {Phys. Rev. Lett.},
  volume = {132},
  issue = {21},
  pages = {210401},
  numpages = {6},
  year = {2024},
  month = {May},
  publisher = {American Physical Society},
  doi = {10.1103/PhysRevLett.132.210401},
  url = {https://link.aps.org/doi/10.1103/PhysRevLett.132.210401}
}

@article{StepanovSangiovanni2024,
  title = {Interconnected renormalization of Hubbard bands and Green's function zeros in Mott insulators induced by strong magnetic fluctuations},
  author = {Stepanov, Evgeny A. and Chatzieleftheriou, Maria and Wagner, Niklas and Sangiovanni, Giorgio},
  journal = {Phys. Rev. B},
  volume = {110},
  issue = {16},
  pages = {L161106},
  numpages = {8},
  year = {2024},
  month = {Oct},
  publisher = {American Physical Society},
  doi = {10.1103/PhysRevB.110.L161106},
  url = {https://link.aps.org/doi/10.1103/PhysRevB.110.L161106}
}

@article{SonVedral2011,
  title={Quantum phase transition between cluster and antiferromagnetic states},
  author={Son, Wonmin and Amico, Luigi and Fazio, Rosario and Hamma, Alioscia and Pascazio, Saverio and Vedral, Vlatko},
  journal={Europhysics Letters},
  volume={95},
  number={5},
  pages={50001},
  year={2011},
  publisher={IOP Publishing}
}

@article{SettySi2024,
  title = {Electronic properties, correlated topology, and Green's function zeros},
  author = {Setty, Chandan and Xie, Fang and Sur, Shouvik and Chen, Lei and Vergniory, Maia G. and Si, Qimiao},
  journal = {Phys. Rev. Res.},
  volume = {6},
  issue = {3},
  pages = {033235},
  numpages = {15},
  year = {2024},
  month = {Sep},
  publisher = {American Physical Society},
  doi = {10.1103/PhysRevResearch.6.033235},
  url = {https://link.aps.org/doi/10.1103/PhysRevResearch.6.033235}
}

@article{SenthilVojta2003,
  title = {Fractionalized Fermi Liquids},
  author = {Senthil, T. and Sachdev, Subir and Vojta, Matthias},
  journal = {Phys. Rev. Lett.},
  volume = {90},
  issue = {21},
  pages = {2S6403},
  numpages = {4},
  year = {2003},
  month = {May},
  publisher = {American Physical Society},
  doi = {10.1103/PhysRevLett.90.216403},
  url = {https://link.aps.org/doi/10.1103/PhysRevLett.90.216403}
}

@article{SuMartin2024,
  title={Global anomalies of Green's function zeros},
  author={Su, Lei and Martin, Ivar},
  journal={arXiv preprint arXiv:2405.08093},
  year={2024}
}

@article{Yates19,
  title = {Almost strong ($0,\ensuremath{\pi}$) edge modes in clean interacting one-dimensional Floquet systems},
  author = {Yates, Daniel J. and Essler, Fabian H. L. and Mitra, Aditi},
  journal = {Phys. Rev. B},
  volume = {99},
  issue = {20},
  pages = {205419},
  numpages = {13},
  year = {2019},
  month = {May},
  publisher = {American Physical Society},
  doi = {10.1103/PhysRevB.99.205419},
  url = {https://link.aps.org/doi/10.1103/PhysRevB.99.205419}
}

@article{SimonMorice2025,
  title = {${\mathbb{Z}}_{2}$ topological invariants from Green function's diagonal zeros},
  author = {Simon, Florian and Morice, Corentin},
  journal = {Phys. Rev. B},
  volume = {112},
  issue = {19},
  pages = {195117},
  numpages = {10},
  year = {2025},
  month = {Nov},
  publisher = {American Physical Society},
  doi = {10.1103/9r5x-rj1x},
  url = {https://link.aps.org/doi/10.1103/9r5x-rj1x}
}

@article{ThakurathiDutta2013,
  title = {Floquet generation of Majorana end modes and topological invariants},
  author = {Thakurathi, Manisha and Patel, Aavishkar A. and Sen, Diptiman and Dutta, Amit},
  journal = {Phys. Rev. B},
  volume = {88},
  issue = {15},
  pages = {155133},
  numpages = {13},
  year = {2013},
  month = {Oct},
  publisher = {American Physical Society},
  doi = {10.1103/PhysRevB.88.155133},
  url = {https://link.aps.org/doi/10.1103/PhysRevB.88.155133}
}

@article{VolovikYakovenko1989,
  title={Fractional charge, spin and statistics of solitons in superfluid $^3${H}e film},
  author={Volovik, GE and Yakovenko, Victor M},
  journal={Journal of Physics: Condensed Matter},
  volume={1},
  number={31},
  pages={5263},
  year={1989},
  publisher={IOP Publishing},
  url={https://iopscience.iop.org/article/10.1088/0953-8984/1/31/025}
}

@article{WangYou2022,
  title={Symmetric mass generation},
  author={Wang, Juven and You, Yi-Zhuang},
  journal={Symmetry},
  volume={14},
  number={7},
  pages={1475},
  year={2022},
  publisher={MDPI}
}

@article{WangZhang2012,
  title = {Strongly correlated topological superconductors and topological phase transitions via Green's function},
  author = {Wang, Zhong and Zhang, Shou-Cheng},
  journal = {Phys. Rev. B},
  volume = {86},
  issue = {16},
  pages = {165116},
  numpages = {10},
  year = {2012},
  month = {Oct},
  publisher = {American Physical Society},
  doi = {10.1103/PhysRevB.86.165116},
  url = {https://link.aps.org/doi/10.1103/PhysRevB.86.165116}
}

@article{WagnerSangiovanni2023,
  title={Mott insulators with boundary zeros},
  author={Wagner, Niklas and Crippa, Lorenzo and Amaricci, Adriano and Hansmann, Philipp and Klett, Marcel and K{\"o}nig, Elio and Sch{\"a}fer, Thomas and Di Sante, Domenico and Cano, Jennifer and Millis, Andrew and others},
  journal={Nat Commun},
volume = {14}, 
pages = {7531},
url = {https://doi.org/10.1038/s41467-023-42773-7},
  year={2023}
}

@article{WagnerSangiovanni2024,
  title = {Edge Zeros and Boundary Spinons in Topological Mott Insulators},
  author = {Wagner, Niklas and Guerci, Daniele and Millis, Andrew J. and Sangiovanni, Giorgio},
  journal = {Phys. Rev. Lett.},
  volume = {133},
  issue = {12},
  pages = {126504},
  numpages = {7},
  year = {2024},
  month = {Sep},
  publisher = {American Physical Society},
  doi = {10.1103/PhysRevLett.133.126504},
  url = {https://link.aps.org/doi/10.1103/PhysRevLett.133.126504}
}

@article{YatesMitra2018,
  title = {Central Charge of Periodically Driven Critical Kitaev Chains},
  author = {Yates, Daniel and Lemonik, Yonah and Mitra, Aditi},
  journal = {Phys. Rev. Lett.},
  volume = {121},
  issue = {7},
  pages = {076802},
  numpages = {5},
  year = {2018},
  month = {Aug},
  publisher = {American Physical Society},
  doi = {10.1103/PhysRevLett.121.076802},
  url = {https://link.aps.org/doi/10.1103/PhysRevLett.121.076802}
}

@article{YehMitra2023,
  title = {Decay rates of almost strong modes in Floquet spin chains beyond Fermi's Golden Rule},
  author = {Yeh, Hsiu-Chung and Rosch, Achim and Mitra, Aditi},
  journal = {Phys. Rev. B},
  volume = {108},
  issue = {7},
  pages = {075112},
  numpages = {13},
  year = {2023},
  month = {Aug},
  publisher = {American Physical Society},
  doi = {10.1103/PhysRevB.108.075112},
  url = {https://link.aps.org/doi/10.1103/PhysRevB.108.075112}
}

@article{YouZhang2022,
  title = {Multiqubit Toffoli Gates and Optimal Geometry with Rydberg Atoms},
  author = {Yu, Dongmin and Wang, Han and Liu, Jin-Ming and Su, Shi-Lei and Qian, Jing and Zhang, Weiping},
  journal = {Phys. Rev. Appl.},
  volume = {18},
  issue = {3},
  pages = {034072},
  numpages = {15},
  year = {2022},
  month = {Sep},
  publisher = {American Physical Society},
  doi = {10.1103/PhysRevApplied.18.034072},
  url = {https://link.aps.org/doi/10.1103/PhysRevApplied.18.034072}
}

@article{YouXu2014,
  title = {Topological number and fermion Green's function for strongly interacting topological superconductors},
  author = {You, Yi-Zhuang and Wang, Zhong and Oon, Jeremy and Xu, Cenke},
  journal = {Phys. Rev. B},
  volume = {90},
  issue = {6},
  pages = {060502},
  numpages = {4},
  year = {2014},
  month = {Aug},
  publisher = {American Physical Society},
  doi = {10.1103/PhysRevB.90.060502},
  url = {https://link.aps.org/doi/10.1103/PhysRevB.90.060502}
}

@article{YouVishwanath2018,
  title = {Symmetric Fermion Mass Generation as Deconfined Quantum Criticality},
  author = {You, Yi-Zhuang and He, Yin-Chen and Xu, Cenke and Vishwanath, Ashvin},
  journal = {Phys. Rev. X},
  volume = {8},
  issue = {1},
  pages = {011026},
  numpages = {14},
  year = {2018},
  month = {Feb},
  publisher = {American Physical Society},
  doi = {10.1103/PhysRevX.8.011026},
  url = {https://link.aps.org/doi/10.1103/PhysRevX.8.011026}
}

@article{ZhaoPhillips2023,
  title = {Failure of Topological Invariants in Strongly Correlated Matter},
  author = {Zhao, Jinchao and Mai, Peizhi and Bradlyn, Barry and Phillips, Philip},
  journal = {Phys. Rev. Lett.},
  volume = {131},
  issue = {10},
  pages = {106601},
  numpages = {7},
  year = {2023},
  month = {Sep},
  publisher = {American Physical Society},
  doi = {10.1103/PhysRevLett.131.106601},
  url = {https://link.aps.org/doi/10.1103/PhysRevLett.131.106601}
}

@article{ZhangWang2022,
  title={Digital quantum simulation of {F}loquet symmetry-protected topological phases},
  author={Zhang, Xu and Jiang, Wenjie and Deng, Jinfeng and Wang, Ke and Chen, Jiachen and Zhang, Pengfei and Ren, Wenhui and Dong, Hang and Xu, Shibo and Gao, Yu and others},
  journal={Nature},
  volume={607},
  number={7919},
  pages={468--473},
  year={2022},
  doi={10.1038/s41586-022-04854-3},
  publisher={Nature Publishing Group UK London}
}

\end{document}